\documentclass[aps,prb,amsmath,amssymb,footinbib,showpacs,twocolumn,superscriptaddress]{revtex4-1}
\usepackage{amsmath}
\usepackage{amssymb}
\usepackage{amsthm}
\usepackage{setspace}
\usepackage{graphicx}
\usepackage{braket}
\usepackage{mathrsfs}
\usepackage{physics}
\usepackage{float}
\usepackage[colorlinks = true,linkcolor = red,urlcolor  = blue,citecolor = blue,anchorcolor = blue]{hyperref}
\usepackage[utf8]{inputenc}
\usepackage[english]{babel}
\usepackage{bm}
\usepackage{xcolor}
\usepackage{ulem}

\begin{document}
	
	\title{Physical mechanisms for zero-bias conductance peaks in Majorana nanowires}
	
	\author{Haining Pan}

	\author{S. Das Sarma}
	\affiliation{Condensed Matter Theory Center and Joint Quantum Institute, Department of Physics, University of Maryland, College Park, Maryland 20742, USA}

\begin{abstract}
	Motivated by the need to understand and simulate the ubiquitous experimentally-observed zero-bias conductance peaks in superconductor-semiconductor hybrid structures, we theoretically investigate the tunneling conductance spectra in one-dimensional nanowires in proximity to superconductors in a systematic manner taking into account several different physical mechanisms producing zero-bias conductance peaks. The mechanisms we consider are the presence of quantum dots, inhomogeneous potential, random disorder in the chemical potential, random fluctuations in the superconducting gap, and in the effective $ g $ factor with the self-energy renormalization induced by the parent superconductor in both short ($L \sim 1~ \mu $m) and long nanowires ($L\sim 3~ \mu $m). We classify all foregoing theoretical results for zero-bias conductance peaks into three types: the good, the bad, and the ugly, according to the physical mechanisms producing the zero-bias peaks and their topological properties. We find that, although the topological Majorana zero modes are immune to weak disorder, strong disorder (``ugly") completely suppresses topological superconductivity and generically leads to trivial zero bias peaks. Compared qualitatively with the extensive existing experimental results in the superconductor-semiconductor nanowire structures, we conclude that most current experiments are likely exploring trivial zero-bias peaks in the ``ugly" situation dominated by strong disorder. We also study the nonlocal end-to-end correlation measurement in both the short and long wires, and point out the limitation of the nonlocal correlation in ascertaining topological properties particularly when applied to short wires. Although we present results for ``good" and ``bad" zero bias peaks, arising respectively from topological Majorana bound states and trivial Andreev bound states, strictly for the sake of direct comparison with the ``ugly" zero bias conductance peaks arising from strong disorder, the main goal of the current work is to establish with a very high confidence level the real physical possibility that essentially all experimentally observed zero bias peaks in Majorana nanowires are most likely ugly, i.e.,purely induced by strong disorder, and are as such utterly nontopological. Our work clearly suggests that an essential prerequisite for any future observation of topological Majorana zero modes in nanowires is a substantial materials improvement of the semiconductor-superconductor hybrid systems leading to much cleaner wires. 
\end{abstract}

\date{\rm\today}
\maketitle

\section{introduction}\label{sec:introduction}
The experimental search for Majorana zero modes (MZM)~\cite{nayak2008nonabelian,sarma2015majorana,alicea2012new,elliott2015colloquium,stanescu2013majorana,leijnse2012introduction,beenakker2013search,lutchyn2018majorana,aguado2017majorana,lutchyn2010majorana,oreg2010helical,sau2010generic,jiang2013nonabelian,sato2017topological,sato2016majorana,plugge2017majorana,karzig2017scalable,wilczek2012quantum,sau2010nonabelian,cheng2009splitting,cheng2010tunneling,lutchyn2011search,potter2011majorana} in the superconductor-semiconductor (SC-SM) hybrid devices has succeeded in observing many of the theoretically predicted apparent topological features, especially the quantized zero-bias conductance peak (ZBCP) in the normal-to-superconductor (NS) tunneling spectroscopy~\cite{churchill2013superconductornanowire,das2012zerobias,finck2013anomalous,albrecht2016exponential,nichele2017scaling,zhang2018quantized,vaitiekenas2018effective,deng2016majorana,zhang2017ballistic,grivnin2019concomitant,moor2018electric,bommer2019spinorbit}. However, there are still some crucial topological features yet to be unambiguously confirmed in experiments; for example, the growing Majorana oscillations with the increasing magnetic field~\cite{dassarma2012splitting}, closing and reopening of bulk superconducting gaps~\cite{das2012zerobias,oreg2010helical,fu2008superconducting,sau2010generic,sau2010nonabelian}, and robust stability of ZBCP over extended regimes of magnetic fields and gate voltages~\cite{deng2012anomalous,gul2018ballistic}. In particular, extensive fine-tuning of various gate voltages applied across the sample appears necessary in the experimental manifestation of the rather fragile quantized tunneling ZBCPs putatively identified as arising from topological Majorana zero modes, seemingly in conflict with the predicted robust and generic nature of the topological phase.  Experiments manifest no signs of any nonlocal correlations, which are difficult to reconcile with the existence of MZMs.  In addition, the fact that no bulk signatures of a topological quantum phase transition (TQPT) (e.g., closing and then re-opening of a gap) have ever been reported in spite of widespread reporting of observed ZBCPs is problematic. All these problems have led to alternative non-topological explanations for the experimental ZBCPs~\cite{liu2012zerobias,bagrets2012class,pikulin2012zerovoltage,lee2012zerobias,akhmerov2009electrically,clarke2017experimentally,liu2017andreev,prada2012transport,lai2019presence,reeg2018zeroenergy,cayao2015sns,fleckenstein2018decayinga}, and there have been theoretical suggestions on how to identify MZMs experimentally as well as to distinguish between topological and trivial (i.e., non-topological) ZBCPs~\cite{akhmerov2009electrically,clarke2017experimentally,liu2017andreev,prada2012transport,lai2019presence}. Unfortunately, a consensus seems to have developed that most, if not all, of the observed ZBCPs are trivial, arising from fermionic (i.e., non-Majorana) subgap states, widely referred to as Andreev bound states (ABS), as opposed to Majorana bound states (MBS). A closely related possibility for trivial ZBCPs is the disorder-induced ABS or anti-localization enhancement of the density of states in class D system where the topological superconductivity is completely suppressed by disorder~\cite{motrunich2001griffiths,brouwer2011topological,brouwer2011probability,sau2013density,pikulin2012zerovoltage}. These alternative explanations focus on one explicit aspect of the system and subsequently modify or insert a specific term in the Hamiltonian introducing some elements of the relevant microscopic physics (e.g., the quantum dot or inhomogeneous potential or disorder, etc.) to explain the development of trivial ZBCPs. For strongly disordered multi-bands platforms~\cite{woods2019subband}, the ZBCP can be simulated and alternatively explained using the random-matrix method in the class D ensemble~\cite{mi2014xshaped,pikulin2012zerovoltage,brouwer1999distribution,dittes2000decay,altland1997nonstandard,beenakker1997randommatrix,guhr1998randommatrix,verbaarschot1985grassmann,pan2019generic}.

In the current work, we take a broad view within the single-subband 1D nanowire model for the SC-SM structure, considering all possibilities, both trivial and topological which produce ZBCPs, critically comparing the results in different situations in order to shed light on how to discern MZM-induced topological ZBCPs from the trivial ones.  We critically compare four distinct physical situations producing ZBCPs within one unified formalism keeping all the SC-SM parameters fixed (except for the specific mechanism leading to the ZBCP in each case), discuss similarities and differences between various cases, and comment on possible methods for distinguishing between trivial and topological phases as a matter of principle. Given the proliferation of many different proposed physical mechanisms leading to trivial ZBCPs in different contexts, it is important to compare them all under one uniform model to understand their relevance and properties. In addition, we carefully study the nonperturbative stability and robustness of pristine MZMs in nanowires to different types of disorder, which leads to the interesting conclusion that, while strong disorder by itself could produce trivial ZBCPs, pristine ZBCPs arising from topological MZMs, if they exist in the system, are immune to weak disorder. Immunity of topological MZMs to weak disorder and  complete suppression of topological superconductivity with the consequent generic appearance of trivial ZBCPs (some of which may be accidentally ``quantized") is the unfortunate dichotomy making it a difficult challenge to interpret the existing tunneling conductance data in SC-SM hybrid nanowires since a priori no quantitative information is available on whether the disorder in the currently existing samples is strong or weak compared with the topological superconducting gap (which has not yet been clearly seen in any experiment). 
 
We consider three distinct physical situations (i.e., ZBCP origins) with ZBCPs in the SC-SM hybrid nanowires, referred to as ``good"/``bad"/``ugly", which differ qualitatively in the way the ZBCPs arise in each case.  All the considered situations apply to the same physical system, namely, a 1D SM (InSb or InAs) nanowire with spin-orbit (SO) coupling in proximity to an ordinary metallic superconductor (Al) subjected to an external magnetic field, described by the same 1D Bogoliubov-de Gennes (BdG) Hamiltonian with the differences among the three cases arising from extra terms in the Hamiltonian representing either inhomogeneous chemical potential (``bad") with or without quantum dots or quenched disorder (``ugly").  The ``good" situation is pristine without these extra terms, and has been the standard model for studying topological superconductivity and Majorana modes in SC-SM structures ever since it was introduced in Refs.~\onlinecite{sau2010generic,sau2010nonabelian}.  The ``bad" situation is further subdivided into two physically distinct cases, depending on how the inhomogeneity in the chemical potential arises in the nanowire.  One bad situation arises from having an unintentional quantum dot in the system, which often happens near the wire end because of the complex materials science of creating the hybrid system~\cite{liu2018distinguishing}. The other bad situation arises from the presence of an inhomogeneous potential along the wire, arising presumably from the presence of charged impurities in the environment~\cite{kells2012nearzeroenergy}. These two bad situations are not qualitatively different as they both give rise to near-zero fermionic subgap states leading to trivial ZBCPs, but their physical origins are different and there are significant quantitative differences between the two, so that considering them separately is sensible. The ``ugly'' is the fluctuation in the nanowire due to the unintentional (mostly) charged impurities invariably present in the nanowire. Since the charge fluctuation is very sensitive to the gate voltages, temperature, subband occupancy, etc., it is intractable when multiple gate voltages are being tuned simultaneously. Thus, it can be treated as a random disorder in the 1D SC-SM nanowire. The random disorder also arises from disorder in the parent superconductor, the dielectric substrates, and the various leads and gates necessary to produce the hybrid system. The precise source of this strong disorder in the ugly case is unimportant for our considerations since we parametrize it simply as a random disorder as described below.
 
We construct the SC-SM nanowire Hamiltonian taking into account various aspects (see Fig.~\ref{fig:fig1}), including the pristine wire, quantum dot, inhomogeneous potential, and disorder in the chemical potential, in the SC gap, and in the effective $ g $ factor. We theoretically calculate the tunneling conductance spectra through an NS junction as a function of the Zeeman field $V_{\text{Z}} $ (magnetic field $ B $) by calculating the S matrix (Fig.~\ref{fig:fig1}). All numerical results are classified into three types: the ``good", the ``bad", and the ``ugly". The ``good" ZBCPs are the true topological MZMs which exist in the pristine nanowire (even with a small amount of disorder). The ``bad" ZBCPs are induced by the quantum dot or the inhomogeneous potential, where a trivial ZBCP emerges from the fermionic subgap ABS with X-shape anitcrossings~\cite{mi2014xshaped}. The ``ugly" ZBCPs arise from the large disorder, especially strong disorder in the chemical potential and the effective $ g $ factor. These kinds of disorder completely alter the pattern of the conductance spectra, suppressing superconductivity completely beyond a disorder-dependent finite magnetic field, leading to trivial and (sometimes) persistent ZBCPs. {We emphasize that ``bad" and ``ugly" mechanisms are completely and fundamentally different from each other as the former arises from a smoothly spatially varying \textit{deterministic} background potential because of the spatially inhomogeneous variation in the chemical potential whereas the latter arises from \textit{strong random} background disorder (which could be in the chemical potential, superconducting gap, or the effective $ g $ factor).  The background potential being deterministic (``bad") versus random (``ugly") makes a qualitative difference.  We emphasize that we also show that weak random disorder preserves the ``good" ZBCPs because of the topological immunity of the MZMs.  Weak versus strong disorder is basically defined by whether the variance fluctuation is larger than the average or not. The goal of this work is to present ZBCP results for all four situations within one unified formalism keeping the system parameters the same throughout so that the experimentalists can judge whether their results fall into one or the other category. Our conclusion is that all currently existing experimental ZBCPs are likely to be ugly ZBCPs induced by strong disorder.

Apart from the tunneling conductance results as a function of the Zeeman field, we additionally calculate the tunneling conductance spectrum at zero Zeeman field as a function of the chemical potential. At zero magnetic field, where everything observed inside the gap should be topologically trivial, the ``bad" and ``ugly" ZBCPs may still manifest fermionic subgap states. This observation of the fermionic subgap states would thus become an indicator of inhomogeneous chemical potential or strong disorder, and therefore, samples showing subgap states at zero field are unsuitable for MZM studies.

In addition, we also study the correlation measurement of tunneling conductance from both ends of the nanowire. In principle, this method, by virtue of the nonlocal nature of the topological MZM, can serve to distinguish Majorana bound states from trivial Andreev bound states in the ``bad" and ``ugly" situations, since the topological state will be correlated but the trivial one will not~\cite{lai2019presence}. However, this proposal will work only if the nanowire is sufficiently long. Therefore, by comparing the long nanowire results ($ L\sim 3 ~\mu$m) with the short ones  ($ L\sim 1 ~\mu$m), we show that in the short nanowire it is not feasible to distinguish between trivial and topological even utilizing the correlation measurement because the end-to-end correlation may be trivially manifested due to wave-function overlaps from the two ends. Unfortunately, most existing experiments seem to be in the ``short wire'' region. Note that ``long" ($ L\sim3~\mu$m) and ``short" ($L \sim1~\mu$m) refer specifically to the actual physical lengths of the InAs and InSb nanowires used in the current experiments--- since the coherence length is unknown in the topological regime, it is possible that all experimental systems so far are in the ``short" wire limit as the topological gap in the experimental nanowires has not been measured or even detected so far experimentally.  Since we use the nominal InAs and InSb parameters in our simulations, our working with a physical wire length is sensible.

We emphasize that the good (pristine MZM)  and bad (smoothly varying background potential) situations have already been discussed in some depth in the recent Majorana nanowire literature because of their perceived importance to experiments.  Most Majorana nanowire theories are obviously based on the pristine nanowire models where there is no background potential at all (neither smooth or nor random) as was originally done~\cite{sau2010generic}.  Very recent work has established rather convincingly that the presence of a smooth non-random background potential could lead generically to nontopological zero bias peaks arising from low-lying ABS, which eventually would become topological Majorana zero modes at high enough magnetic fields~\cite{liu2018distinguishing,huang2018metamorphosis,lai2019presence,vuik2018reproducing,stanescu2018illustrated}. These almost zero energy Andreev bound states can be construed as two spatially overlapping (and thus, only partially separated) MBS, which would eventually convert into isolated MBS at high enough magnetic field above the topological quantum phase transition point.  It has earlier been suggested that it is likely that most experimentally observed ZBCPs in nanowires are trivial and arise from the presence of smooth background potential, perhaps because of the existence of unintentional quantum dots at the wire ends~\cite{liu2017andreev,stanescu2018illustrated}.
	
Our current work differs from these earlier conclusions in the sense that our extensive simulations, as compared with the latest (and presumably the best) experimental ZBCPs presented in Refs.~\onlinecite{zhang2018quantized, nichele2017scaling}, presented in this paper strongly suggest that most experimentally observed ZBCPs, including even the ones claimed to manifest the expected Majorana conductance quantization, are in fact ``ugly" ZBCPs arising purely from strong random disorder present in the currently available semiconductor nanowire Majorana platforms.  Our claim of the experimental ZBCPs most likely being random disorder induced (and not induced by smoothly varying deterministic background potential) distinguishes our work from the earlier recent theoretical work, which attributes the experimental ZBCPs to be inhomogeneous potential induced (i.e., ``bad" in our terminology).  Our conclusion is based on a detailed comparison with the best available experimental results which appeared in the literature only very recently (2017 and 2018) as we will discuss in Sec.~\ref{sec:disscussion}.
	
The main purpose of our showing extensive ``good" and ``bad" results on the current paper is simply to compare with the ``ugly" results and draw a contrast among the three ZBCP mechanisms so that the veracity of our claim of most observed ZBCP being ``ugly" can be critically evaluated by the reader directly by looking at the results presented here using similar parameters and models with the only difference being ``bad" arises from a smooth deterministic background potential and ``ugly" arises from a background random disorder potential with ``good" being the usual pristine case.

The remainder of this paper is organized as follows. In Sec.~\ref{sec:theory}, we start with a pristine nanowire and modify each term according to the corresponding possible dominant physical mechanism (for the ``bad" and the ``ugly" cases) in SC-SM hybrid structures as schematically shown in Fig.~\ref{fig:fig1} and explicitly write down their Hamiltonians. In Sec.~\ref{sec:results}, we show the representative numerical results of the conductance spectra as a function of the Zeeman field for all three types as well as their correlation measurements from both ends. In addition, we present tunneling conductance spectra at zero magnetic field, where the pristine (i.e., ``good") system should not have fermionic subgap states, but the other two cases (i.e., ``bad" and ``ugly") may have fermionic subgap states. In Sec.~\ref{sec:disscussion}, we discuss the resemblance of our theoretical simulations to the current experimental results, and compare the conductance spectra for long nanowires with short nanowires. Our conclusion is presented in Sec.~\ref{sec:conclusion}. We present only limited representative numerical results in the main text, deferring our detailed results for the appendices. In Appendix~\ref{app:A}, we provide detailed correlation properties of the calculated ZBCPs for short and long wires both with and without the self-energy of the parent SC while the main text being devoted only to the presentation of the experimentally relevant tunneling conductance spectra (i.e., including the self-energy of the parent SC). Appendix~\ref{app:B} provides all the corresponding energy spectra and wave functions to help unequivocally determine the ABS or MBS in the theory. We emphasize that the vast amount of our simulation results presented in the appendices are integral parts of our theory, and are relegated to Appendix only to enable a seamless reading of the main text without being burdened by too many results.
\section{theory}\label{sec:theory}
The general form of the Hamiltonian to describe a SC-SM hybrid nanowire is~\cite{stanescu2013majorana}
\begin{equation}\label{eq:tot}
H_{\text{tot}}=H_{\text{SM}}+H_{\text{Z}}+H_{\text{V}}+H_{\text{SC}}+H_{\text{SC-SM}},
\end{equation}
where $ H_{\text{SM}} $ is the Hamiltonian for SM component, $ H_{\text{Z}} $ describes the contribution from the applied magnetic field (entering as the Zeeman splitting energy), $ H_{\text{V}} $ contains various effects of disorder and gate potentials, $H_{\text{SC}}  $ quantifies the parent SC, and $H_{\text{SC-SM}}  $ is the SC-SM coupling. This model has been studied extensively since its introduction in Refs.~\onlinecite{sau2010generic,sau2010nonabelian}, but usually with some of the terms (e.g., $ H_{\text{V}} $) left out to emphasize one or other physical mechanisms.  In the current work, we keep all the terms to study and contrast the different situations within one comprehensive framework.
\subsection{Minimal effective model}
We start with the minimal effective Hamiltonian of a pristine nanowire without any quantum dot, inhomogeneous potential, or disorder, which implies $H_{\text{V}}=0 $. (This, by definition, corresponds to the ``good" case where isolated topological MZMs arise at two wire ends for sufficiently large Zeeman splitting and sufficiently long wires, i.e., above the TQPT.) The pristine nanowire is then described by the ``standard" minimal BdG Hamiltonian~\cite{sau2010generic,lutchyn2010majorana,oreg2010helical}
$\hat{H}=\frac{1}{2}\int dx ~\hat{\Psi}^\dagger(x)H_{\text{tot}}\hat{\Psi}(x) $
, with 
\begin{equation}\label{eq:bdg-pristine}
H_{\text{tot}} = \left( -\frac{\hbar^2}{2m^*} \partial^2_x -i \alpha \partial_x \sigma_y - \mu \right)\tau_z + V_{\text{Z}}\sigma_x + \Delta \tau_x.
\end{equation} 
Here, $\hat{\Psi}(x)=\left(\hat{\psi}_{\uparrow}(x),\hat{\psi}_{\downarrow}(x),\hat{\psi}_{\downarrow}^\dagger(x),-\hat{\psi}_{\uparrow}^\dagger(x)\right)^T$ represents a position-dependent spinor;  $\vec{\bm{\sigma}}$ and $\vec{\bm{\tau}}$ denote Pauli matrices in the spin and particle-hole space, respectively. The magnetic field is applied along the longitudinal direction of the nanowire providing a Zeeman term $H_{\text{Z}}=V_{\text{Z}}\sigma_x$, where $ V_{\text{Z}}=\frac{1}{2}g \mu_B B $ and $ \mu_B $ is Bohr magneton. Rashba spin-orbit coupling with strength $ \alpha $ is assumed to be perpendicular to the wire length~\cite{bychkov1984oscillatory}. We emphasize the pristine nanowire aspect by imposing a spatially constant chemical potential $ \mu $ with an effective $ g $ factor and a SC proximitized gap $ \Delta $ in the weak SC-SM coupling limit~\cite{stanescu2010proximity,sau2010robustness}. Thus, $ H_{\text{SC-SM}} $ here is given simply by the last term $ \Delta\tau_x $ in Eq.~\eqref{eq:bdg-pristine}. Unless otherwise specified, the values of effective parameters in Eq.~\eqref{eq:bdg-pristine} are~\cite{lutchyn2018majorana,gul2018ballistic,kammhuber2016conductance,kammhuber2017conductance,zhang2018quantized,chen2019ubiquitous,woods2019zeroenergy}  $m^*=0.015~m_e$ (for the effective mass), where $ m_e $ is the electron rest mass,  $\Delta=0.2 $ meV (for the proximity-induced SC gap), $ \mu=1 $ meV (for the chemical potential), $ \alpha=0.5 $~eV\AA{} (for the spin-orbit coupling), and the length of the nanowire $ L=1~\mu $m~\cite{grivnin2019concomitant,vaitiekenas2018effective,moor2018electric,nichele2017scaling} (for the short wire) or $ 3~\mu $m (for the long wire). (This choice of parameters corresponds approximately to the InSb-Al hybrid SC-SM systems.) We calculate all the energy spectra numerically by discretizing the continuum Hamiltonian into a finite difference tight-binding model~\cite{dassarma2016how} and then exactly diagonalizing the corresponding Hamiltonian matrix. The tight-binding model is diagonalized for different values of $ V_{\text{Z}} $ to obtain the corresponding eigenvalues and eigenvectors utilizing Arnoldi iteration technique~\cite{arnoldi1951principle} for sparse matrices (except for the Hamiltonian in the presence of the self-energy discussed next). The schematic of a pristine nanowire (``good") model is shown in Fig.~\ref{fig:fig1}(a).

\subsection{Self-energy}
Under real experimental conditions, the weak SC-SM coupling limit, i.e., $ H_{\text{SC-SM}}=\Delta \tau_x $ as in Eq.~\eqref{eq:bdg-pristine}, may not be sufficient to describe the system, especially for those involving epitaxial aluminum (Al) as the parent SC~\cite{chang2015hard,das2012zerobias,krogstrup2015epitaxy} Therefore, we consider the proximity effect in an intermediate regime within a Green's function approach.~\cite{stanescu2010proximity,sau2010robustness,sau2010nonabelian}. Note that the SC proximity effect is due to the electrons in the SM nanowire penetrating into the covering parent SC segment and vice versa. To approximate this effect, one can construct a microscopic tight-binding model between the SC and the SM, integrate out the SC degrees of freedom, and replace the parent SC by a self-energy~\cite{stanescu2010proximity,sau2010robustness,reeg2017transport,liu2017andreev,stanescu2013majorana,stanescu2017proximityinduced,stanescu2013majorana}
\begin{equation}\label{eq:selfenergy}
\Sigma(\omega)=-\gamma\frac{\omega+\Delta_0\tau_x}{\sqrt{\Delta_0^2-\omega^2}},
\end{equation}
 where $ \gamma $ is the effective SC-SM coupling (tunneling) strength, $ \omega $ is the energy, and $ \Delta_0 $ is the bulk parent SC gap. Unless otherwise specified, these values of parameters are used throughout: $ \gamma=0.2 $ meV and $ \Delta_0=0.2 $ meV~\cite{stanescu2017proximityinduced}. Explicitly, the Hamiltonian then becomes energy-dependent including the self-energy
\begin{equation}\label{eq:SE}
H_{\text{SE}}(\omega)=\left( -\frac{\hbar^2}{2m^*} \partial^2_x -i \alpha \partial_x \sigma_y - \mu \right)\tau_z + V_{\text{Z}}\sigma_x +\Sigma(\omega).
\end{equation}
Since the Hamiltonian is $ \omega $-dependent in the presence of the self-energy, it can be solved self-consistently in an iterative manner for each energy state~\cite{pan2019curvature}. Note that the self-energy term $ \Sigma(\omega) $ in Eq.~\eqref{eq:SE} represents the coupling term $ H_{\text{SC-SM}} $ of Eq.~\eqref{eq:tot}.

One of the practical problems encountered in experiments is that the bulk SC gap of the parent superconductor is suppressed by the applied magnetic field, and often in fact vanishes for sufficiently large Zeeman field~\cite{lee2012zerobias}. To better simulate the real experimental situation~\cite{deng2016majorana,zhang2018quantized,zhang2017ballistic,bommer2019spinorbit,moor2018electric,vaitiekenas2018effective,grivnin2019concomitant,nichele2017scaling}, we therefore further consider a $ V_{\text{Z}} $-dependent bulk SC gap, where it collapses at some experimentally determined non-universal
 $ V_{\text{C}} $, namely, the constant $ \Delta_0 $ in Eq.~\eqref{eq:selfenergy} is then replaced by~\cite{liu2017andreev}
\begin{equation}\label{eq:gapcollapse}
\Delta_0(V_{\text{Z}})=\Delta_0(V_{\text{Z}}=0)\sqrt{1-\left(\frac{V_{\text{Z}}}{V_{\text{C}}}\right)^2}\theta(V_{\text{C}}-V_{\text{Z}}),
\end{equation} 
where $ \theta(\dots) $ is the Heaviside-step function indicating that the SC gap will never reopen once it has collapsed since the parent bulk SC gap has vanished causing a complete disappearance of the proximity effect in the SM. As such regimes of gap collapsing $ V_{\text{Z}} $ are not of our interest, we do not extend Zeeman field $ V_{\text{Z}} $ in the numerical calculated tunneling conductance spectra beyond the SC collapse field $ V_{\text{C}} $ (i.e., the theory throughout the paper is only discussed within $ V_{\text{Z}}<V_{\text{C}} $, and should not be applied to the regime of $ V_{\text{Z}}>V_{\text{C}} $). Thus the Hamiltonian with the self-energy Eq.~\eqref{eq:selfenergy} then becomes
\begin{equation}\label{eq:SE2}
\tilde{H}_{\text{SE}}(\omega)=\left( -\frac{\hbar^2}{2m^*} \partial^2_x -i \alpha \partial_x \sigma_y - \mu \right)\tau_z + V_{\text{Z}}\sigma_x +\Sigma(\omega,V_{\text{Z}}),
\end{equation}
where 
\begin{equation}\label{eq:SE3}
\Sigma(\omega,V_{\text{Z}})=-\gamma\frac{\omega+\Delta_0(V_{\text{Z}})\tau_x}{\sqrt{\Delta_0^2(V_{\text{Z}})-\omega^2}}.
\end{equation}
Equation~\eqref{eq:SE2} along with Eqs.~\eqref{eq:SE3} and \eqref{eq:gapcollapse} are the Hamiltonian to produce most of the numerical results in the main text. In essence, the reason for including the self-energy with the bulk $ V_{\text{Z}} $-dependent SC gap collapse is to introduce the renormalization effects by the parent SC. The functional form of the SC gap collapse Eq.~\eqref{eq:gapcollapse} is chosen merely because it phenomenologically simulates well the real experimental situation -- any other smooth form of the parent SC gap collapse does not change any aspect of our results or conclusions. 
 
 We will show results both with and without the self-energy term to distinguish weak- and intermediate-coupling SC-SM systems in Appendix~\ref{app:A}. In the main text, only results with the self-energy are presented since the self-energy effect is crucial under real experimental conditions. (Note that we call the results with self-energy ``intermediate-coupling" rather than ``strong-coupling" since the strongly coupled SC-SM represents the situation where the SC completely overwhelms the SM nanowire, leading to very unfavorable conditions for the creation of MZMs--  weak-coupling and intermedate-coupling situations, without and with the self-energy respectively, are the experimentally relevant situations.)

\subsection{Quantum dot}
The previous Hamiltonian in Eq.~\eqref{eq:bdg-pristine} describes a pristine ``good" nanowire without any disorder, i.e., $ H_{\text{V}}=0 $. However, the presence of an unintentional quantum dot at the end of the nanowire may be inevitable under real experimental conditions due to the mismatch of Fermi energy between the normal lead and the SM nanowire by creating a Schottky barrier~\cite{liu2017andreev,moore2018twoterminal}. Therefore, although the quantum dot may not be intentionally introduced in experiments, it is expected to be quite ubiquitous in many SC-SM nanowire experimental setups~\cite{deng2016majorana,zhang2018quantized,chen2019ubiquitous,nichele2017scaling}. Theoretically, the ``quantum dot" is a potential fluctuation at the end of the nanowire which is a short segment uncovered by the parent SC. Since it is a zero-dimensional object, the quantum dot usually appears at the contact point connecting the SM nanowire to the lead. Thus, the quantum dot will play a role in $ H_{\text{V}}=V(x) $, where $ V(x) $ is simply chosen as a Gaussian barrier. Namely, the quantum dot potential is given by
\begin{equation}\label{eq:qdpot}
V(x)=V_{\text{D}} \exp\left(-\frac{x^2}{l^2}\right)\theta(l-x),
\end{equation} 
where $ V_{\text{D}} $ defines the peak of the dot barrier and $ l $ is the length of the quantum dot. Here $ V_{\text{D}} $ and $ l $ are the parameters modeling the quantum dot. By intensive numerical calculations, we ensure that the specific form of the quantum dot potential does not qualitatively modify the results~\cite{liu2017andreev,moore2018quantized}. Consequently, the BdG Hamiltonian of SC-SM nanowire with a quantum dot then becomes
\begin{eqnarray}\label{eq:qd}
H_{\text{QD}}&=&\left( -\frac{\hbar^2}{2m^*} \partial^2_x -i \alpha \partial_x \sigma_y - \mu+V_{\text{D}} e^{-\frac{x^2}{l^2}}\theta(l-x) \right)\tau_z\nonumber\\
 &+& V_{\text{Z}}\sigma_x + \Delta\theta(x-l) \tau_x,
\end{eqnarray}
where $ \theta(x-l) $ is included to account for the partially covering parent SC (i.e., the SC is absent over a length $ l $ at the end of the nanowire). For the same reason, we incorporate the self-energy Eq.~\eqref{eq:selfenergy} as well for the finer simulation of experimental results in the presence of the quantum dot. Thus, the Hamiltonian in the presence of the quantum dot and the self-energy then becomes
\begin{eqnarray}\label{eq:SEQD}
H_{\text{QD,SE}}(\omega)&=&\left( -\frac{\hbar^2}{2m^*} \partial^2_x -i \alpha \partial_x \sigma_y - \mu+V_{\text{D}} e^{-\frac{x^2}{l^2}}\theta(l-x)  \right)\tau_z\nonumber\\
& +& V_{\text{Z}}\sigma_x -\gamma\frac{\omega+\Delta_0(V_{\text{Z}})\tau_x}{\sqrt{\Delta_0^2(V_{\text{Z}})-\omega^2}}\theta(x-l).
\end{eqnarray}
The schematic of the nanowire with a quantum dot is shown in Fig.~\ref{fig:fig1}(b). This is one of our ``bad" situations, with the possibility of ZBCPs arising from the quantum dot. The second ``bad" situation with an inhomogeneous potential along the whole wire, in contrast to a potential fluctuation just at the end, is discussed below.

\subsection{Inhomogeneous potential}
The inhomogeneous potential is an alternative mechanism producing ZBCP in the topologically-trivial regime~\cite{kells2012nearzeroenergy,stanescu2018illustrated,moore2018twoterminal,degottardi2013majoranaa,rainis2013realistic,stanescu2014nonlocality}. This is the second type of ``bad" situation we consider. To be specific, the inhomogeneous potential is a smooth confining potential in the SM due to charged impurities in the environment or the gate voltage~\cite{mourik2012signatures,das2012zerobias,deng2016majorana,zhang2018quantized,chen2019ubiquitous}. In the theoretical model, we use, similar to the quantum dot case above, a Gaussian smooth confining potential~\cite{kells2012nearzeroenergy,stanescu2018illustrated}
\begin{equation}\label{eq:inhompot}
V(x)=V_{\text{max}}\exp\left(-\frac{x^2}{2\sigma^2}\right),
\end{equation}
where $ V_{\text{max}} $ defines the height of confining potential and $ \sigma $ controls the linewidth of the inhomogeneous potential. Therefore, the BdG Hamiltonian of the nanowire with an inhomogeneous potential is
\begin{eqnarray}\label{eq:inhom}
H_{\text{inhom}}&=&\left( -\frac{\hbar^2}{2m^*} \partial^2_x -i \alpha \partial_x \sigma_y - \mu+V_{\text{max}}e^{-\frac{x^2}{2\sigma^2}} \right)\tau_z\nonumber\\
& +& V_{\text{Z}}\sigma_x + \Delta \tau_x.
\end{eqnarray}
We note that both types of ``bad" situations, the quantum dot and inhomogeneous potential, can be construed to produce an effective spatially varying chemical potential $ \mu-V(x) $ in the BdG equations defined by Eqs.~\eqref{eq:qd} and \eqref{eq:inhom} respectively, with the only difference between the two ``bad" cases being the way inhomogeneous potential in $ V(x) $ arises. A slight difference in the theoretical model between the quantum dot and the inhomogeneous potential case lies in the spatial extent of the parent SC segment covering the nanowire. Unlike the quantum dot case, the parent SC fully covers the SM nanowire in the ``bad" situation of the inhomogeneous potential. We may also incorporate the self-energy Eq.~\eqref{eq:selfenergy} here, and the Hamiltonian then becomes
\begin{eqnarray}\label{eq:seinhom}
H_{\text{inhom,SE}}(\omega)&=&\left( -\frac{\hbar^2}{2m^*} \partial^2_x -i \alpha \partial_x \sigma_y - \mu+V_{\text{max}}e^{-\frac{x^2}{2\sigma^2}}\right)\tau_z \nonumber\\
&+& V_{\text{Z}}\sigma_x  -\gamma\frac{\omega+\Delta_0(V_{\text{Z}})\tau_x}{\sqrt{\Delta_0^2(V_{\text{Z}})-\omega^2}}.
\end{eqnarray}
The schematic of the nanowire with the inhomogeneous potential is shown in Fig.~\ref{fig:fig1}(c). This is our second type of the ``bad" situation.
\subsection{Disorder}
There are two completely distinct aspects of disorder we study in our work.  We show that the pristine MZM-induced topological ZBCPs, if they exist in the system, are to a large extent immune to the effects of disorder by virtue of their topological robustness.  Thus, the ``good" ZBCPs are robust to disorder effects.  By contrast, disorder by itself can produce trivial ZBCPs, which mimic MZM-induced ZBCPs, complicating the interpretation of experimentally-observed ZBCPs.

Under real experimental conditions, unintentional disorder is unavoidable, and therefore, disorder may also play an important role in the emergence of topologically-trivial ZBCP~\cite{sau2013density,lin2012zerobias,brouwer2011topological,stanescu2014nonlocality,liu2012zerobias,neven2013quasiclassical,pikulin2012zerovoltage,sau2013bound,bagrets2012class,degottardi2013majorana,adagideli2014effects,roy2013topologically,hui2015bulk}. In essence, the superconducting nanowire which hosts the Majorana modes acts like an effective $ p $-wave SC~\cite{kopnin1991mutual,rice1995sr2ruo4,dassarma2006proposal} which is not necessarily immune to nonmagnetic disorder~\cite{motrunich2001griffiths,potter2011engineering}. We first introduce disoder in the chemical potential as  $ V_{\text{imp}}(x) $ in Eq.~\eqref{eq:tot}~\cite{liu2012zerobias}, i.e., $ H_{\text{V}}=V_{\text{imp}}(x) $.  $ V_{\text{imp}}(x) $ is a random potential represented by an uncorrelated Gaussian distribution with zero mean value and standard deviation $ \sigma_{\mu} $, i.e., $ V_{\text{imp}}(x)\sim\mathcal{N}(0,\sigma_{\mu}^2) $, where $ \mathcal{N}(\mu,\sigma^2) $ denotes a Gaussian distribution with mean value of $ \mu $ and variance of $ \sigma^2 $. We clarify that the impurity potential is randomly generated and the results in Sec.~\ref{sec:results} are shown for a specific configuration of randomness without averaging over disorder. Thus, the Hamiltonian Eq.~\eqref{eq:bdg-pristine} then becomes
\begin{eqnarray}\label{eq:muVar}
H_{\text{disorder},\mu}&=&\left( -\frac{\hbar^2}{2m^*} \partial^2_x -i \alpha \partial_x \sigma_y - \mu+V_{\text{imp}}(x) \right)\tau_z \nonumber\\
&+& V_{\text{Z}}\sigma_x + \Delta \tau_x,
\end{eqnarray}
and the Hamiltonian in the presence of the self-energy Eq.~\eqref{eq:selfenergy} then becomes
\begin{eqnarray}\label{eq:semuVar}
H_{\text{disorder,}\mu\text{,SE}}(\omega)&=&\left( -\frac{\hbar^2}{2m^*} \partial^2_x -i \alpha \partial_x \sigma_y - \mu+V_{\text{imp}}(x) \right)\tau_z\nonumber\\
&+& V_{\text{Z}}\sigma_x  -\gamma\frac{\omega+\Delta_0(V_{\text{Z}})\tau_x}{\sqrt{\Delta_0^2(V_{\text{Z}})-\omega^2}}.
\end{eqnarray}
The schematic of disorder in the chemical potential is shown in Fig.~\ref{fig:fig1}(d). This is the ``ugly" situation in the presence of a large amount of disorder. Here we can think of the chemical potential itself having random spatial disorder with the effective random chemical potential being $ \mu - V_{\text{imp}}(x) $.

For completeness, we additionally introduce disorder in the effective $ g $ factor and the SC gap in our theoretical model. Since the Zeeman field is related to the applied magnetic field and the definite value of $ g $ in experiments is unknown~\cite{pan2019curvature}, we avoid directly handling the random $ g $ factor by transferring its randomness to $ V_{\text{Z}} $. Thus we define a dimensionless factor $ \tilde{g}(x)={g(x)}/{\bar{g}} $, where $ g(x) $ is the random $ g$ factor and $ \bar{g} $ stands for its mean value. Since $ V_{\text{Z}} $ is linearly proportional to $ g $, $ \tilde{g}(x) $  also equals $ {V_{\text{Z}}(x)}/\bar{V}_Z $. We randomize $ \tilde{g}(x) $ in the form of Gaussian distribution $ \mathcal{N}(1,\sigma_{g}^2) $ as before. Note that, to avoid the possibility of a physically meaningless negative $ g $ factor, the standard deviation $ \sigma_{g} $ cannot be set too large. With the random $ V_{\text{Z}}(x)=\tilde{g}(x)\bar{V}_Z $, the Hamiltonian Eq.~\eqref{eq:bdg-pristine} becomes
\begin{equation}\label{eq:gVar}
H_{\text{disorder,}g}=\left( -\frac{\hbar^2}{2m^*} \partial^2_x -i \alpha \partial_x \sigma_y - \mu \right)\tau_z \nonumber+ V_{\text{Z}}(x)\sigma_x + \Delta \tau_x
\end{equation}
and the Hamiltonian with the self-energy Eq.~\eqref{eq:SE} then becomes
\begin{eqnarray}\label{eq:segVar}
H_{\text{disorder,}g\text{,SE}}(\omega)&=&\left( -\frac{\hbar^2}{2m^*} \partial^2_x -i \alpha \partial_x \sigma_y - \mu \right)\tau_z\nonumber\\
&+& V_{\text{Z}}(x)\sigma_x  -\gamma\frac{\omega+\Delta_0(V_{\text{Z}})\tau_x}{\sqrt{\Delta_0^2(V_{\text{Z}})-\omega^2}}.
\end{eqnarray}
The schematic of disorder in the effective $ g $ factor is shown in Fig.~\ref{fig:fig1}(e). This type of Zeeman disorder in the effective $ g $ factor is the second mechanism leading to create ``ugly" ZBCPs. 

The last type of disorder we consider is in the SC gap. It can be defined as $ \Delta(x)\sim\mathcal{N}(\Delta,\sigma_{\Delta}^2) $ in Eq.~\eqref{eq:bdg-pristine} without the self-energy or $ \Delta_0(x)\sim\mathcal{N}(\Delta_0,\sigma_{\Delta_0}^2)  $ in Eq.~\eqref{eq:SE} with the self-energy. Again, to avoid any unphysical negative SC gap, the standard deviation should not be too large.  Thus the Hamiltonian~\eqref{eq:bdg-pristine} then becomes
\begin{equation}\label{eq:DeltaVar}
H_{\text{disorder},\Delta}=\left( -\frac{\hbar^2}{2m^*} \partial^2_x -i \alpha \partial_x \sigma_y - \mu \right)\tau_z \nonumber+ V_{\text{Z}}\sigma_x + \Delta(x) \tau_x
\end{equation}
and the Hamiltonian utilizing the self-energy term~\eqref{eq:selfenergy} becomes
\begin{eqnarray}\label{eq:seDeltaVar}
H_{\text{disorder,}\Delta_0\text{,SE}}(\omega)&=&\left( -\frac{\hbar^2}{2m^*} \partial^2_x -i \alpha \partial_x \sigma_y - \mu \right)\tau_z\nonumber\\
&+& V_{\text{Z}}\sigma_x -\gamma\frac{\omega+\Delta_0(x;V_{\text{Z}})\tau_x}{\sqrt{\Delta_0^2(x;V_{\text{Z}})-\omega^2}}.
\end{eqnarray}
The schematic of disorder in the SC gap is shown in Fig.~\ref{fig:fig1}(f). In Sec.~\ref{sec:results}, we will show that neither the topological MBS-induced ZBCP is destroyed due to this gap disorder nor any trivial ABS-induced ZBCP is created in the presence of disorder in the SC gap. Thus this is another subcategory of the ``good" ZBCPs in contrast to both chemical potential and Zeeman disorder which lead to ugly ZBCPs. Although the topological MZMs are protected against some gap disorder~\cite{lutchyn2012momentum}, a very large gap disorder obviously destroys the MZMs since it suppresses the topological superconductivity itself~\cite{sau2013density,hui2015bulk}.

\subsection{Spatial wave function}
Since all of our foregoing models, after discretization, are based on the tight-binding approximation, we can obtain the wave functions straightforwardly by diagonalizing the BdG Hamiltonian. Specifically, for an $ N $-site system, the dimension of the Hamiltonian is $ 4N$-by-$ 4N $ where the factors of 4 are due to the Nambu spinor basis $ \left(c_\uparrow,c_\downarrow,c_\downarrow^\dagger,-c_\uparrow^\dagger\right)^T $. Therefore the corresponding component on the same site should be summed up to obtain an $ N $-component wave function $ \abs{\psi(x_i)}^2 $. The trivial ABSs are distinct from the topological MBSs in the spatial separation between the localized bound states in the nanowire, where ABS in the topologically-trivial regime are two highly-overlapping (or only partially-separated) Majorana modes at one end of the nanowire; MBS in the topological regime are two well-separated Majorana modes at both ends of the  nanowire~\cite{stanescu2018illustrated,vuik2018reproducing,huang2018metamorphosis}. Thus, the trivial ABSs here are in fact quasi-MZMs except that the localized modes overlap too strongly for them to be considered in isolation. Therefore, to identify the ABS-induced ZBCP, we need to convert the wave function from the particle-hole basis to the Majorana basis. To properly address the self-antiparticle property of Majorana fermion, the creation (equivalently, annihilation) operator $ \gamma $ can be considered as one half of the ``regular" fermion. Thus we combine two Majorana zero modes to form one well-defined regular fermion~\cite{kitaev2001unpaired,alicea2012new}, i.e.,
\begin{eqnarray}\label{eq:majorana}
  c&=&(\gamma_1+i\gamma_2)/2 \nonumber\\
  c^\dagger&=&(\gamma_1-i\gamma_2)/2.
\end{eqnarray}
Thus the two wave functions in the Majorana basis can be represented by the spatial wave function $ \psi(x) $ as~\cite{stanescu2018illustrated}
\begin{eqnarray}\label{eq:majoranabasis}
\phi_1(x)&=&\frac{1}{\sqrt{2}}\left(\psi_\epsilon(x)+\psi_{-\epsilon}(x)\right),\nonumber\\
\phi_2(x)&=&\frac{i}{\sqrt{2}}\left(\psi_\epsilon(x)-\psi_{-\epsilon}(x)\right),
\end{eqnarray}
where $ \psi_\epsilon(x) $ ($ \psi_{-\epsilon}(x) $) is the wave function in the Nambu spinor basis with energy $ \epsilon  $ ($ -\epsilon $) in the electron (hole) channel and $ \phi_{1,2}(x) $ are two wave functions in the Majorana basis in the same energy level. Note that $ \phi_{1,2}(x)$ are generally not the eigenstates of Hamiltonian Eq.~\eqref{eq:tot} unless $ \epsilon=0 $ when they in fact represent the zero-energy Majorana mode.  The quasi-Majorana ABSs have small, but finite energy, which is close to zero, but not exactly zero. The detailed results of wave function calculations in the Majorana basis are presented in Appendix~\ref{app:B}. The energy spectra (eigenvalues), which are also shown in Appendix~\ref{app:B}, can be obtained along with the corresponding wave function (eigenvector) when the BdG Hamiltonian at each Zeeman field is diagonalized.

\subsection{Differential conductance spectrum}
To simulate the experimental measurement of tunneling conductance $ G=dI/dV $~\cite{deng2016majorana,zhang2018quantized,zhang2017ballistic,bommer2019spinorbit,moor2018electric,vaitiekenas2018effective,grivnin2019concomitant,chen2019ubiquitous}, we attach a normal lead to the end of the nanowire and numerically calculate the tunneling conductance through the NS junction using the S matrix method. The normal lead has the same Hamiltonian as the SC-SM nanowire except for the absent SC term, i.e., 
\begin{equation}\label{eq:lead}
H_{\text{lead}}=\left( -\frac{\hbar^2}{2m^*} \partial^2_x -i \alpha \partial_x \sigma_y - \mu+E_{\text{lead}}\right)\tau_z + V_{\text{Z}}\sigma_x
\end{equation}
where $ E_\text{lead}\sim-25 $ meV is an additional on-site energy in the lead controlled by the voltage of the tunnel gate~\cite{liu2017andreev}. The tunneling barrier $ Z\sim 10 $ meV is added on the first site in the nanowire at the interface of the NS junction.~\cite{setiawan2017electron} We use KWANT to compute the S matrix~\cite{groth2014kwant}. Since the calculation technique is well-established, we refer the reader to existing references for technical details~\cite{dassarma2016how,liu2017role,groth2014kwant,blonder1982transition,anantram1996current,setiawan2015conductance,rainis2013realistic,stanescu2011majorana,prada2012transport,stanescu2014soft,roy2012majorana,qu2011signature,roy2013nature,rosdahl2018andreev,dumitrescu2015majorana,stenger2017tunneling,buttiker1992scattering}. The schematic for the simulated model is in Fig.~\ref{fig:fig1} under six distinct aforementioned situations (from (a) to (f)): the pristine nanowire, the nanowire in the presence of the quantum dot, the nanowire in the presence of the inhomogeneous potential, the nanowire in the presence of disorder in the chemical potential, the nanowire in the presence of disorder in the effective $ g $ factor, and the nanowire in the presence of disorder in the SC gap.

We insert a set of discrete $ V_{\text{Z}} $ into the Hamiltonian and calculate the differential conductance as a function of $ V_{\text{bias}} $ from -0.3 to 0.3 mV. The conductance varies between $ G=0 $ and $ 4e^2/h $ because of two spin channels in general~\cite{setiawan2015conductance}. We present two-dimensional color plots, where the two axes are $ V_{\text{Z}} $ and $ V_{\text{bias}} $, to visualize the pattern of conductance spectra, with red indicating quantized conductance $ 2e^2/h $ and blue indicating zero conductance. The numerical results for the tunneling conductance are presented in Sec.~\ref{sec:results}.

\begin{figure}[htbp]
	\includegraphics[width=3.4in]{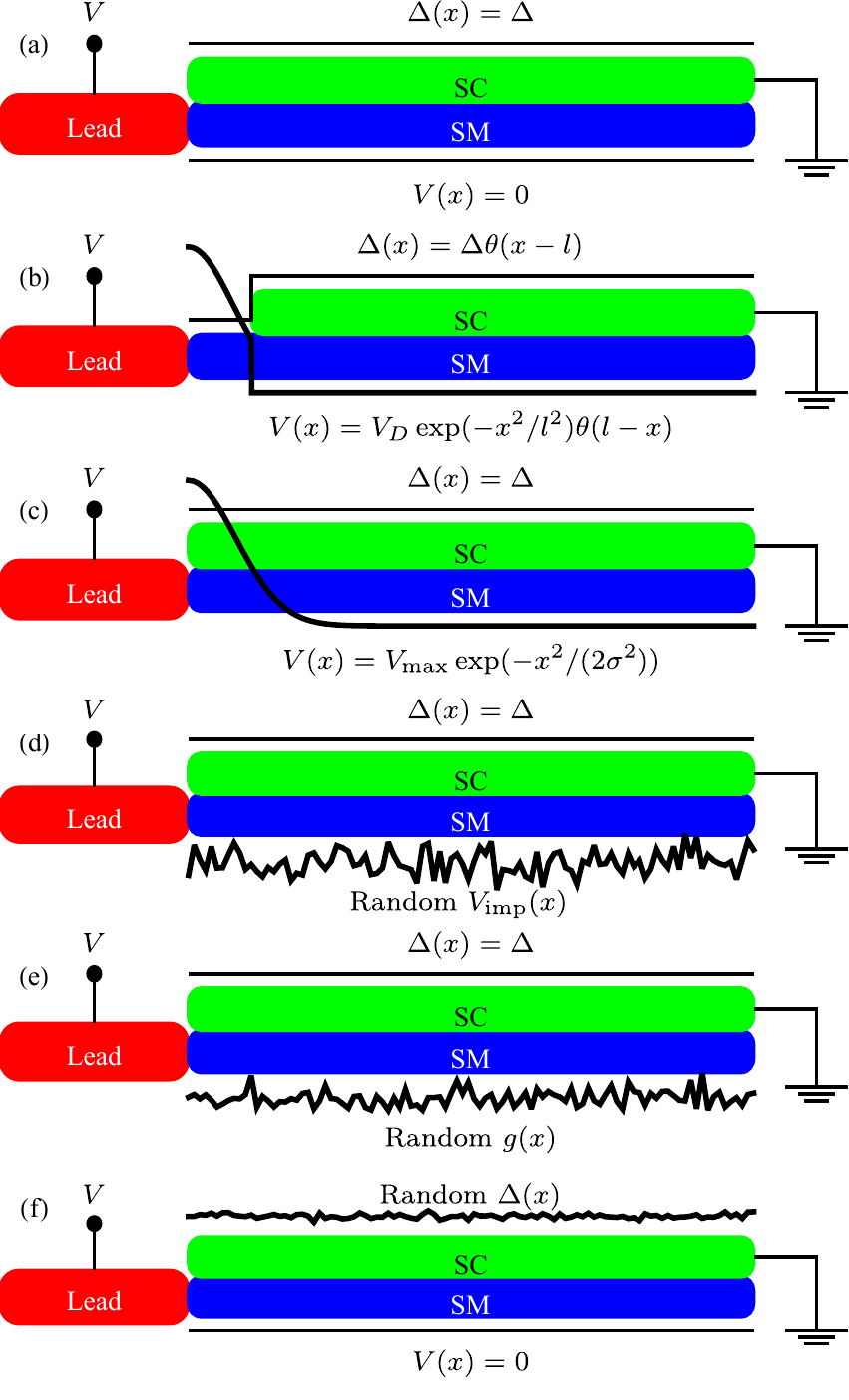}
	\caption{The schematic of the NS junction composed of a lead and (a) pristine nanowire with a constant SC gap $ \Delta $ in the clean limit $ V(x)=0 $; (b) nanowire with a quantum dot $ V(x) $ and a partially-covered parent SC; (c) nanowire with an inhomogeneous potential $ V(x) $ and a constant SC gap $ \Delta $; (d) nanowire with disorder $ V(x) $ in the chemical potential; (e) nanowire with disorder $ \tilde{g}(x) $ in the effective $ g $ factor; and (f) nanowire with disorder in the SC gap $ \Delta(x) $.}
	\label{fig:fig1}	
\end{figure}

\subsection{Dissipation and temperature}
In the experimental situation, there is invariably some dissipation in the nanowire because of coupling to the environment, which we simulate phenomenologically by adding a dissipative term to the diagonal part of the BdG Hamiltonian~\cite{liu2017role}. Dissipation also introduces a particle-hole asymmetry in the observed tunneling conductance at finite voltages which is not present in the dissipationless BdG formalism by virtue of the exact particle-hole symmetry~\cite{liu2017role}. In reality, the experiments are at the temperature $ T\sim20$ mK~\cite{vaitiekenas2018effective,grivnin2019concomitant}. To include finite temperature effect, the conductance spectrum is calculated as a convolution with the derivative of Fermi distribution at finite temperature. The dissipation and finite temperature effects are already taken into account by following recent works in the literature~\cite{liu2017phenomenology,liu2017role,dassarma2016how,setiawan2017electron,lin2012zerobias,moore2018quantized,liu2017role,sau2010nonabelian,liu2012zerobias,pikulin2012zerovoltage,rainis2013realistic,setiawan2017electron}. Thus we do not intend to discuss the effect of the dissipation and finite temperature throughout the paper by sticking to zero temperature and small dissipation ($ \Gamma=10^{-4} $ meV) in all numerical results.

\section{results}\label{sec:results}
We emphasize that our definitions for good, bad, and ugly physical mechanisms are both mathematically and physically sharply defined with no ambiguity as shown clearly in Fig.~\ref{fig:fig1}.  Physically, the good situation is pristine MZM with little background disorder and a constant chemical potential; the bad situation has a spatially varying (but deterministic) chemical potential with no random disorder; the ugly case has strong random disorder.  Mathematically, the three situations are distinguished by the term $ H_\text{V} $ in the Hamiltonian defining the BdG equation [see Eq.~\eqref{eq:tot}] with $ H_\text{V} $ being a constant (``good"), spatially varying in a deterministic manner (``bad"), and strongly random (``ugly"). Thus the three situations, good/bad/ugly, are both physically and mathematically distinct.

In this section, we show representative numerical results for the calculated differential tunneling conductance as a function of $ V_\text{bias} $ and $ V_{\text{Z}} $ in Figs.~\ref{fig:fig2}-\ref{fig:fig6}. The complete correlation conductance measurements from both ends of the nanowire are shown in Appendix~\ref{app:A}. Our goal is to simulate stable ZBCPs as observed experimentally, taking into account various possible experimental situations, including the pristine nanowire, the nanowire in the presence of the quantum dot, in the presence of the inhomogeneous potential, in the presence of disorder in the chemical potential, in the presence of disorder in the effective $ g $ factor, and in the presence of disorder in the SC gap, within a unified formalism keeping all system parameters the same except for the specific mechanism leading to that ZBCP. Based on the nature of the ZBCP sticking to zero energy (as well as the underlying physical mechanism), we classify the conductance results into three types: the good (in Sec.~\ref{sec:good}), the bad (in Sec.~\ref{sec:bad}), and the ugly (in Sec.~\ref{sec:ugly}).  We emphasize that all ZBCPs other than the good ones are topologically trivial since the ZBCPs begin to stick to zero energy in these trivial cases before the nominal TQPT. This triviality is reinforced from the wave functions in the Majorana basis in Appendix~\ref{app:B}, where the two Majorana modes are not well-separated for the bad and the ugly cases in spite of the occurrence of ZBCPs. 

In addition, we notice that by including the self-energy with a gradual-collapsing SC gap (as happens experimentally), the amplitude of the ZBCP oscillation is significantly suppressed as $ V_{\text{Z}} $ increases. For each type of ZBCP, the left-right correlation conductance measurements are also discussed. Although the end-to-end correlation measurement can be, in principle, used to distinguish MBS from ABS in long wires, we show that the nonlocal end-to-end measurements in short wires can trivially manifest such correlations, which renders the current end-to-end measurement experiments at best inconclusive~\cite{anselmetti2019endtoend,menard2019conductancematrix}. Besides presenting the conductance spectrum as a function $ V_{\text{Z}} $, we also present conductance results for zero magnetic field in Sec.~\ref{sec:B0}, which qualitatively reproduce the experiments in Ref.~\onlinecite{anselmetti2019endtoend}. Obviously, the observed existence of subgap states at zero magnetic field indicates the presence of substantial disorder in the system which casts serious doubt on the topological nature of the corresponding finite field ZBCPs.

\subsection{The good ZBCP}\label{sec:good}

The good ZBCP arises from the genuine topological Majorana mode which occurs beyond the TQPT. First, we present the results of good ZBCPs in the pristine nanowire model in Figs.~\ref{fig:fig2}(a) and~\ref{fig:fig2}(b). The schematic of the pristine model is shown in Fig.~\ref{fig:fig1}(a). In Figs.~\ref{fig:fig2}(a) and~\ref{fig:fig2}(b), the chemical potential and the SC gap are all simply constant without any disorder. The identical nonlocal conductance correlated between the two ends as shown in Fig.~\ref{fig:fig2} manifests the most ideal theoretical instance of the good ZBCP, where the ZBCP is completely topological and appears only beyond the TQPT~\cite{huang2018metamorphosis}.

The good ZBCP arising from MZM remains immune to some finite amount of disorder as shown in Figs.~\ref{fig:fig2}(c) and~\ref{fig:fig2}(d). In Figs.~\ref{fig:fig2}(c) and~\ref{fig:fig2}(d), we provide an example of the good ZBCP in the presence of weak disorder in the chemical potential with a Gaussian distribution of variance $ \sigma_{\mu}=0.4 $ meV, which accounts for $ 40\% $ of the chemical potential. The corresponding schematic is in Fig.~\ref{fig:fig1}(d). We find no ZBCP emerging in the trivial regime below TQPT, and the topological ZBCP with the Majorana oscillation emerging beyond the TQPT in the usual manner. The nonlocal conductance measurements are almost identical from both ends exhibiting the expected Majorana correlations from the two ends.

Another type of disorder is also found to have a modest impact on the good ZBCP as in Figs.~\ref{fig:fig2}(e)(f), where we show the calculated conductance for SC gap disorder. The corresponding schematic is in Fig.~\ref{fig:fig1}(f). The strength of the random gap disorder is parameterized by the standard deviation of 0.06 meV, which accounts for $ 30\% $ of the mean SC gap. Note that, we avoid using a very large strength of disorder to preserve the SC gap, otherwise, the SC gap has a possibility to be negative which would be unphysical. In the presence of disorder in the SC gap, we again find that the topological ZBCP, occurring beyond the TQPT, is relatively immune to disorder, and no trivial ZBCP is induced below the TQPT. To show that we are not deliberately choosing particular random configurations, we provide more disorder-averaged conductance spectra in Appendix~\ref{app:A} , where we observe a robust ZBCP beyond the TQPT. Thus, the good ZBCP survives weak disorder in the chemical potential and the SC gap.

\begin{figure}[htbp]
	\centering
	\includegraphics[width=3.4in]{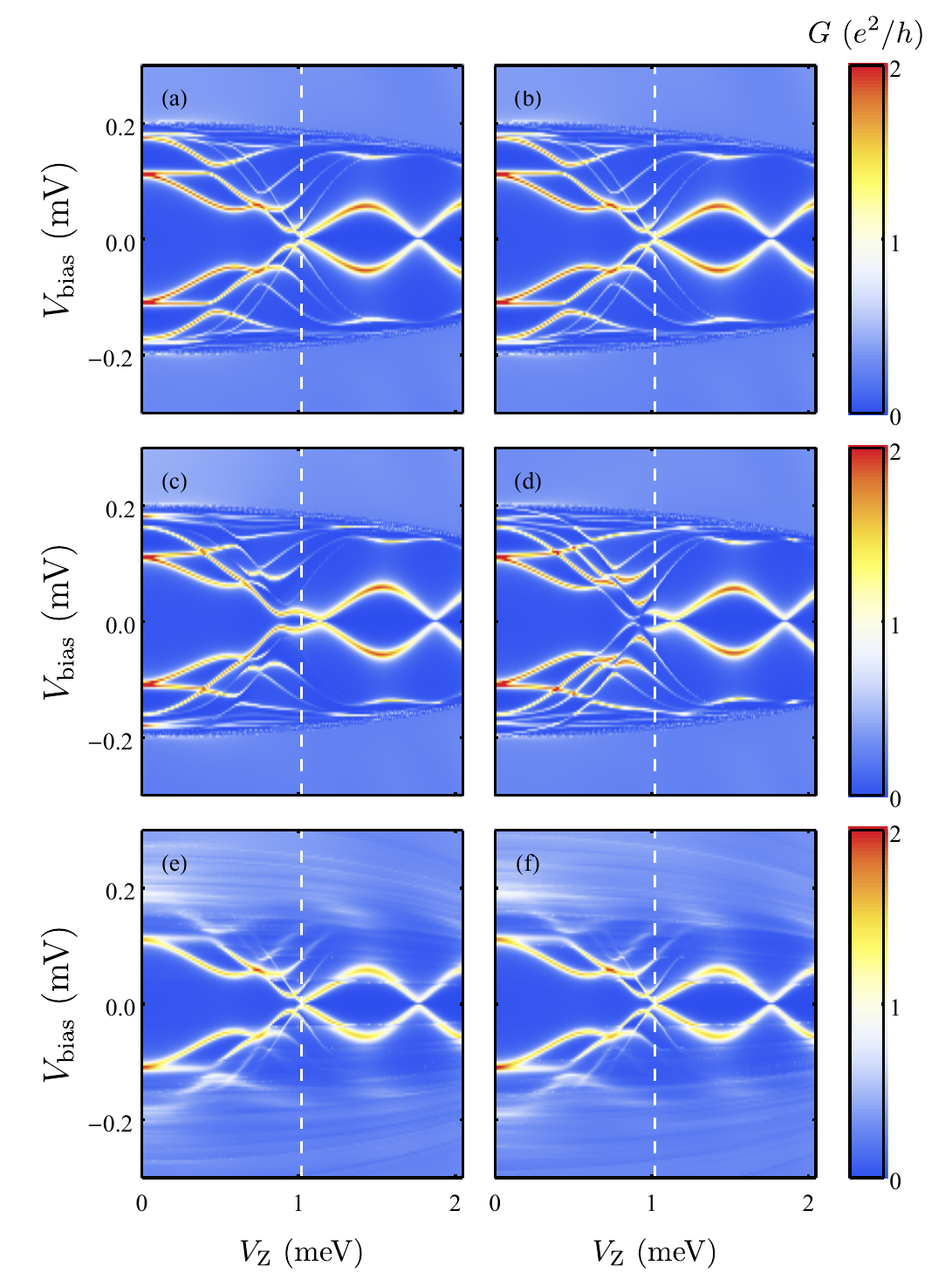}
	\caption{(a) and (b) show an example of the good ZBCP in a pristine nanowire with the self-energy in a $ 1~\mu $m wire. The color plots show the differential tunneling conductance $ G $ as a function of $ V_{\text{Z}} $ ($ x $ axis) and $ V_{\text{bias}} $ ($ y $ axis) from the left lead (left column) and the right lead (right column). The SC gap collapse $ V_{\text{C}}=3 $ meV. The TQPT is labeled in the white dashed line at $ V_{\text{Z}}=1.02 $ meV. The complete correlation conductance measurements are shown in Fig.~\ref{fig:good}; 
 	(c) and (d) show an example of the good ZBCP in the presence of a small amount of disorder in the chemical potential in a $ 1~\mu $m wire. The parameters are: standard deviation of disorder in the chemical potential $ \sigma_\mu=0.4 $ meV, SC gap collapse $ V_{\text{C}}=3 $ meV. The TQPT is labeled in the white dashed line at $ V_{\text{Z}}=1.02 $ meV. The complete correlation conductance measurements are shown in Fig.~\ref{fig:small}; 
	(e) and (f) show an example of the good ZBCP in the presence of disorder in the SC gap in a $ 1~\mu $m wire. The parameters are: standard deviation of disorder in the gap $ \sigma_\Delta=0.06 $ meV, mean parent SC gap $ \Delta_0=0.2 $ meV, and SC gap collapse $ V_{\text{C}}=3 $ meV. The TQPT is labeled in the white dashed line at $ V_{\text{Z}}=1.02 $ meV. The complete correlation conductance measurements are shown in Fig.~\ref{fig:DeltaVar}.
	}
	\label{fig:fig2}
\end{figure}

\subsection{The bad ZBCP}\label{sec:bad}
The bad ZBCP is topologically trivial because it exists below the TQPT. In Fig.~\ref{fig:fig3}, we present the calculated conductance spectra for the nanowire in the presence of a quantum dot at its end, as shown in Fig.~\ref{fig:fig1}(b). In Figs.~\ref{fig:fig3}(a)(b), we find that two ABSs coalesce into a zero-energy bound state producing a stable ZBCP from $ V_{\text{Z}}=0.6 $ to 0.9 meV. These two ABSs anitcross at zero energy for several times before $ V_\text{Z} $ reaches the TQPT. If the amplitudes of anitcrossings are tiny, within the finite energy resolution scale in experiments (where thermal broadening also provides a finite energy resolution around zero energy), these anitcrossings may be incorrectly identified as ZBCPs although they arise from almost-zero-energy trivial ABSs, not from isolated MBSs. Apart from the fact that the trivial ZBCPs arise below the TQPT, the trivial ZBCPs also differ from the topological ZBCPs in the amplitude of the ZBCP oscillation. In short nanowires ($ L=1~\mu $m), the true Majorana-induced ZBCP should have a prominent oscillation in the topological regime [as shown in the right of the white dashed line in Figs.~\ref{fig:fig2}(a)(b)]. However, in Figs.~\ref{fig:fig3}(a) and~\ref{fig:fig3}(b), the ZBCP only has a small amplitude of the ZBCP oscillation. Admittedly, one could go to a very high magnetic field to measure the amplitude of the ZBCP oscillation, but this may not be feasible because the SC gap may collapse at such a high magnetic field. Thus, if the SC gap collapses even below the TQPT (e.g., $ V_{\text{C}}=1 $ meV shown in Fig.~\ref{fig:fig3} is smaller than the nominal TQPT $ 1.02 $ meV), one will never expect to observe the real Majorana mode under such a situation. We believe that in most of the current experimental samples, the bulk SC gap collapse happens before the TQPT is reached, dooming any manifestation of the MZMs.

Besides the quantum dot, the inhomogeneous potential [as shown in Fig.~\ref{fig:fig1}(c)] can also induce the bad ZBCP as shown in Figs.~\ref{fig:fig3}(c) and~\ref{fig:fig3}(d). We take the same Gaussian form of $ V(x) $ in the inhomogeneous potential case as in the quantum dot case except that the potential is now extended over the bulk of the nanowire instead of being confined to the end as it is for the quantum dot. Thus, both quantum dots and inhomogeneous potential induce bad ZBCPs below the TQPT.

\begin{figure}[htbp]
	\centering
	\includegraphics[width=3.4in]{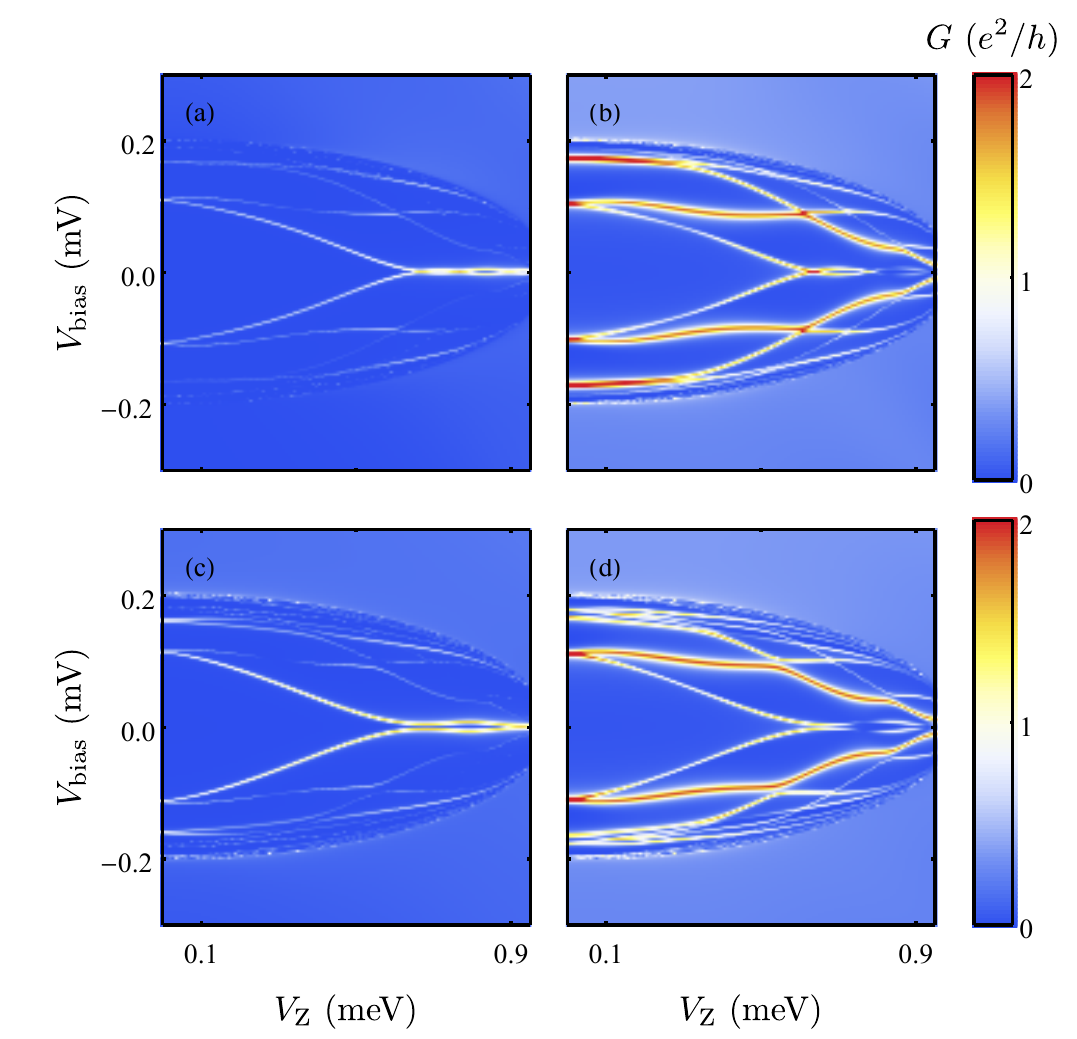}
	\caption{Two examples of the bad ZBCP due to the quantum dot in (a) and (b) and the inhomogeneous potential in (c) and (d) respectively with the self-energy in a $ 1~\mu $m wire. The left (right) column shows the conductance measured from the left (right) lead. For the quantum dot case [(a) and (b)], the parameters are: SC gap collapse $ V_{\text{C}}=1 $ meV, the peak value of the Gaussian-shaped quantum dot $ V_{\text{D}}=1.7 $ meV, and the size of the quantum dot $ l=0.2~\mu $m. For the inhomogeneous potential case [(c) and (d)], the parameters are: SC gap collapse $ V_{\text{C}}=1 $ meV, the peak value of the Gaussian-shaped potential confinement $ V_{\text{max}}=1.4 $ meV, and the linewidth $ \sigma=0.15~\mu $m. The complete correlation conductance measurements are shown in Fig.~\ref{fig:qd} for the quantum dot and Fig.~\ref{fig:inhom} for the inhomogeneous potential respectively.
	} 
	\label{fig:fig3}
\end{figure}

\subsection{The ugly ZBCP}\label{sec:ugly}
The ugly ZBCP induced by disorder is also topologically trivial. In Fig.~\ref{fig:fig4}, we present two distinct configurations of the random disorder in the chemical potential, where the schematic is shown in Fig.~\ref{fig:fig1}(d). Figures~\ref{fig:fig4}[(a) and (b)], which are calculated conductance from the left and right lead, respectively, share a common disorder configuration; Figs.~\ref{fig:fig4}(c) and (d) share another common configuration. The disorder-induced ugly ZBCPs are ubiquitous. We note that the disorder configuration in a given sample is not necessarily fixed and most likely changes as various gate voltages are tuned to optimize the zero-bias peaks, as is the common experimental practice. (The same happens also in thermal cycling.) For example, the occurrence of the disorder-induced ZBCP in Fig.~\ref{fig:fig4}(a) could shift from the left lead to the right lead as shown in Fig.~\ref{fig:fig4}(d). In addition, under the same configuration of disorder [e.g., Fig.~\ref{fig:fig4}(a) versus (b), and Fig.~\ref{fig:fig4}(c) versus (d)], we also find the end-to-end correlation from both ends, although this arises here simply due to the shortness of the wire. Thus, ugly disorder is capable, particularly when gate voltages are tuned so as to modify the disorder configuration in a given sample, of producing well-correlated ZBCPs in nanowires although these ZBCPs are completely trivial. Of course, it is possible that the end-to-end correlations are absent for ugly ZBCPs in a given situation (even for a short wire) since the correlations in the trivial ZBCPs depend on many details and are not a universal nonlocal property.  More examples are provided in  Appendix~\ref{app:A}.

For completeness, we also study the nanowire in the presence of disorder in the effective $ g $ factor and obtain qualitatively similar results as presented in Fig.~\ref{fig:fig5}. This corresponds to the schematic shown in  Fig.~\ref{fig:fig1}(e). Again, Figs.~\ref{fig:fig5}(a) and~\ref{fig:fig5}(b) share a common disorder configuration; Figs.~\ref{fig:fig5}(c) and~\ref{fig:fig5}(d) share another common configuration. Therefore, we conclude that, in the short wire, the disorder-induced trivial ABS not only resembles Majorana-induced ZBCP, but also manifests the pseudo end-to-end correlation from two ends, which could be very misleading in experiments. We emphasize again that whether end-to-end correlations are present for ugly ZBCPs depend on many details, and short wires may or may not manifest end-to-end correlations for ugly ZBCPs in specific instances. The important point is that the existence of end-to-end conductance oscillations cannot be construed to be a smoking gun evidence for good ZBCPs since ugly ZBCPs manifest them do (as do the bad ZBCPs also) in many instances.

\begin{figure}[htbp]
	\centering
	\includegraphics[width=3.4in]{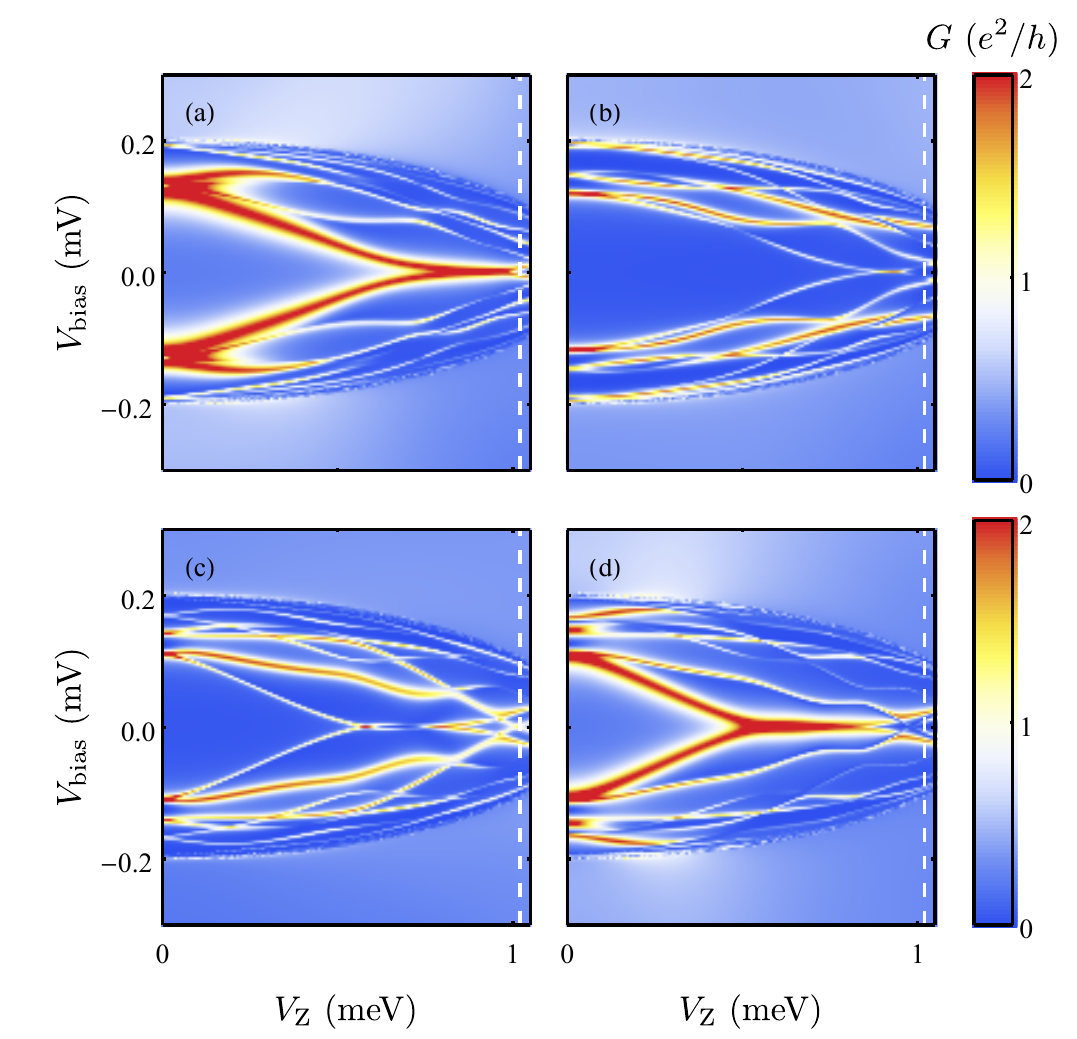}
	\caption{Two examples of the ugly ZBCP in the presence of a large amount of disorder in the chemical potential with the self-energy in the $ 1~\mu $m wire. (a) and (b) share a common configuration of disorder; (c) and (d) share another common one. The left(right) column shows the conductance measured from the left(right) lead. The parameters are: standard deviation of disorder in the chemical potential $ \sigma_{\mu}=1 $ meV, SC gap collapse $ V_{\text{C}}=1.2 $ meV. The nominal TQPT is labeled in the white dashed line at $ V_{\text{Z}}=1.02 $ meV. The complete correlation conductance measurements are shown in Fig.~\ref{fig:muVar}.
	}
	\label{fig:fig4}
\end{figure}
\begin{figure}[htbp]
	\centering
	\includegraphics[width=3.4in]{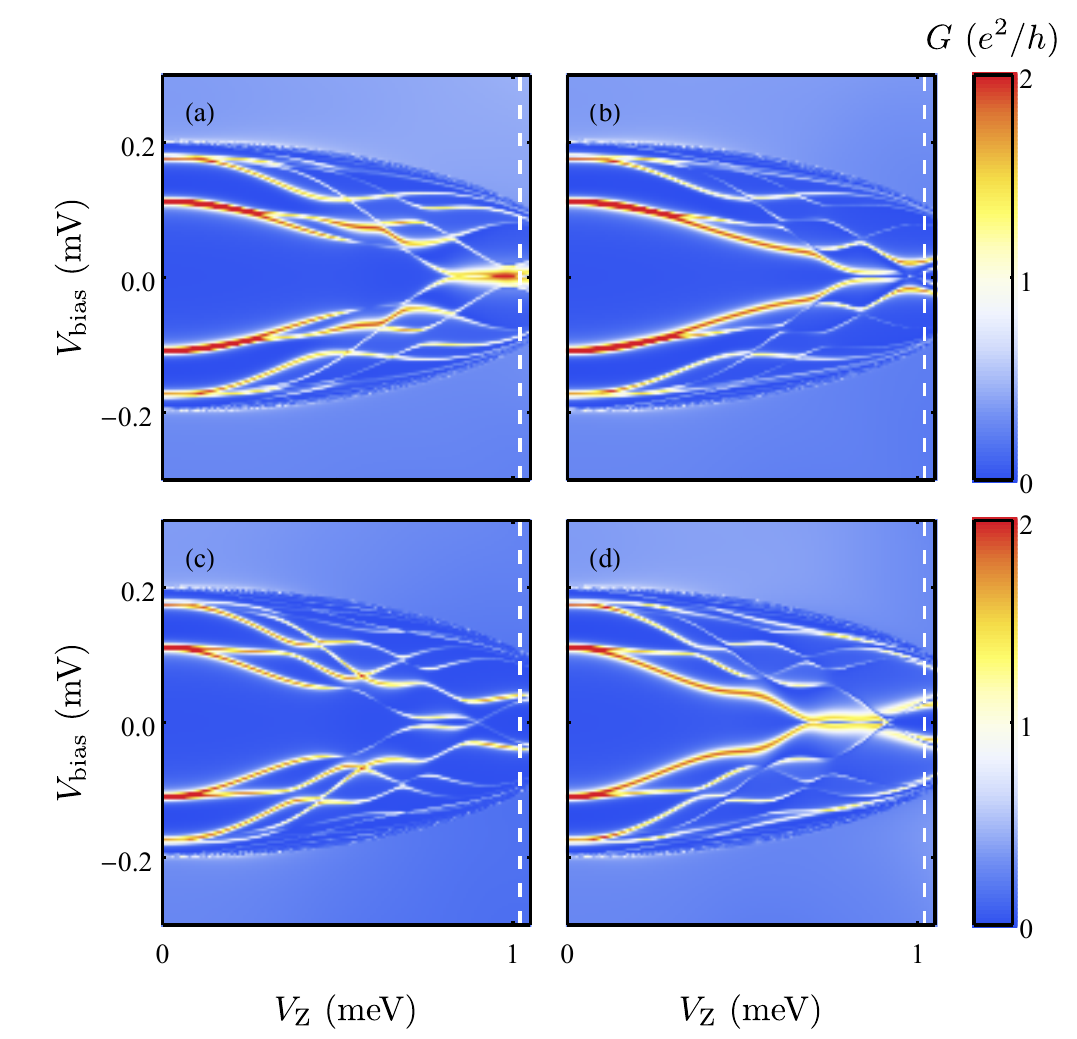}
	\caption{Two examples of the ugly ZBCP in the presence of disorder in the effective $ g $ factor with the self-energy in the $ 1~\mu $m wire. (a) and (b) share a common configuration of disorder; (c) and (d) share another common one. The left (right) column shows the conductance measured from the left (right) lead. The parameters are: standard deviation of disorder in the effective $ g $ factor is $ \sigma_g =0.8$, SC gap collapse $ V_{\text{C}}=1.2 $ meV. The nominal TQPT is labeled in the white dashed line at $ V_{\text{Z}}=1.02 $ meV. The complete correlation conductance measurements are shown in Fig.~\ref{fig:gVar}.		
	}
	\label{fig:fig5}
\end{figure}

\subsection{Zero magnetic field}\label{sec:B0}

All preceding conductance spectra are calculated for a fixed chemical potential $ \mu $ as a function of $ V_{\text{Z}} $; however, we additionally show the nonlocal end-to-end conductance measurement at zero magnetic field as a function of the chemical potential in Fig.~\ref{fig:fig6} to theoretically reproduce the experiment in Ref.~\onlinecite{anselmetti2019endtoend}. In Fig.~\ref{fig:fig6}, the left (right) lead measurements are shown in the first (second) row. All three mechanisms (good, bad, ugly) discussed in this article are presented in Fig.~\ref{fig:fig6}. The first column is for the pristine nanowire; the second and third column are in the presence of the quantum dot and inhomogeneous potential respectively; the fourth and fifth columns are both in the presence of disorder in the chemical potential. Two separate conductance spectra in the ugly case due to two different configurations are presented here again to demonstrate that the specific disorder choice is not important for the physics being discussed. Since the nanowire is short ($ L=1~\mu $m), the nonlocal conductance measurements are trivially correlated. In addition, we notice that the bad and ugly cases will bring down the fermionic subgap states to lower energies as opposed to the good case. This is particularly noticeable for the bad case in Fig.~\ref{fig:fig6} where the subgap trivial states at zero field happen to be almost near zero energy although the system is simply a non-topological $ s $-wave BCS superconductor by construction. Therefore, whenever there is strong disorder in the nanowire, there could be prominent fermionic subgap bound states at both ends of the wire, even at zero magnetic field. This further implies that if one already finds fermionic subgap states in the system, the chance of seeing an ABS mimicking MBS will be highly enhanced at finite magnetic fields, because those fermionic subgap states could move to zero, and then anitcross with each other, which could produce trivial ZBCPs within the finite experimental energy resolution. Thus, it is important to ascertain that there are no low energy subgap states in the nanowire at zero field before embarking on the $ V_\text{Z} $-dependent search for ZBCPs in the hybrid system.

\begin{figure*}[ht]
	\centering
	\includegraphics[width=6.8in]{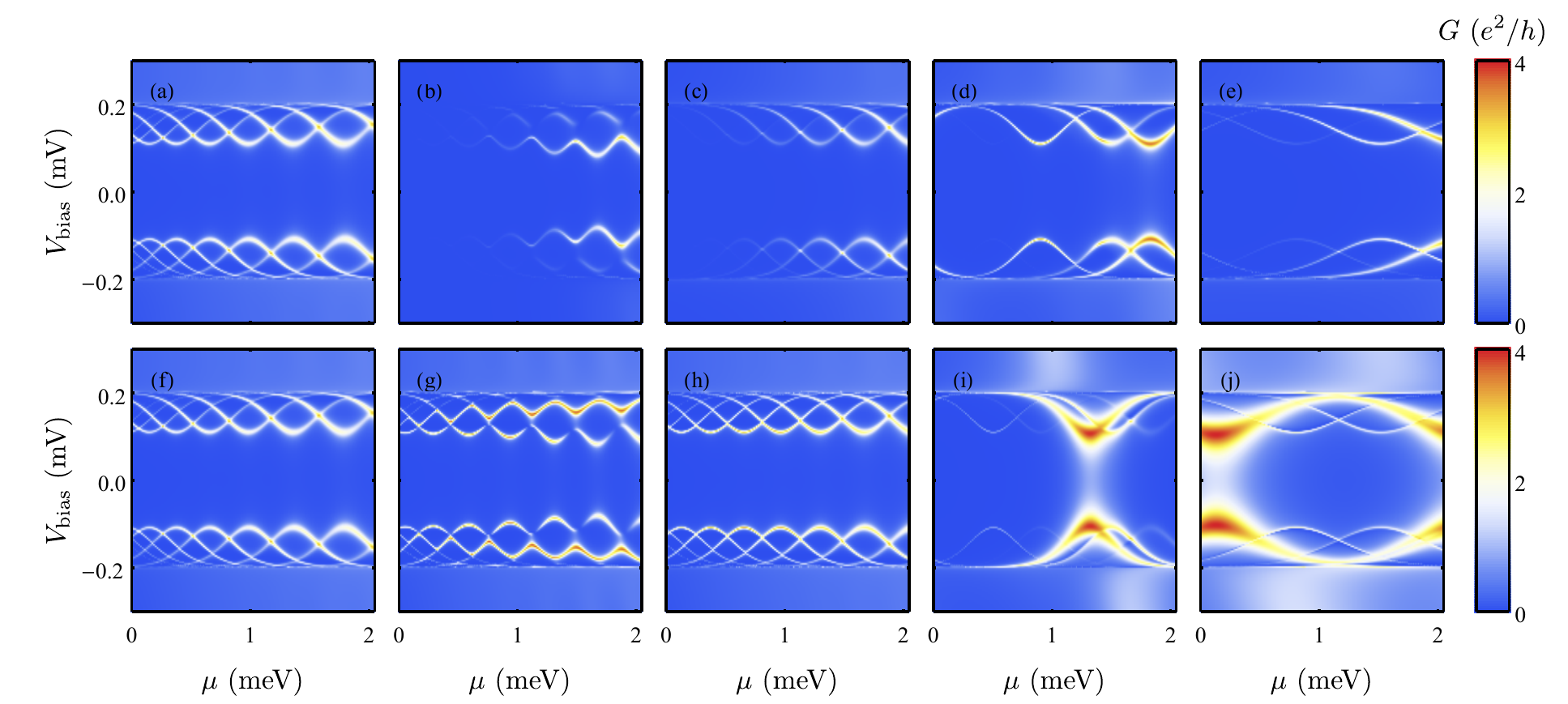}
	\caption{The conductance spectra as a function of the chemical potential $\mu  $ and $ V_{\text{bias}} $ at zero magnetic field in the $ 1~\mu $m wire with the self-energy. The first (second) row shows the conductance measured from the left (right) lead. Note that the range of the conductance is $ 0\sim 4 e^2/h$ here. (a) and (f) are in the pristine nanowire case. (b) and (g) are in the presence of a quantum dot with the peak value of $V_{\text{D}}= 1.7 $ meV and the size of $ l=0.2~\mu $m. (c) and (h) are in the presence of an inhomogeneous potential with the peak value of $ V_{\text{max}}=1.4 $ meV  and the linewidth  $ \sigma=0.15~\mu $m. (d), (i) and (e), (j) are in the presence of disorder in the chemical potential with two distinct configurations. The standard deviation of disorder is $ \sigma_{\mu}=3 $ meV. }
	\label{fig:fig6}
\end{figure*}

\section{discussion}\label{sec:disscussion}

In this section, we focus on the experimental results and attempt to fine-tune parameters to fit them. Figures.~\ref{fig:fig7}(a) and~\ref{fig:fig7}(c) are from experimental Refs.~\onlinecite{nichele2017scaling} and \onlinecite{zhang2018quantized} respectively; Figs.~\ref{fig:fig7}(b) and~\ref{fig:fig7}(d) are the corresponding theoretical reproduction using our results after fitting and fine-tuning. Both experimental observations are qualitatively reproduced by the trivial ZBCPs in the presence of a certain disorder configuration in the chemical potential through fine-tuning. Since the exact experimental magnetic field for the TQPT is unknown~\cite{cole2015effects}, we are not able to directly determine whether the experimental ZBCP is topological or not, but the fact that we can reproduce the experimental ZBCP fairly well by using ``ugly" ZBCPs in our simulations establishes that the experimental ZBCPs may very well arise simply from disorder.  This is also consistent with the experiment not observing any gap reopening or ZBCP oscillations which should be concomitant with the TQPT if the ZBCP is indeed arising from topological MZMs. We find that most experimentally-observed features are qualitatively reproduced by the disorder-induced ugly ZBCP, including the vanishing amplitude of the ZBCP oscillation with increasing magnetic field and the instability of ZBCP over regimes of high magnetic fields. Namely, the ZBCP will vanish approximately beyond $ B=3 $ T in Fig.~\ref{fig:fig7}(a) and $ B=1 $ T in Fig.~\ref{fig:fig7}(c).

To be specific, we choose the experimental result in Fig.~\ref{fig:fig7}(c) from the most compelling experimental paper~\cite{zhang2018quantized} in the subject entitled \textit{Quantized Majorana conductance} and present measured conductance from Ref.~\onlinecite{zhang2018quantized} at zero bias voltage as a function of the magnetic field in Fig.~\ref{fig:fig9}(a): The conductance grows from zero up to a quantized value of $ 2e^2/h $, persists for a very short plateau before it drops. In Fig.~\ref{fig:fig9}(b), we also show the measured conductance cut as a function of the bias voltage at a fixed magnetic field: It is a quantized peak at $ B=0.88 $ T, where the maximal peak in Fig.~\ref{fig:fig9}(a) is. However, this quantized value does not indicate the topological state--- we can easily reproduce the same scenario with the manifestation of all these features by the ugly ZBCPs shown in Fig.~\ref{fig:fig10}. In Figs.~\ref{fig:fig10}(e) and ~\ref{fig:fig10}(f), we choose an instance of ugly ZBCP from Fig.~\ref{fig:fig4}(a): The conductance at zero-bias voltage also grows from zero to quantized $ 2e^2/h $ and then drops. We also see a quantized peak in Fig.~\ref{fig:fig10}(f), which plots the conductance as a function of bias voltage at the maximal peak in Fig.~\ref{fig:fig10}(e). Using another disorder profile in Fig.~\ref{fig:fig4}(d) does not change the scenario qualitatively, which is shown in Figs.~\ref{fig:fig10}(g) and~\ref{fig:fig10}(h). In Figs.~\ref{fig:fig10}(i) and~\ref{fig:fig10}(j), we show the conductance from the random matrix theory in Ref.~\onlinecite{pan2019generic}, which also manifests this unstable ``quantization plateau''. In Ref.~\onlinecite{pan2019generic}, we use the random matrix theory to describe the statistical features of this highly-disordered system in a class-D ensemble, which only requires the particle-hole symmetry. Obviously, the random matrix theory of Ref.~\onlinecite{pan2019generic} produces only 'ugly' conductance results since the physics is by definition driven entirely by strong disorder. All the ZBCP we generated are guaranteed to be topologically trivial because of the even number of channels in the disordered system. Thus both the phenomenological random matrix theory of Ref. \onlinecite{pan2019generic}, which has only disorder in the model and nothing else, and our fully microscopic theory with random disorder both reproduce the observed ``quantized conductance" behavior remarkably well, reinforcing our claim that the observed ZBCPs, even the apparently quantized ones, may easily arise from background random disorder. We believe that the results shown in Figs.~\ref{fig:fig7},~\ref{fig:fig9}, and ~\ref{fig:fig10}, taken together, establish compellingly that the best current experimentally observed ZBCPs most likely arise from strong disorder, and fall into our ``ugly" nontopological category.
 
For direct comparison between ugly and good/bad results, we also plot the typical conductance cuts for the good ZBCP in Figs.~\ref{fig:fig10}(a) and~\ref{fig:fig10}(b): The conductance remains quantized as $ V_\text{Z} $ increases, where the non-quantized deviation is simply due to the Majorana oscillation. For the bad ZBCP as shown in Figs.~\ref{fig:fig10}(c) and~\ref{fig:fig10}(d), we find the conductance is typically not quantized. Therefore the ``quantization plateau" does not appear--- only the spikes appear as $ V_\text{Z} $ increases. This turns out to be a huge distinction between the bad ZBCP, where conductance quantization is typically unstable, and the ugly ZBCP, where the ``quantization plateau" will persist at least for a noticeable segment before it disappears. We find that it is impossible to fine-tune parameters (e.g., quantum dot confinement potentials) of the bad ZBCP case to reproduce the experimental behavior shown in Fig.~\ref{fig:fig9} whereas for the ugly case, the necessary fine-tuning is no more involved than necessary for the experimentalists to obtain conductance quantization.

Based on our direct comparison between our ugly results and the experimental conductance quantization, as shown in Figs.~\ref{fig:fig7},~\ref{fig:fig9}, and ~\ref{fig:fig10}, we are led to assert that the even the best currently measured ZBCPs are likely to be ``ugly" ZBCPs, i.e., trivial ZBCPs induced by strong random disorder in the system. We emphasize, however, that if the random disorder is suppressed in future better samples, our results in Figs.~\ref{fig:fig2},~\ref{fig:good} and~\ref{fig:small} in the appendix show that topological MZMs should emerge in Majorana nanowires.  The disorder must be reduced well below the average quantities (e.g., chemical potential, SC gap, and Zeeman splitting) in order for the MZMs to manifest themselves.

\begin{figure}[htbp]
	\centering
	\includegraphics[width=3.4in]{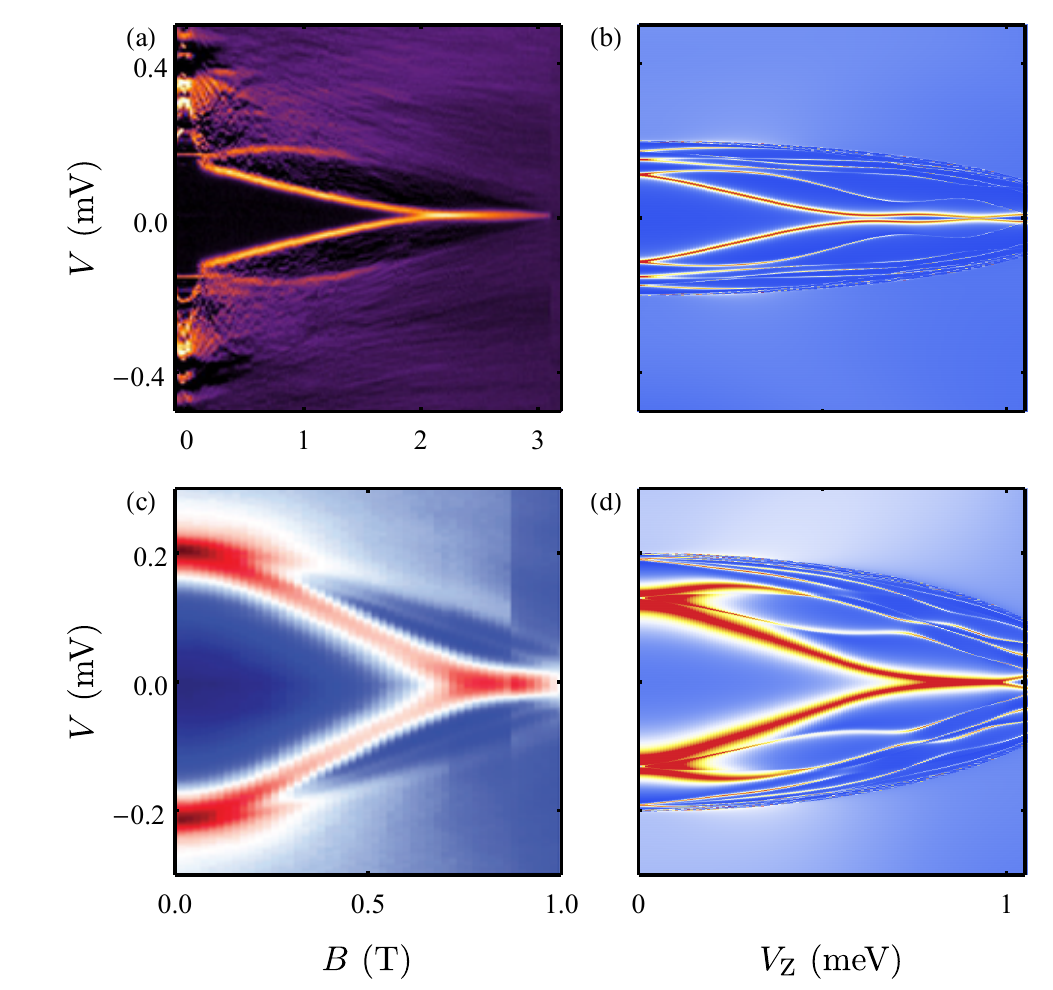}
	\caption{(a) Tunneling conductance as a function of the magnetic field at a small transmission rate to the lead. The darker color indicates the smaller conductance. This experimental result is from Ref.~\onlinecite{nichele2017scaling}; (b) fine-tuning parameters to fit (a); The ZBCP is the ugly one with $ \sigma_{\mu}=1 $ meV; (c) tunneling conductance as a function of the magnetic field. The redder color indicates the larger conductance. This experimental result is from Ref.~\onlinecite{zhang2018quantized}; (d) fine-tuning parameters to fit (d). The ZBCP is the ugly one with $ \sigma_{\mu}=1 $ meV.}
	\label{fig:fig7}
\end{figure}
\begin{figure*}[ht]
	\centering
	\includegraphics[width=6.8in]{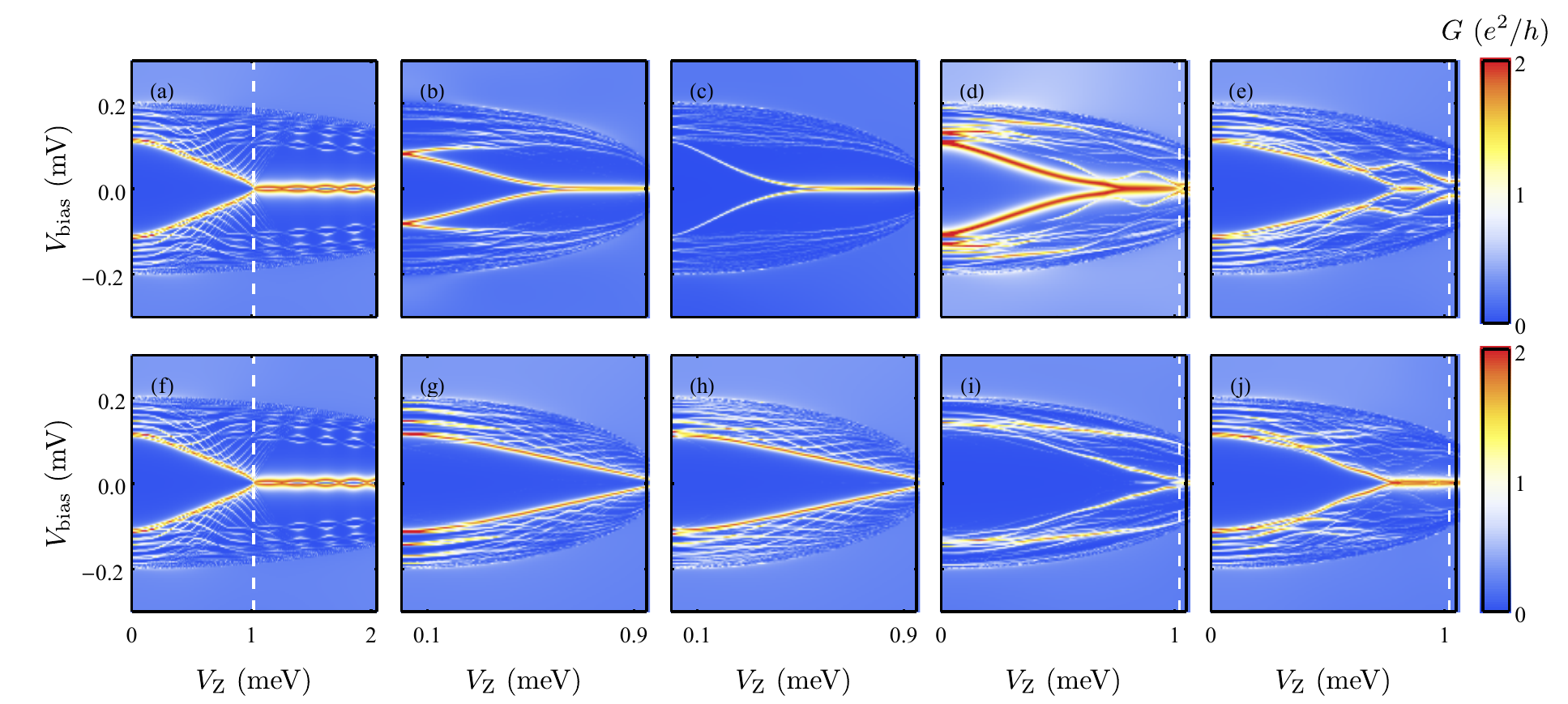}
	\caption{Conductance spectra measured from the left lead (the first row) and the right lead (the second row) in the long wire $ L=3~\mu $m. (a) and (f) are the good ZBCP in the pristine nanowire with the SC gap collapse $ V_{\text{C}}=3  $ meV. (b) and (g) are the bad ZBCP in the presence of the quantum dot with the peak value of $ V_{\text{D}}=0.6 $ meV and the size of $ l=0.4~\mu $m. The SC gap collapse is $ V_{\text{C}}=1 $ meV. (c) and (h) are the bad ZBCP in the presence of the inhomogeneous potential with the peak value of $ V_{\text{max}}=1.2 $ meV and the linewidth of $ \sigma=0.4\mu $m. The SC gap collapse is $ V_{\text{C}}=1 $ meV. (d) and (i) are the ugly ZBCP in the presence of disorder in the chemical potential, where $ \sigma_{\mu}=1 $ meV. The SC gap collapse is $ V_{\text{C}}=1.2 $ meV. (e) and (j) are the ugly ZBCP in the presence of disorder in the effective $ g $ factor, where $ \sigma_{g}=0.6 $. The SC gap collapse is $ V_{\text{C}}=1.2 $ meV.}
	\label{fig:fig8}
\end{figure*}

In addition, we compare the nonlocal correlation measurements in the short wire ($ L=1~\mu $m) with the one in the long wire ($ L=3~\mu $m). We note that the long and short here refer only to the actual physical length of the wire, and nothing else. The nonlocal correlation measurements for each case (``good"/``bad"/``ugly") in the short wire are shown in Figs.~\ref{fig:fig2}-\ref{fig:fig6}. We additionally provide the nonlocal conductance measurements in the long wire in Fig.~\ref{fig:fig8}. The left lead and right lead measurements are in the first and second row respectively. In Fig.~\ref{fig:fig8}, the first column is for the pristine nanowire; the second to the fifth column are in the presence of the quantum dot, inhomogeneous potential, disorder in the chemical potential, and disorder in the effective $ g $ factor respectively. These nonlocal measurements inform us of the properties of ZBCPs and the corresponding likely mechanisms; for instance, in Fig.~\ref{fig:fig8}(a) and~\ref{fig:fig8}(f) (the good case), the left and right measurements show conclusively identical conductance spectra. For the bad and ugly cases in the long wire, the ABS-induced ZBCPs are completely uncorrelated as they are determined by the detailed shape of the quantum dot, inhomogeneous potential or disorder profile at both ends of the nanowire. However, it is a different scenario in the short wire limit; for instance in Fig.~\ref{fig:fig4} (the ugly case), the ZBCPs measured from both ends will be trivially correlated just because of the short wire. Imagine a scenario where none of the physics being discussed here was known theoretically and the very first experimental paper reported results like Fig.~\ref{fig:fig4}, everything would be temptingly deemed to be well-established as the discovery of topological MZMs since it is a quantized zero-bias conductance peak and it is nonlocal. Unfortunately, this conclusion would be most likely incorrect as we know from the results presented in the current work where we find that disorder induced ZBCPs mimic many features of the MZM-induced ZBCPs, particularly in short wires. The same is true for Fig.~\ref{fig:fig5}, where Figs.~\ref{fig:fig5}(a) and~\ref{fig:fig5}(b) as well as Fig.~\ref{fig:fig5}(c) and~\ref{fig:fig5}(d) appear to manifest similar ZBCPs from both ends although the ZBCPS are purely ugly and nontopological--- thus reinforcing the conclusion that the mere fine-tuned observation of end-to-end ZBCP correlations by itself cannot be construed to be a signature or evidence for topological MZMs.  Purely ugly disorder induced ZBCPs could manifest end-to-end correlations just accidentally. A key problem is that there is no way to know \textit{a priori} whether the experimental wires are long or short since the nanowire coherence length is completely unknown (and long or short is defined with respect to the coherence length) although the current experimental samples with $ L \sim 1 $ micron are most likely in the short wire regime. For the results of long wires in Fig.~\ref{fig:fig8}, which are longer than the superconducting coherence length, only the real MZM would have the perfect end-to-end correlation. The trivial ABS, on the other hand, may have a pseudo end-to-end correlation in the short wire (typically shorter than $ 1~\mu $m). This leads to the conundrum that although the nonlocal correlation measurement, in principle, can serve as a reliable diagnostic for MZMs, the prerequisite for this indicator being the long nanowire limit may not be satisfied in the experimental samples. Unless sufficiently long nanowires (at least longer than the SC coherence length) can be fabricated, the observation of the end-to-end correlation can never prove the existence of topological MZMs. In fact, as a cautionary note, we emphasize that such accidental end-to-end correlations could happen for ugly ZBCPs even in the long wire limit as shown in Figs.~\ref{fig:muVar} and~\ref{fig:gVar} of Appendix. This is of course purely accidental with no significance except that if one post-selects and fine-tunes experimental results, a certain fraction of ugly ZBCPs will manifest apparent nontopological end to end correlations which could be mistaken for the nonlocal correlation of real topological MZMs. Of course, generically, short wires do not manifest any end-to-end conductance correlations arising from trivial ZBCPs as shown in Appendix~\ref{app:A}, but the important point here is that trivial ZBCPs in short (or even, long) wires can be correlated from the two ends under suitable conditions, making the correlation test not conclusive unless one can be sure that the experiment is indeed being carried out in the no-disorder limit.

Many more numerical simulations for good, bad, and ugly ZBCPs are presented in the appendices along with correlation results.  We also present wave functions and energy spectra in the appendices.

\begin{figure}[ht]
	\centering
	\includegraphics[width=3.4in]{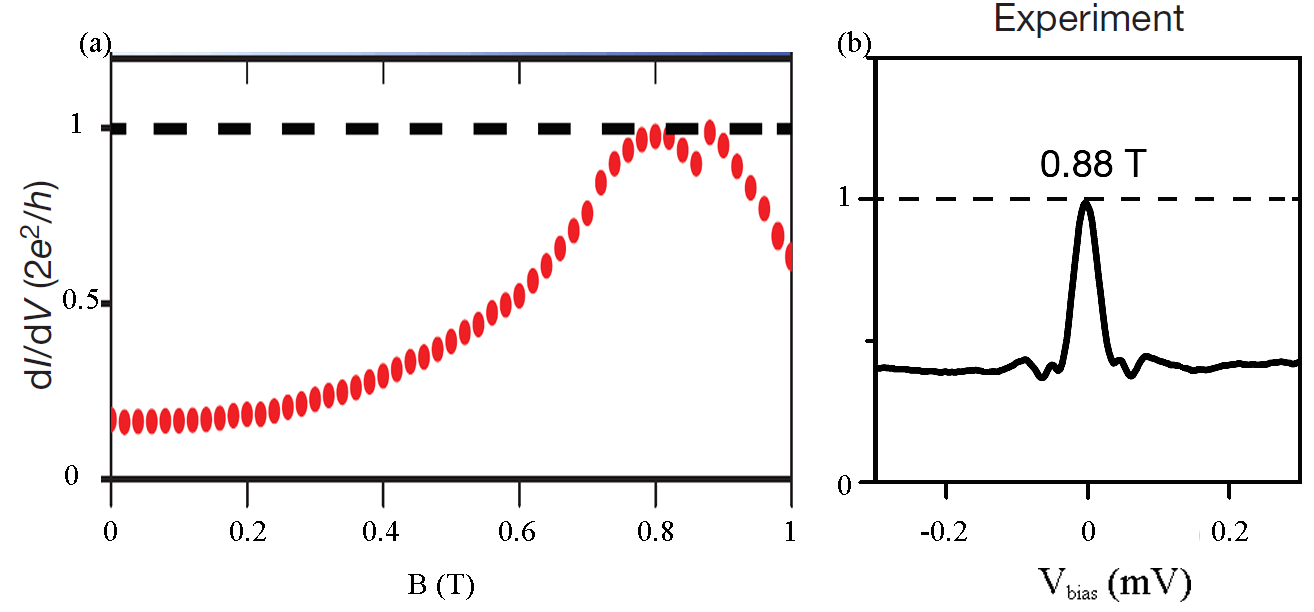}
	\caption{{(a) The conductance at the zero bias voltage as a function of the magnetic field. (b) The conductance cut as a function of the bias voltage at the magnetic field of the maximal peak. Both of these experimental results are from Ref.~\onlinecite{zhang2018quantized}.}}
	\label{fig:fig9}
\end{figure}

\begin{figure}
	\centering
	\includegraphics[width=3.4in]{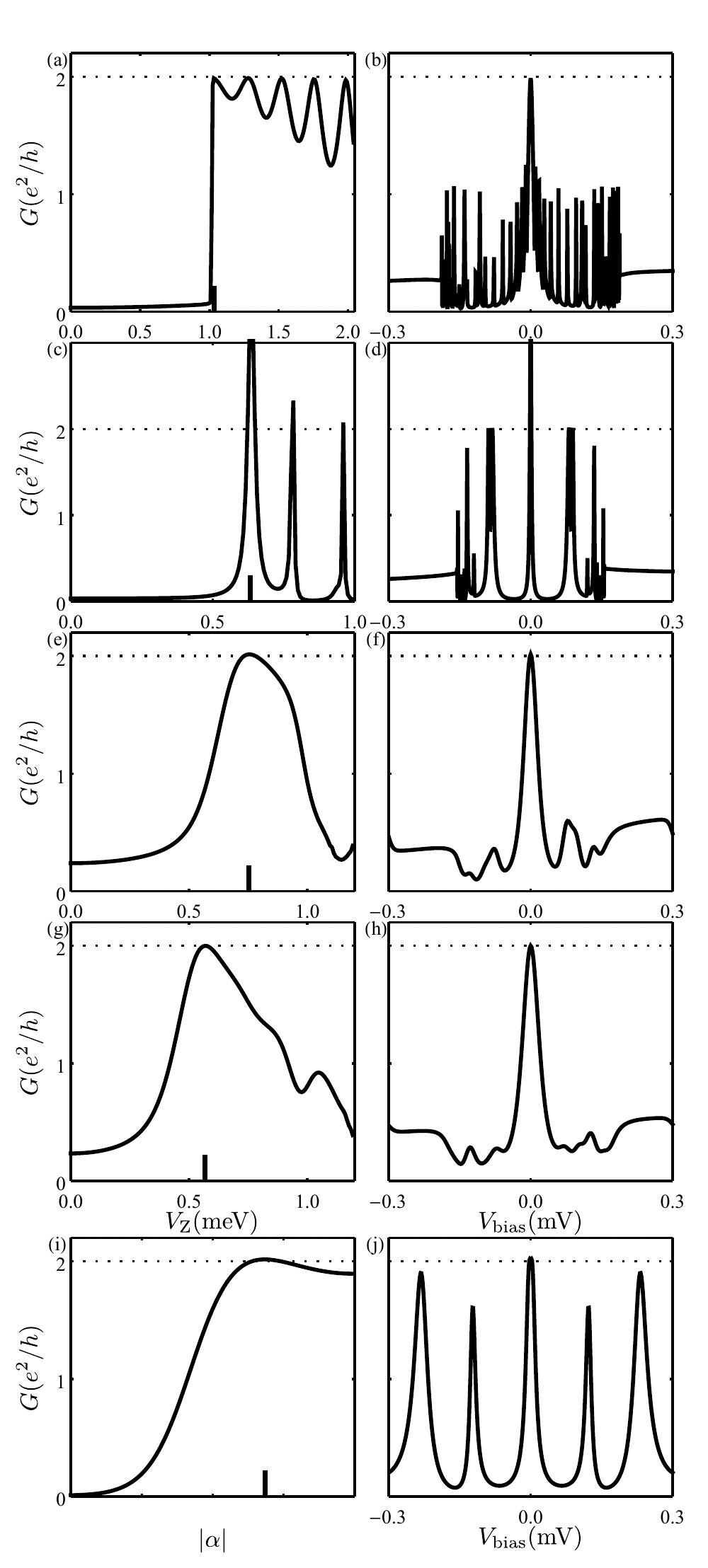}
	\caption{{[(a) and (b)] The conductance at zero bias voltage and at the Zeeman field of the maximal peak (labeled in black) respectively in an instance of good ZBCP shown in Fig.~\ref{fig:good}(c). 
			[(c) and (d)] The conductance at zero bias voltage and at the Zeeman field of the maximal peak (labeled in black) respectively in an instance of bad ZBCP shown in Fig.~\ref{fig:fig3}(b). 
			[(e) and (f)] The conductance at zero bias voltage and at the Zeeman field of the maximal peak (labeled in black) respectively in an instance of ugly ZBCP shown in Fig.~\ref{fig:fig4}(a) at the temperature $ T=58 $ mK. 
			[(g) and (h)] The conductance at zero bias voltage and at the Zeeman field of the maximal peak (labeled in black) respectively in an instance of ugly ZBCP shown in Fig.~\ref{fig:fig4}(d) at the temperature $ T=75 $ mK.
			[(i) and (j)] The conductance at zero bias voltage and at the Zeeman field of the maximal peak (labeled in black) respectively from the random matrix theory in Ref.~\onlinecite{pan2019generic}.}}
	\label{fig:fig10}
\end{figure}
\section{conclusion}\label{sec:conclusion}

We have provided extensive numerical simulations for the Majorana properties of semiconductor nanowires in the SC-SM hybrid structures, taking into account the essential effects of disorder, including quantum dots and inhomogeneous potentials along the wire as well as random disorder in the chemical potential or the SC gap or the effective $ g $ factor.  We find three different types of tunneling zero-bias conductance peaks:  the good, the bad, and the ugly.  The good ZBCPs arise from the intrinsic topological properties of the system for the Zeeman field above the topological quantum phase transition point, with the ZBCPs from the two ends of the wire showing a high level of correlations even in long wires by virtue of the nonlocal topological properties of the system.  We show that good ZBCPs are immune to weak disorder in the chemical potential and the superconducting gap, and are robust to system parameters such as the chemical potential or Zeeman field provided one is the topological regime (i.e., Zeeman field above the TQPT value).  The bad ZBCPs arise from quantum dots or other inhomogeneous potentials in the nanowire, and are essentially quasi-Majorana modes where the two MZMs, instead of being well-separated as in the good case, overlap with each other giving rise to near-zero-energy Andreev bound states.  These ABSs produce trivial ZBCPs for Zeeman field values below TQPT, mimicking many properties of good ZBCPs, including even the end-to-end correlation properties in short wires.  Since experimentally neither the TQPT critical field nor the SC coherence length is known, the mere observation of ZBCPs by themselves (or even the observation of end-to-end correlations) hardly could be construed to be evidence supporting the existence of MZMs in nanowires since bad ZBCPs are capable of mimicking the properties of the good ZBCPs.  The situation becomes worse when strong random disorder is considered leading to ``ugly" ZBCPs, which are trivial, but may mimic all the properties of good ZBCPs, including end-to-end correlations. Our direct comparison with the available experimental data indicates that most experimental ZBCP observations are consistent with the ZBCPs being ugly although one can never be sure without knowing what the TQPT field is and whether the nanowire is long or short from a topological viewpoint.  The fact that subgap conductance and even some end-to-end conductance correlations have been observed already for zero magnetic field in nanowires~\cite{anselmetti2019endtoend} suggests that strong disorder is playing a key role in the existing SC-SM samples, and the observed ZBCPs are likely to be of the undesirable ugly type.

A key difference between the bad and the ugly ZBCPs is the fact that the system manifesting bad ZBCPs should, in principle, eventually manifest good ZBCPs at larger magnetic field values above the TQPT.  By contrast, the strongly disordered systems manifesting ugly ZBCPs cannot manifest topological properties at any Zeeman field since disorder has eliminated the TQPT.  It may therefore appear that one should be able to observe good ZBCPs in a system manifesting bad ZBCPs simply by increasing the magnetic field so that the bad ZBCPs below TQPT transmute to good ZBCPs above TQPT.  The same can also be achieved in principle by tuning the chemical potential through the TQPT.  Although theoretically appealing, this crossover of trivial ZBCPs arising from ABS to topological ZBCPs arising from MBS has never been experimentally achieved because of the SC bulk gap collapse problem in real nanowires, where with increasing field, the bulk gap eventually collapses at some characteristic field ($ \sim 1 $T), thus severely restricting the field range of the topological regime.  In particular, if the gap collapse happens at a field lower than the TQPT field, there is no hope ever of observing the topological regime with true Majorana modes.  Current experiments suggest that this is the likely scenario, making the gap collapse a very serious problem preventing the existence of topological Majorana modes.

An equally serious problem is that most experimental nanowires may be in the ``short wire" ($ \sim 1~\mu $m) regime, where the concept of topology simply does not apply even if the system is fairly disorder-free.  In such a situation, the MBSs overlap producing near-zero-energy ABSs which then produce bad ZBCPs.  The fact that experimentally Majorana oscillations are never seen, however, indicates that this situation may not be the dominant scenario in the current experimental samples, where strong random disorder and the associated ugly ZBCPs arising purely from random disorder is the dominant physical mechanism.

Our most important finding in this paper is that although weak disorder does not adversely affect the topological properties of the MZMs, strong random disorder, with the root mean square fluctuation of disorder being comparable or larger than the average system parameter such as the SC gap or the chemical potential, not only suppresses all topological properties, but also introduces relatively stable ZBCPs with conductance values  $\sim 2e^2/h $, closely mimicking recent experimental results. We have recently come to the same conclusion using a phenomenological random matrix theory, where we find disorder always produces zero bias conductance peaks and a certain fraction of these peaks have values relatively close to $ 2e^2/h $.  Our current microscopic theory for ugly ZBCPs should be considered complementary to the random matrix theory of Ref.~\onlinecite{pan2019generic}.  We therefore believe that the existing experimentally observed ZBCPs in nanowires all arise from strong disorder in the system, particularly since the observed experimental behavior of these ZBCPs as functions of magnetic field and gate voltage is very similar to what we find in our ``ugly" ZBCPs calculations.  Our work clearly indicates that further progress in the field can only be achieved by reducing disorder in the system.  Any effort to build topological qubits out of the currently available nanowires based on their apparent ``quantized conductance" properties is doomed to failure unless the background disorder is reduced substantially, bringing the random variance well below the average parameter values.

Our work reinforces the need for much cleaner wires for progress in the field.  In addition, one must control the gap collapse problem so that higher Zeeman fields can be applied to the system without suppressing the bulk superconductivity completely.  It would also be desirable to obtain estimates for the actual coherence length in nanowires so that short versus long wire regimes can be discerned quantitatively in the experimental systems.  We believe that without improvement in these three directions (i.e. less disorder, longer wires, no bulk gap collapse) it would be difficult to establish the existence of topological Majorana modes.

This work is supported by Laboratory for Physical Sciences and Microsoft. We also acknowledge the support of the University of Maryland supercomputing resources (hpcc.umd.edu).

\onecolumngrid
\appendix
\setcounter{secnumdepth}{3}
\renewcommand\figurename{Figure}

\section{Correlation of conductance spectra}\label{app:A}
In this section, the complete nonlocal correlation of conductance spectra are presented including the pristine nanowire in Fig.~\ref{fig:good}, a small amount of disorder in the chemical potential in Fig.~\ref{fig:small}, disorder in the SC gap in Fig.~\ref{fig:DeltaVar}, the presence of a quantum dot in Fig.~\ref{fig:qd}, the presence of the inhomogeneous potential in Fig.~\ref{fig:inhom}, a large disorder in the chemical potential in Fig.~\ref{fig:muVar}, disorder in the effective $ g $ factor in Fig.~\ref{fig:gVar}, and the short but uncorrelated instances in Fig.~\ref{fig:uncorr}.

\begin{figure*}[h]
	\centering
	\includegraphics[width=6.8in]{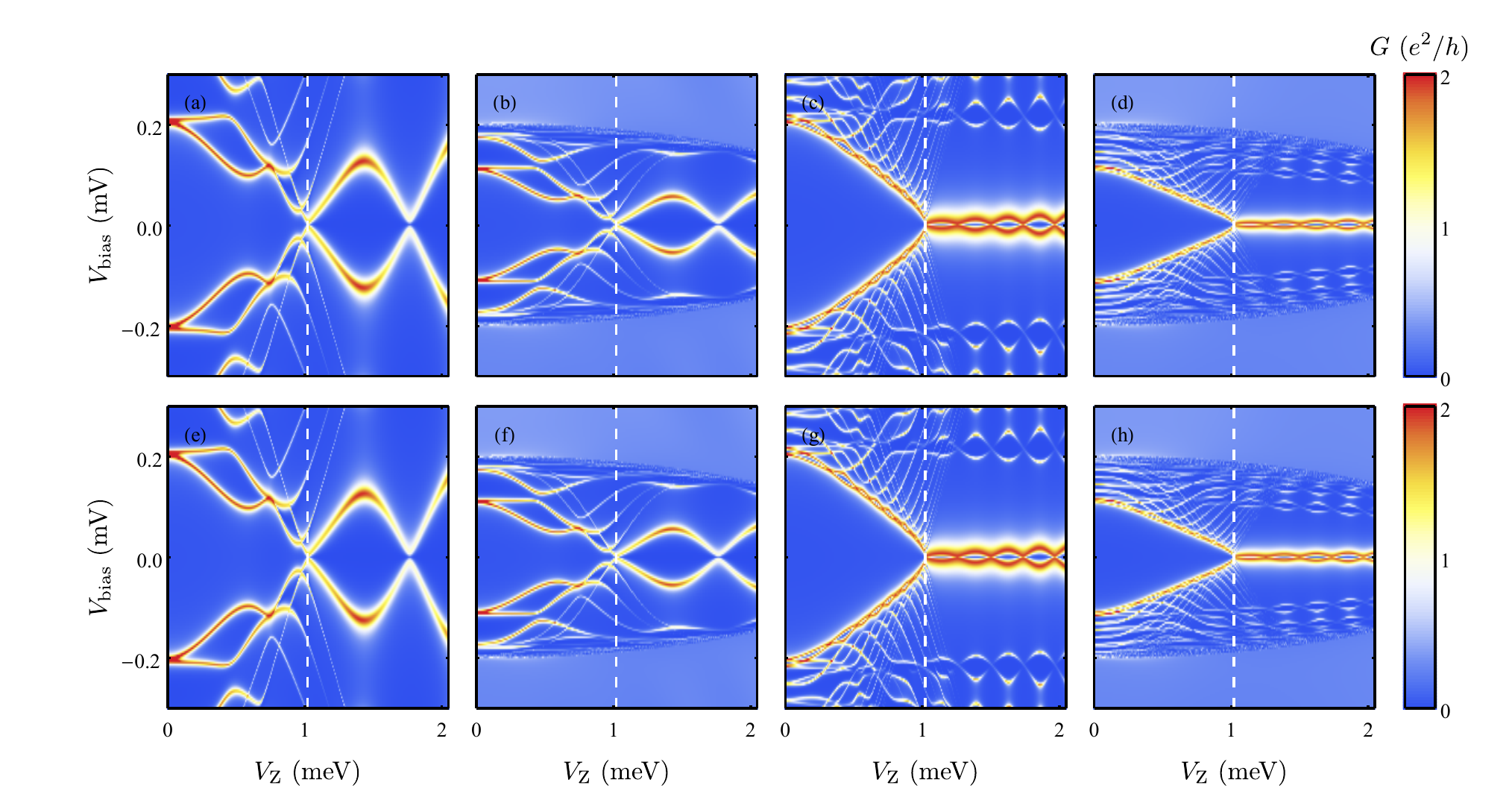}
	\caption{The good ZBCP in two $ 1~\mu $m pristine wires [shown in (a)-(d)] and two $ 3~\mu $m pristine wires [shown in (c)-(h)]. The color plots show the differential tunneling conductance $ G $ as a function of $ V_{\text{Z}} $ ($ x $ axis) and $ V_{\text{bias}} $ ($ y $ axis) measured from the left lead (in the first row) and the right lead (in the second row). Nanowires with the self-energy are shown in (b), (f), (d), and (h) and without the self-energy are shown in (a), (c), (e), and (g). The SC gap collapse $ V_{\text{C}}=3 $ meV for the self-energy case. The TQPT is labeled in the white dashed line at $ V_{\text{Z}}=1.02 $ meV. The corresponding wave functions and energy spectra are shown in Fig.~\ref{fig:wfgood}.}
	\label{fig:good}
\end{figure*}
\begin{figure*}[h]
	\centering
	\includegraphics[width=6.8in]{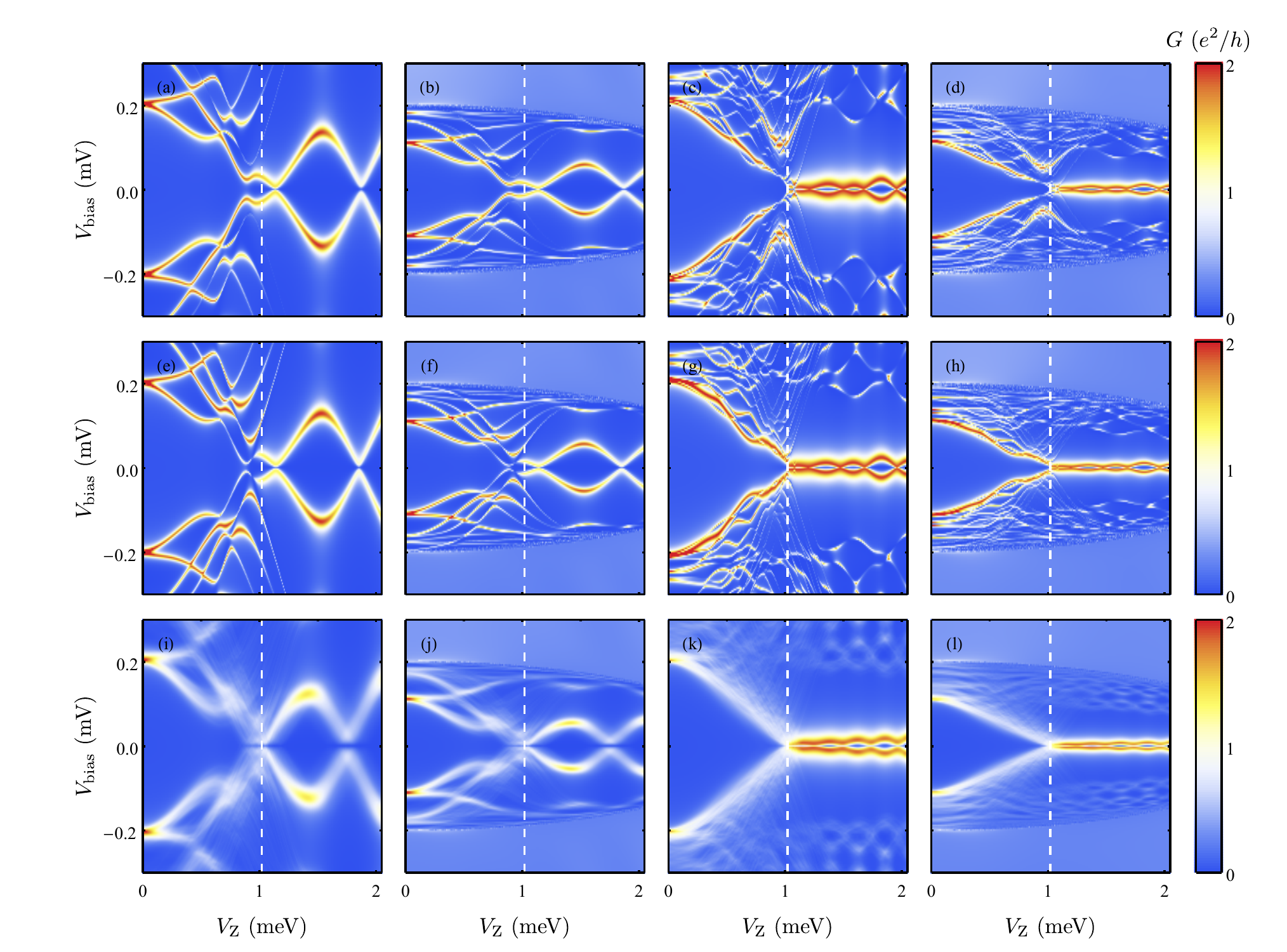}
	\caption{The good ZBCP in the presence of a small amount of disorder in the chemical potential for $ 1~\mu $m wires [shown in (a), (b), (e), (f), (i), and (j)] and $ 3~\mu $m wires [shown in (c), (d), (g), (h), (k), and (l)]. The color plots show the differential tunneling conductance $ G $ as a function of $ V_{\text{Z}} $ ($ x $ axis) and $ V_{\text{bias}} $ ($ y $ axis) measured from the left lead (in the first row) and the right lead (in the second row). The third row shows the disorder-averaged conductance over 200 samples; the first two rows are the conductance spectra under one specific configuration of disorder. The standard deviation of disorder in the chemical potential $ \sigma_{\mu}=0.4 $ meV for wires both with the self-energy shown in (b), (d), (f), (h), (j), and (l) and without the self-energy shown in (a), (c), (e), (g), (i), and (k). The SC gap collapse $ V_{\text{C}}=3$ meV for the self-energy case. The TQPT is labeled in the white dashed line at $ V_{\text{Z}}=1.02 $ meV. The corresponding wave functions and energy spectra are shown in Fig.~\ref{fig:wfsmall}.}
	\label{fig:small}
\end{figure*}
\begin{figure*}[h]
	\centering
	\includegraphics[width=6.8in]{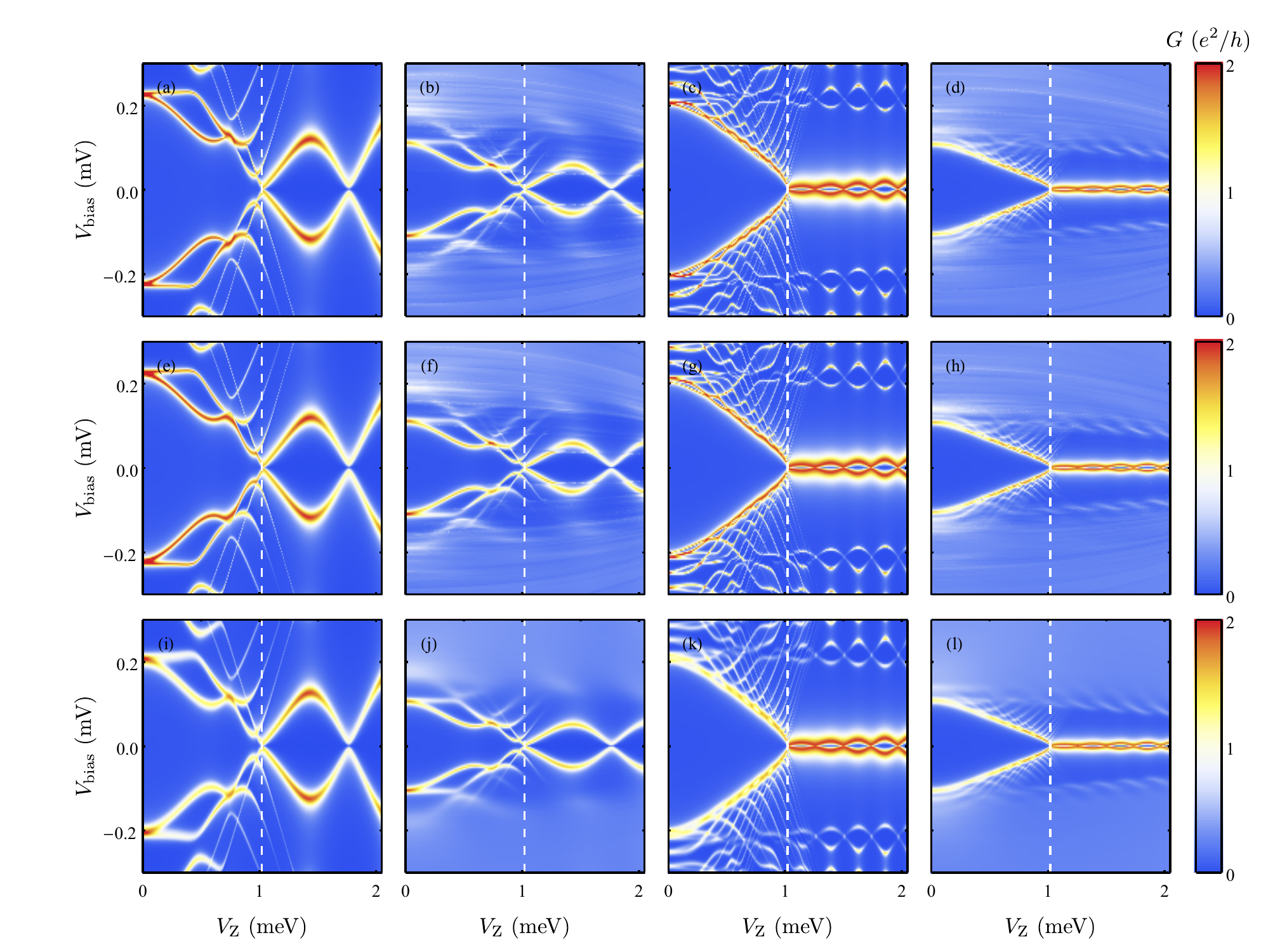}
	\caption{The good ZBCP in the presence of a small amount of disorder in the SC gap for  $ 1~\mu $m wires [shown in (a), (b), (e), (f), (i), and (j)] and $ 3~\mu $m wires [(shown in (c), (d), (g), (h), (k), and (l))]. The color plots show the differential tunneling conductance $ G $ as a function of $ V_{\text{Z}} $ ($ x $ axis) and $ V_{\text{bias}} $ ($ y $ axis) measured from the left lead (in the first row) and the right lead (in the second row). The third row shows the disorder-averaged conductance over 200 samples; the first two rows are the conductance spectra under one specific configuration of disorder. Mean proximity-indcued SC gap $ \Delta=0.2 $ meV/parent SC gap $ \Delta_0=0.2 $meV and the standard deviation of disorder in the SC gap $ \sigma_{\Delta}=0.06 $ meV are for wires both with the self-energy shown in (b), (d), (f), (h), (j), and (l) and without the self-energy shown in (a), (c), (e), (g), (i), and (k). The SC gap collapse $ V_{\text{C}}=3$ meV for the self-energy case. The TQPT is labeled in the white dashed line at $ V_{\text{Z}}=1.02 $ meV. The corresponding wave functions and energy spectra are shown in Fig.~\ref{fig:wfDeltaVar}.}
	\label{fig:DeltaVar}
\end{figure*}
\begin{figure*}[h]
	\centering
	\includegraphics[width=6.8in]{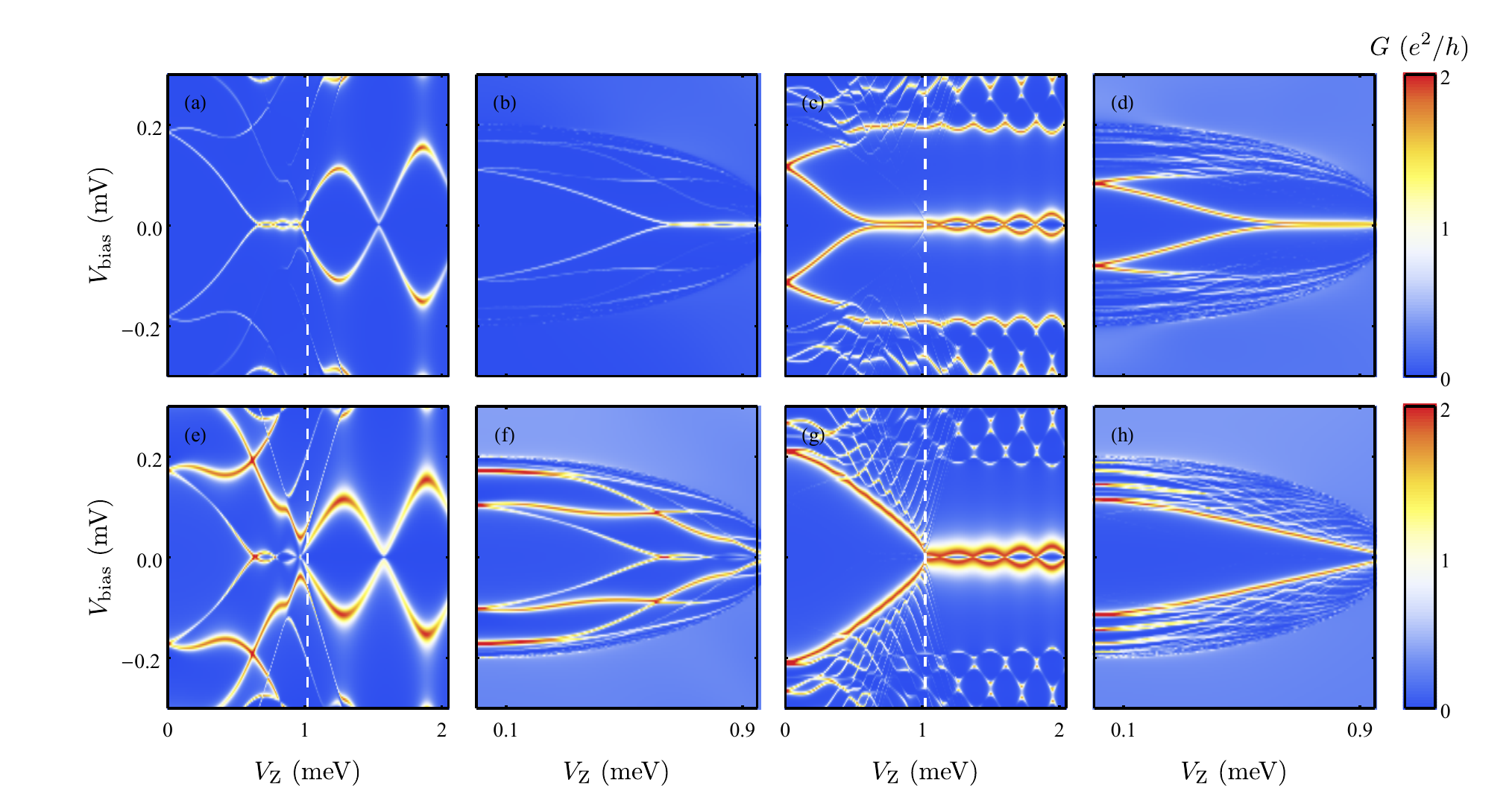}
	\caption{		
		The bad ZBCP due to the quantum dot in two $ 1~\mu $m pristine wires [shown in (a), (b), (e), and (f)] and two $ 3~\mu $m pristine wires [shown in (c), (d), (g), and (h)]. The color plots show the differential tunneling conductance $ G $ as a function of $ V_{\text{Z}} $ ($ x $ axis) and $ V_{\text{bias}} $ ($ y $ axis) measured from the left lead (in the first row) and the right lead (in the second row). For the short wire $ L=1~\mu $m, the peak value of the quantum dot $ V_{\text{D}}=1.7 $ meV and size $ l=0.2~\mu $m for wires both with the self-energy shown in (b) and (f) and without the self-energy shown in (a) and (e). For the long wire $ L=3~\mu $m,  the peak value of the quantum dot $ V_{\text{D}}=0.6 $ meV and size $ l=0.4~\mu $m for wires both with the self-energy shown in (d) and (h) and without the self-energy shown in (c) and (g). The SC gap collapse $ V_{\text{C}}=1 $ meV for the self-energy case. The TQPT is labeled in the white dashed line at $ V_{\text{Z}}=1.02 $ meV. The corresponding wave functions and energy spectra are shown in Figs.~\ref{fig:wfqd} and \ref{fig:wfqd3}.} 
	\label{fig:qd}
\end{figure*}
\begin{figure*}[h]
	\centering
	\includegraphics[width=6.8in]{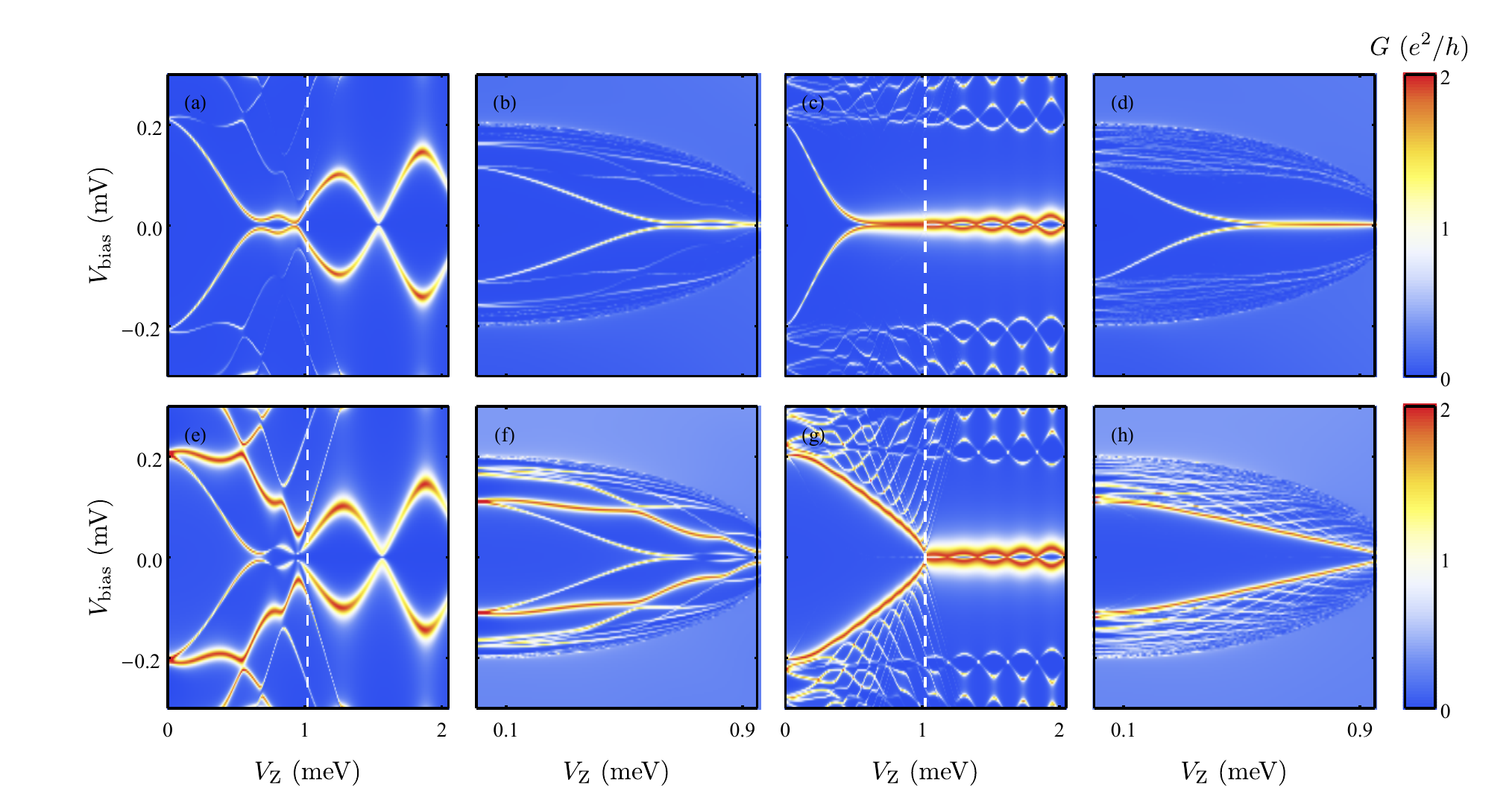}
	\caption{	
		The bad ZBCP due to the inhomogeneous potential in two $ 1~\mu $m pristine wires [shown in (a), (b), (e), and (f)] and two $ 3~\mu $m pristine wires [shown in (c), (d), (g), and (h)]. The color plots show the differential tunneling conductance $ G $ as a function of $ V_{\text{Z}} $ ($ x $ axis) and $ V_{\text{bias}} $ ($ y $ axis) measured from the left lead (in the first row) and the right lead (in the second row). For the short wire $ L=1~\mu $m, the peak value of the inhomogeneous potential $ V_{\text{max}}=1.4 $ meV and linewidth $ \sigma=0.15~\mu $m for wires both with the self-energy shown in (b) and (f) and without the self-energy shown in (a) and (e). For the long wire $ L=3~\mu $m, the peak value of the inhomogeneous potential $ V_{\text{max}}=1.2 $ meV and linewidth $ \sigma=0.4~\mu $m for wires both with the self-energy shown in (d) and (h) and without the self-energy shown in (c) and (g). The SC gap collapse $ V_{\text{C}}=1 $ meV for the self-energy case. The TQPT is labeled in the white dashed line at $ V_{\text{Z}}=1.02 $ meV. The corresponding wave functions and energy spectra are shown in Figs.~\ref{fig:wfinhom} and \ref{fig:wfinhom3}.
	}
	\label{fig:inhom}
\end{figure*}

\begin{figure*}[h]
	\centering
	\includegraphics[width=6.8in]{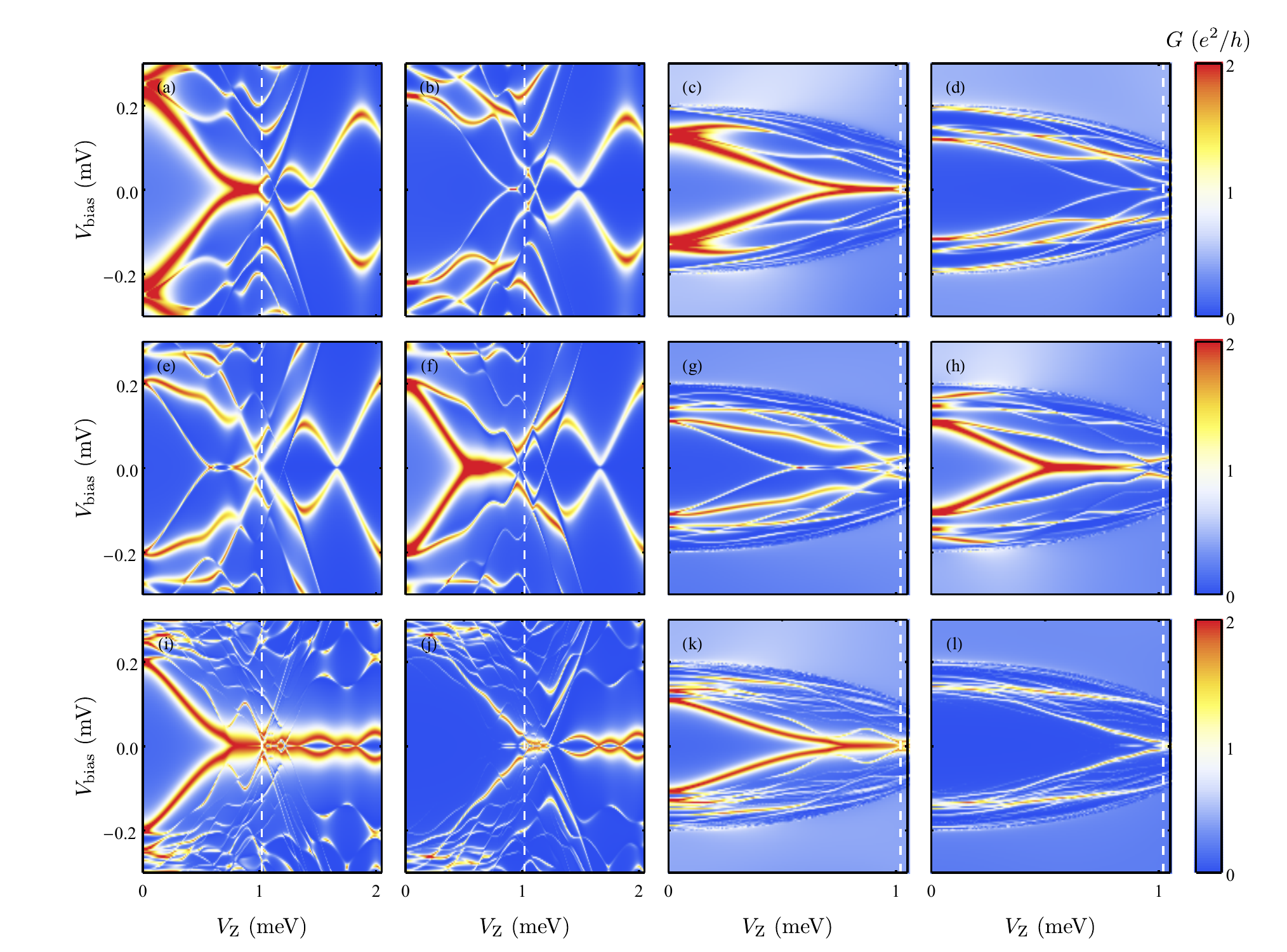}
	\caption{The ugly ZBCP in the presence of a large amount of disorder in the chemical potential for two $ 1~\mu $m wires (shown in the first two rows) and for one $ 3~\mu $m wire (shown in the third row). The color plots show the differential tunneling conductance $ G $ as a function of $ V_{\text{Z}} $ ($ x $ axis) and $ V_{\text{bias}} $ ($ y $ axis) measured from the left lead (shown in the first and third columns) and the right lead (shown in the second and fourth columns). The conductance spectra in the first row share a common configuration of disorder; the ones in the second row share another. The standard deviation of the chemical potential $ \sigma_{\mu}=1 $ meV for wires both with the self-energy shown in (c), (d), (g), and (h) and without the self-energy shown in (a), (b), (e), and (f). For $ L=3~\mu $m wire, the standard deviation of the chemical potential $ \sigma_{\mu}=1 $ for wires both with the self-energy shown in (k) and (l) and without the self-energy shown in (i) and (j).  The SC gap collapse $ V_{\text{C}}=1.2 $ meV for the self-energy case. The TQPT is labeled in the white dashed line at $ V_{\text{Z}}=1.02 $ meV. The corresponding wave functions and energy spectra are shown in Figs.~\ref{fig:wfmuVar} and \ref{fig:wfmuVar3}.}
	\label{fig:muVar}
\end{figure*}

\begin{figure*}[h]
	\centering
	\includegraphics[width=6.8in]{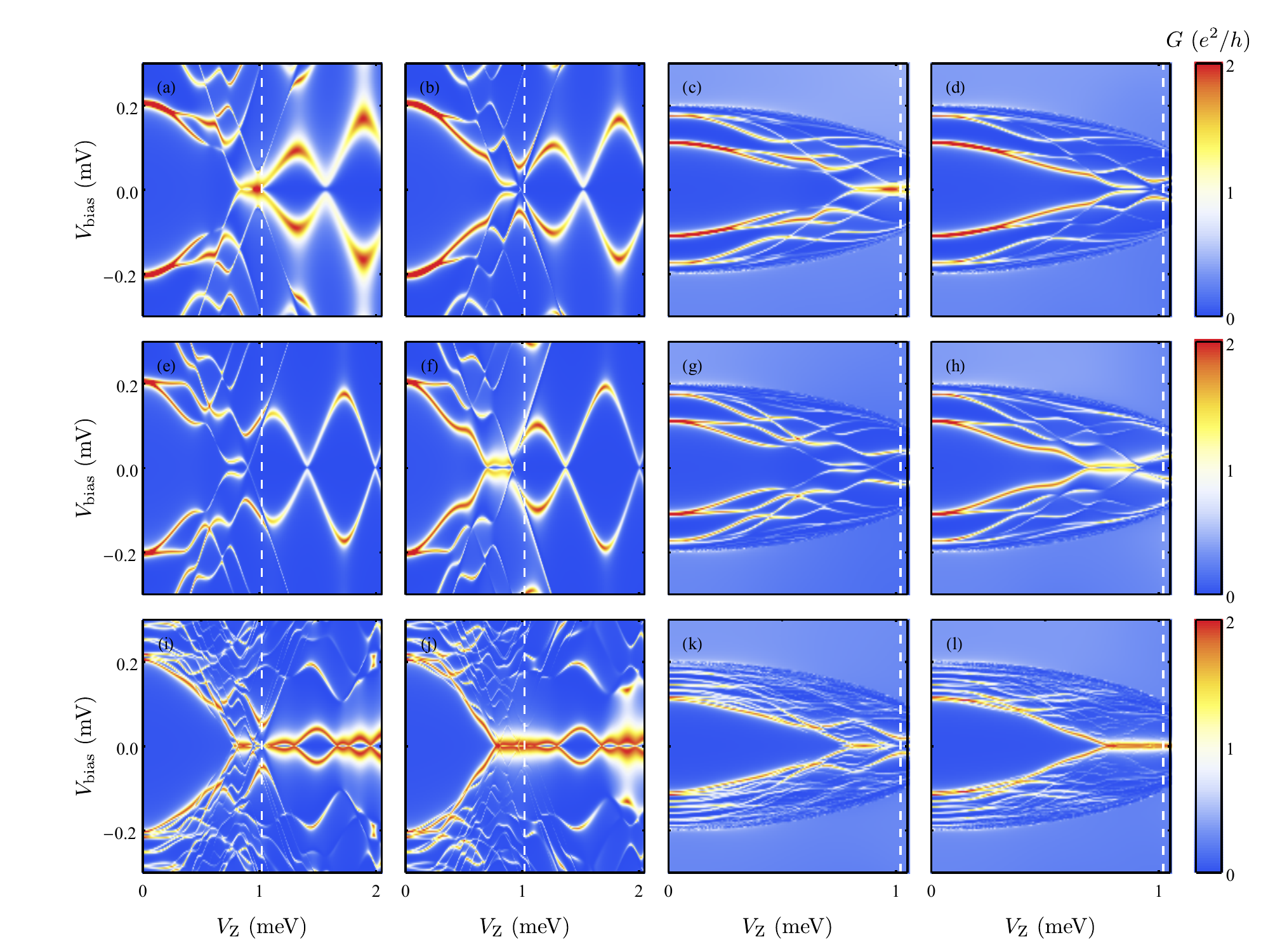}
	\caption{The ugly ZBCP in the presence of disorder in the effective $ g $ factor for two $ 1~\mu $m wires (shown in the first two rows) and for one $ 3~\mu $m wire (shown in the third row). The color plots show the differential tunneling conductance $ G $ as a function of $ V_{\text{Z}} $ ($ x $ axis) and $ V_{\text{bias}} $ ($ y $ axis) measured from the left lead (shown in the first and third columns) and the right lead (shown in the second and fourth columns). The conductance spectra in the first row share a common configuration of disorder; the ones in the second row share another. The standard deviation of disorder in the effective $ g $ factor is $ \sigma_g=0.8 $ for wires both with self-energy shown in (c), (d), (g), and (h) and without the self-energy shown in  (a), (b), (e), and (f). For $ L=3~\mu $m wire, the effective $ g $ factor is $ \sigma_g=0.6 $ for wires both with the self-energy shown in (k) and (l) and without the self-energy shown in (i) and (j). The SC gap collapse $ V_{\text{C}}=1.2 $ meV for the self-energy case. The TQPT is labeled in the white dashed line at $ V_{\text{Z}}=1.02 $ meV. The corresponding wave functions and energy spectra are shown in Figs.~\ref{fig:wfgVar} and \ref{fig:wfgVar3}.}
	\label{fig:gVar}
\end{figure*}

\begin{figure*}[h]
	\centering
	\includegraphics[width=6.8in]{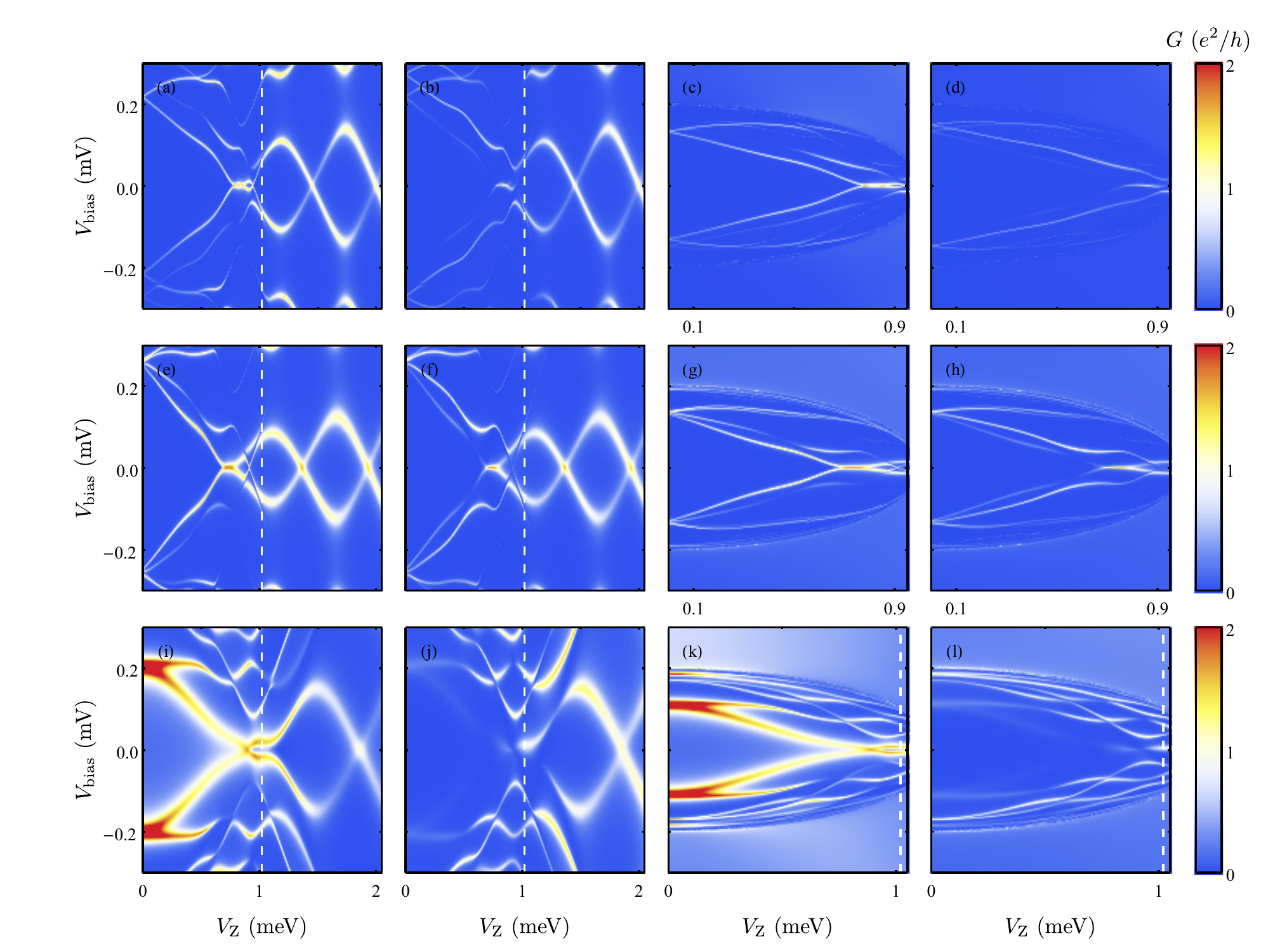}
	\caption{The short but uncorrelated occasions in bad (the first and second rows) and ugly (the third row) ZBCPs. The color plots show the differential tunneling conductance $ G $ as a function of $ V_{\text{Z}} $ ($ x $ axis) and $ V_{\text{bias}} $ ($ y $ axis) measured from the left lead (in the first and third columns) and the right lead (in the second and fourth columns). The bad ZBCPs due to the two quantum dots at both ends in $ 1~\mu $m wires are shown in (a)-(d). The peak value of the quantum dot is $ V_{\text{D,L}}=1.7 $ meV on the left and $ V_{\text{D,R}}=2.3 $ meV on the right, the size of the quantum dot is $ l_{\text{L}}=0.2~\mu $m on the left  and $ l_{\text{R}}=0.15~\mu $m on the right for both wires with the self-energy shown in (c) and (d) and without the self-energy shown in (a) and (b). The SC collapse $ V_{\text{C}}=1 $ meV. The bad ZBCPs due to the inhomogeneous potential with the Gaussian barriers on both ends are shown in (e), (f), (g), and (h). The peak value of the Gaussian barrier is $ V_{\text{max,L}}=1.4 $ meV on the left and $ V_{\text{max,R}}=1.9 $ meV on the right, the linewidth of the Gaussian barrier is $ \sigma_{\text{L}}=0.15~\mu$m on the left and $ \sigma_{\text{R}}=0.1~\mu $m on the right for both wires with the self-energy shown in (g) and (h) and without the self-energy shown in (e) and (f). The SC collapse $ V_{\text{C}}=1 $ meV. The ugly ZBCPs due to the disorder in the chemical potential are shown in (i), (j), (k), and (l). The standard deviation of the chemical potential $ \sigma_{\mu}=1 $ meV for wires both with the self-energy shown in (k) and (l) and without the self-energy shown in (i) and (j). The SC collapse $ V_{\text{C}}=1.2 $ meV. The corresponding wave functions and energy spectra are shown in Fig.~\ref{fig:wfuncorr}.}
	\label{fig:uncorr}
\end{figure*}

\clearpage
\onecolumngrid
\section{Energy spectra and wave functions}\label{app:B}
In this section, the energy spectra as a function of the Zeeman field $ V_{\text{Z}} $ are shown in the first column and the corresponding wave functions at several representative $ V_{\text{Z}} $ in the Majorana basis defined in Eq.~\eqref{eq:majoranabasis} are presented in the second to the fourth columns. In energy spectra, the energies have identical ranges as those in conductance spectra and the red dashed lines are for the nominal TQPT. The two Majoranas are labeled with blue and cyan in the lowest state while red and orange in the second state.
\begin{figure*}[h]
	\centering
	\includegraphics[width=6.8in]{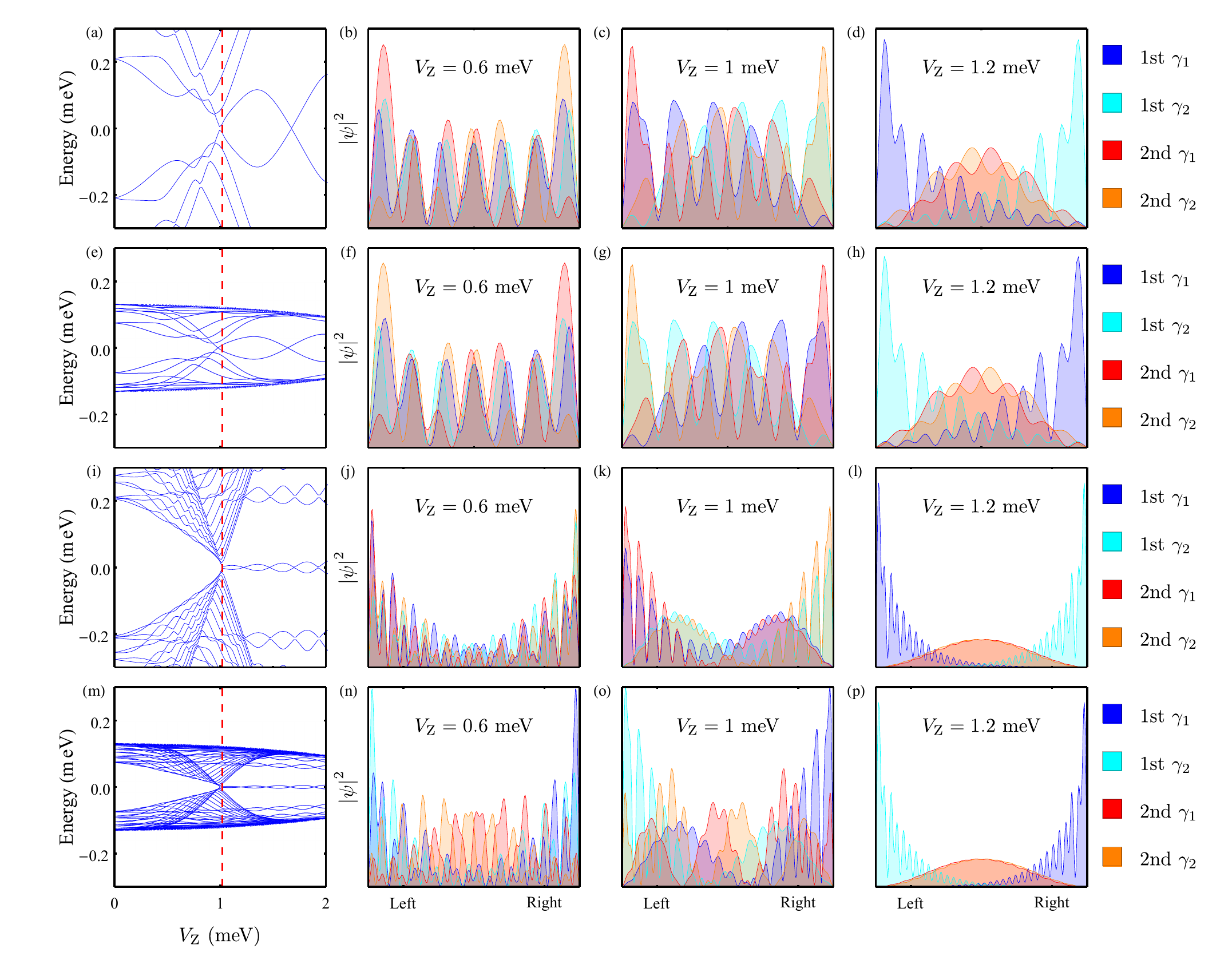}
	\caption{(a), (b), (c), and (d) correspond to Figs.~\ref{fig:good}(a) and~\ref{fig:good}(e). (e), (f), (g), and (h) correspond to Figs.~\ref{fig:good}(b) and~\ref{fig:good}(f). (i), (j), (k), and (l) correspond to Figs.~\ref{fig:good}(c) and~\ref{fig:good}(g). (m), (n), (o), and (p) correspond to Figs.~\ref{fig:good}(d) and~\ref{fig:good}(h).}
	\label{fig:wfgood}	
\end{figure*}
\begin{figure*}[h]
	\centering
	\includegraphics[width=6.8in]{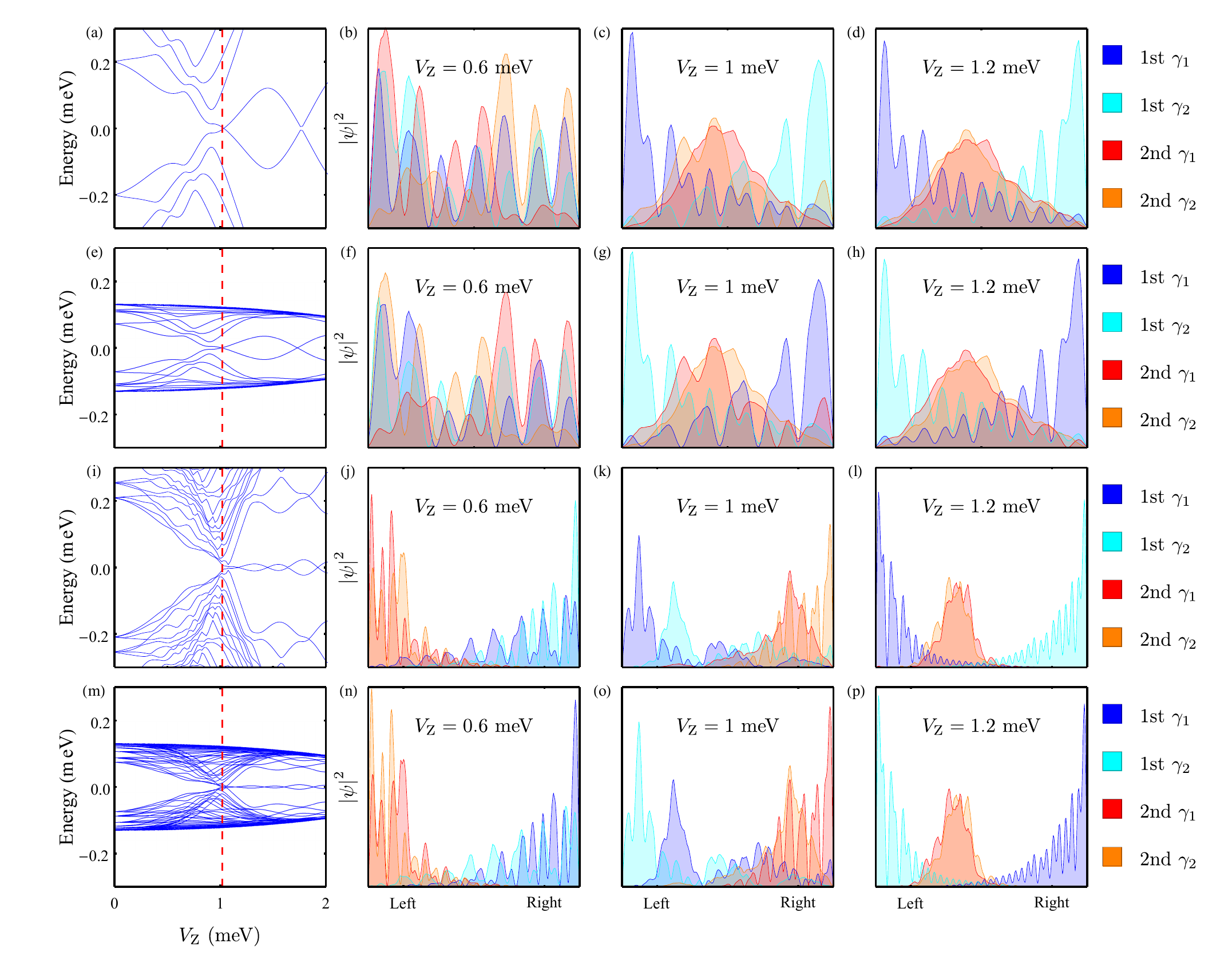}
	\caption{(a), (b), (c), and (d) correspond to Figs.~\ref{fig:small}(a) and~\ref{fig:small}(e). (e), (f), (g), and (h) correspond to Figs.~\ref{fig:small}(b) and~\ref{fig:small}(f). (i), (j), (k), and (l) correspond to Figs.~\ref{fig:small}(c) and~\ref{fig:small}(g). (m), (n), (o), and (p) correspond to Figs.~\ref{fig:small}(d) and~\ref{fig:small}(h).}
	\label{fig:wfsmall}	
\end{figure*}
\begin{figure*}[h]
	\centering
	\includegraphics[width=6.8in]{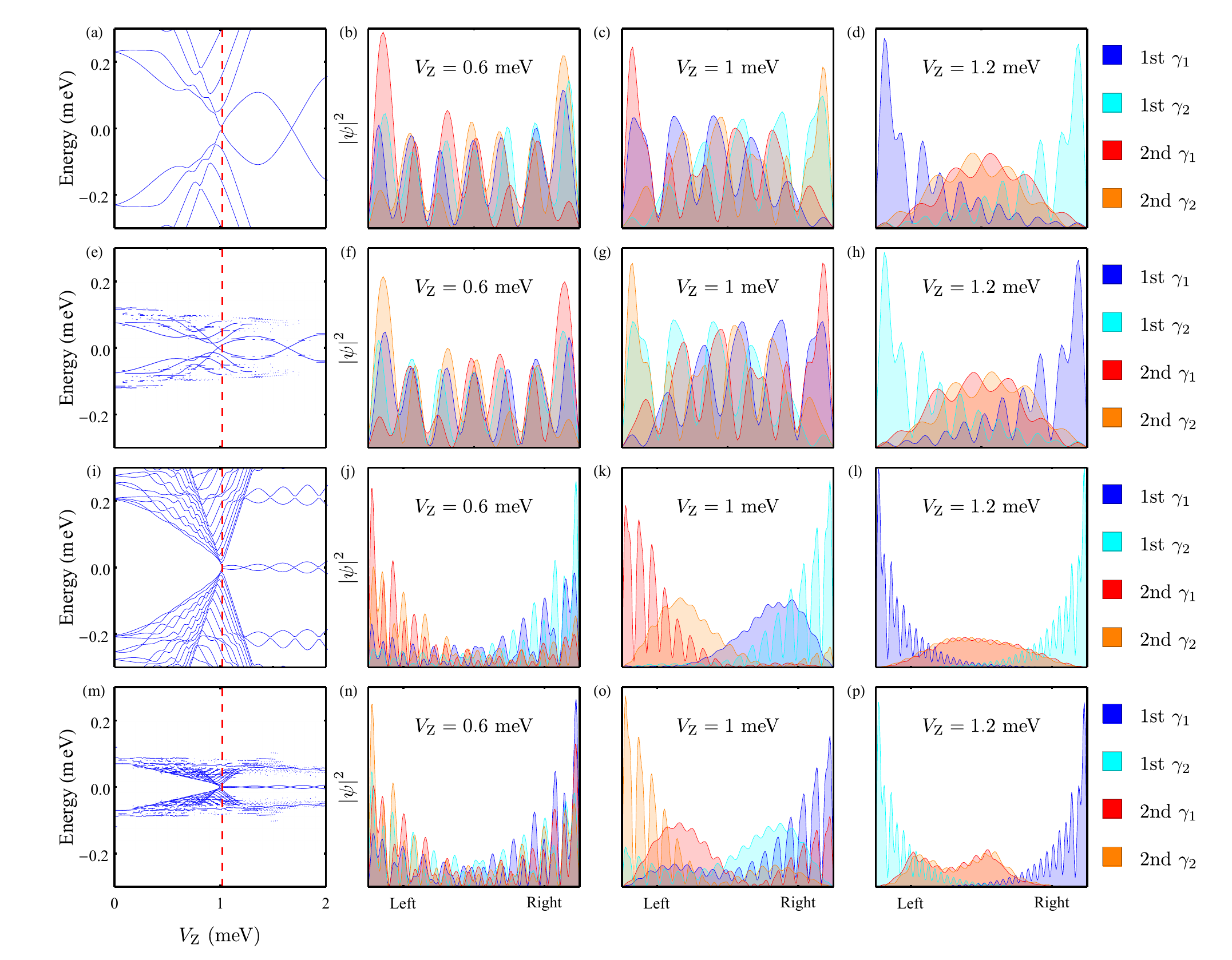}
	\caption{(a), (b), (c), and (d) correspond to Figs.~\ref{fig:DeltaVar}(a) and~\ref{fig:DeltaVar}(e). (e), (f), (g), and (h) correspond to Figs.~\ref{fig:DeltaVar}(b) and~\ref{fig:DeltaVar}(f). (i), (j), (k), and (l) correspond to Figs.~\ref{fig:DeltaVar}(c) and~\ref{fig:DeltaVar}(g). (m), (n), (o), and (p) correspond to Figs.~\ref{fig:DeltaVar}(d) and~\ref{fig:DeltaVar}(h).}
	\label{fig:wfDeltaVar}	
\end{figure*}
\begin{figure*}[h]
	\centering
	\includegraphics[width=6.8in]{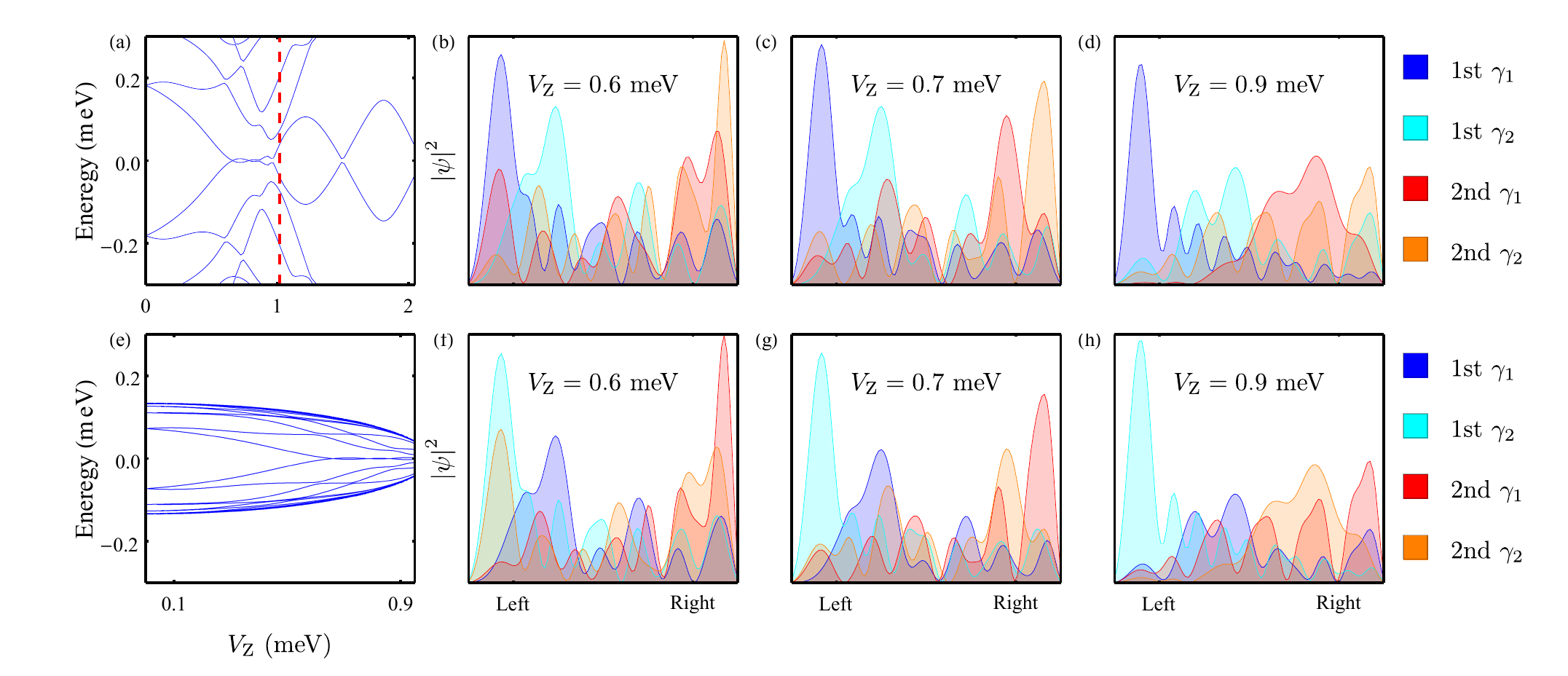}
	\caption{(a), (b), (c), and (d) correspond to Figs.~\ref{fig:qd}(a) and~\ref{fig:DeltaVar}(e). (e), (f), (g), and (h) correspond to Figs.~\ref{fig:qd}(b) and~\ref{fig:DeltaVar}(f).}
	\label{fig:wfqd}	
\end{figure*}
\begin{figure*}[h]
	\centering
	\includegraphics[width=6.8in]{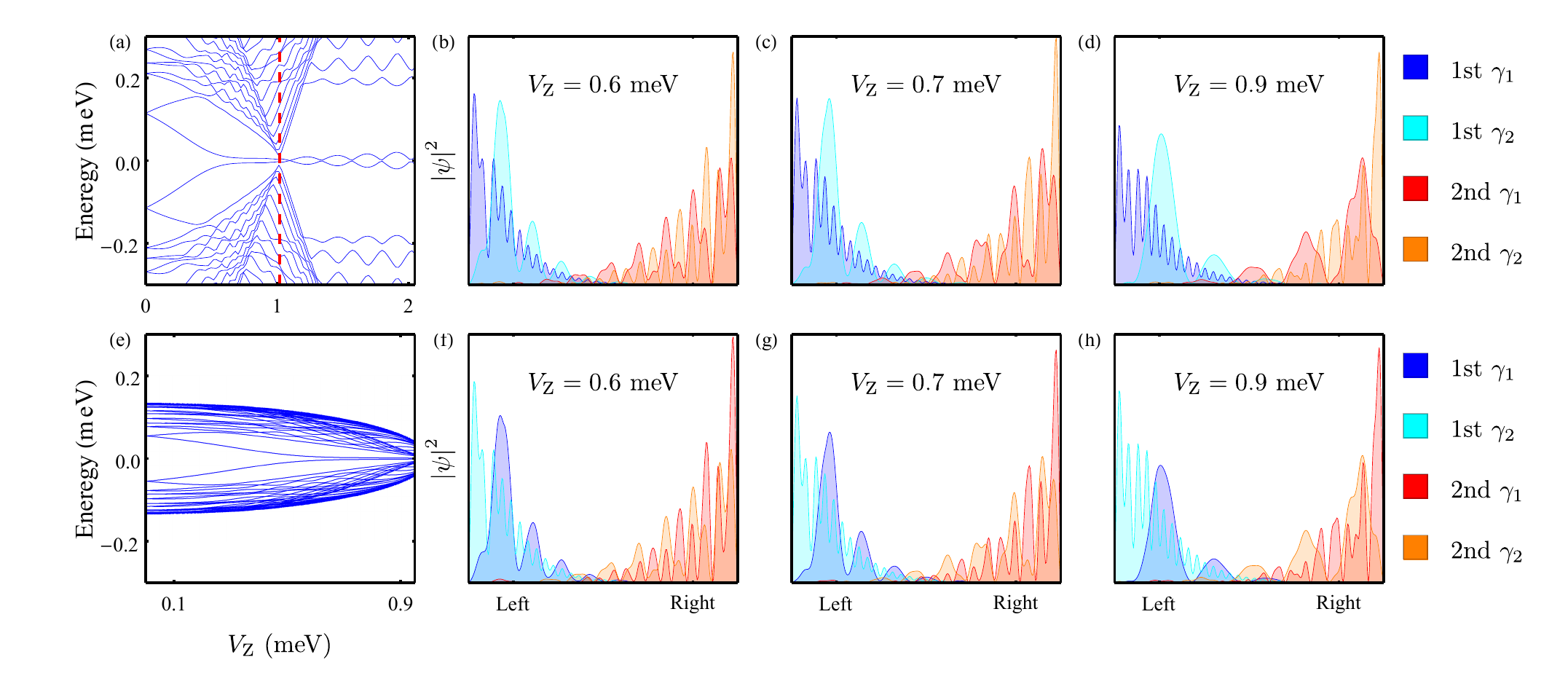}
	\caption{(a), (b), (c), and (d) correspond to Figs.~\ref{fig:qd} (c) and~\ref{fig:qd}(g). (e), (f), (g), and (h) correspond to Figs.~\ref{fig:qd}(d) and~\ref{fig:qd}(h).}
	\label{fig:wfqd3}	
\end{figure*}
\begin{figure*}[h]
	\centering
	\includegraphics[width=6.8in]{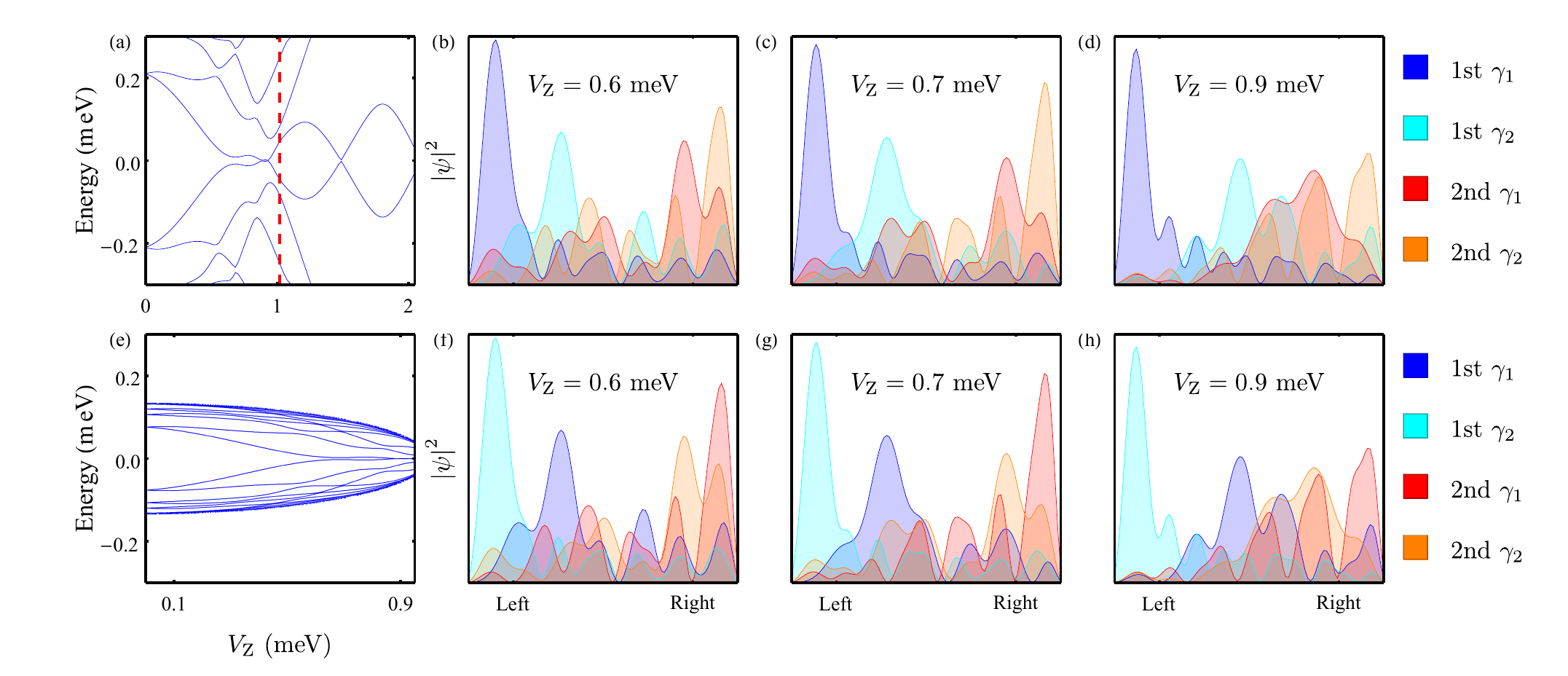}
	\caption{(a), (b), (c), and (d) correspond to Figs.~\ref{fig:inhom}(a) and~\ref{fig:inhom}(e). (e), (f), (g), and (h) correspond to Figs.~\ref{fig:inhom}(b) and~\ref{fig:inhom}(f).}
	\label{fig:wfinhom}	
\end{figure*}
\begin{figure*}[h]
	\centering
	\includegraphics[width=6.8in]{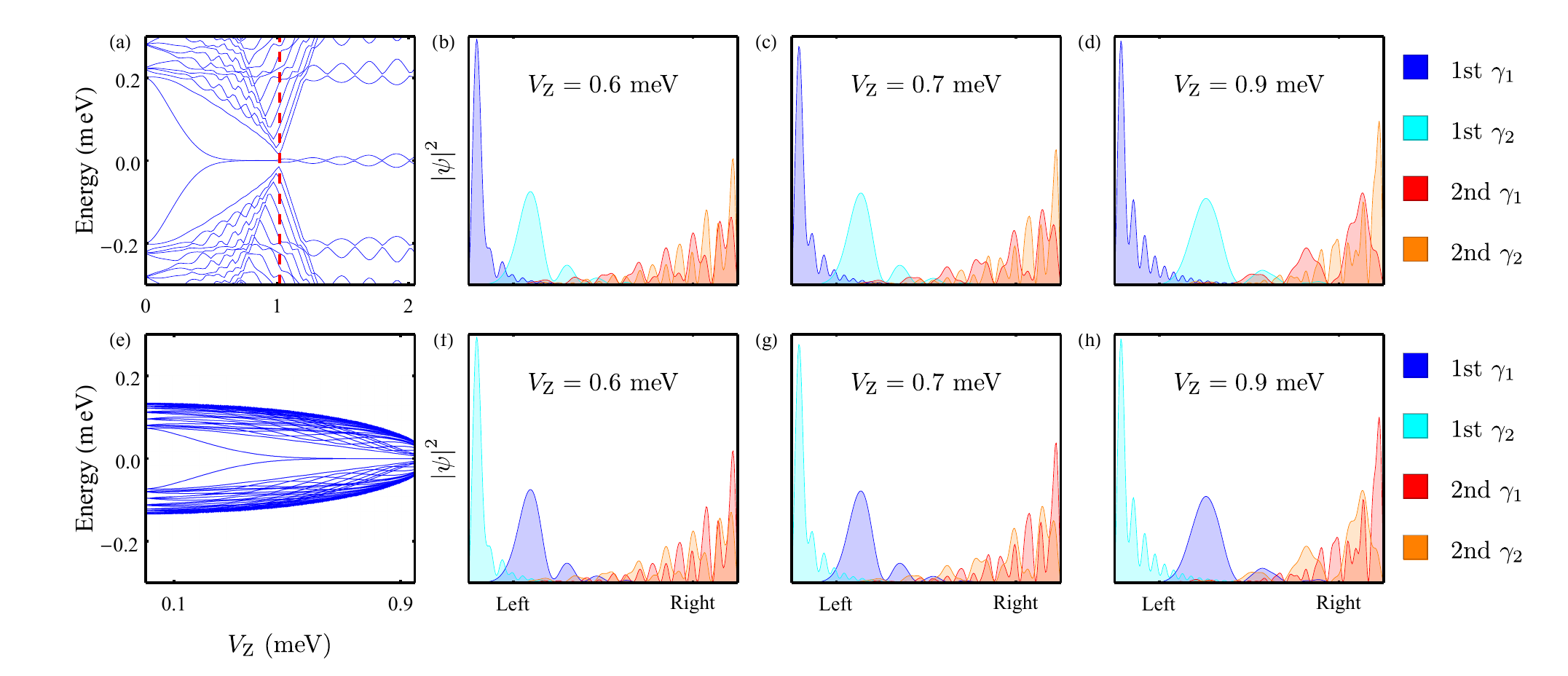}
	\caption{(a), (b), (c), and (d) correspond to Figs.~\ref{fig:inhom}(c) and~\ref{fig:inhom}(g). (e), (f), (g), and (h) correspond to Figs.~\ref{fig:inhom}(d), and (h).}
	\label{fig:wfinhom3}	
\end{figure*}
\begin{figure*}[h]
	\centering
	\includegraphics[width=6.8in]{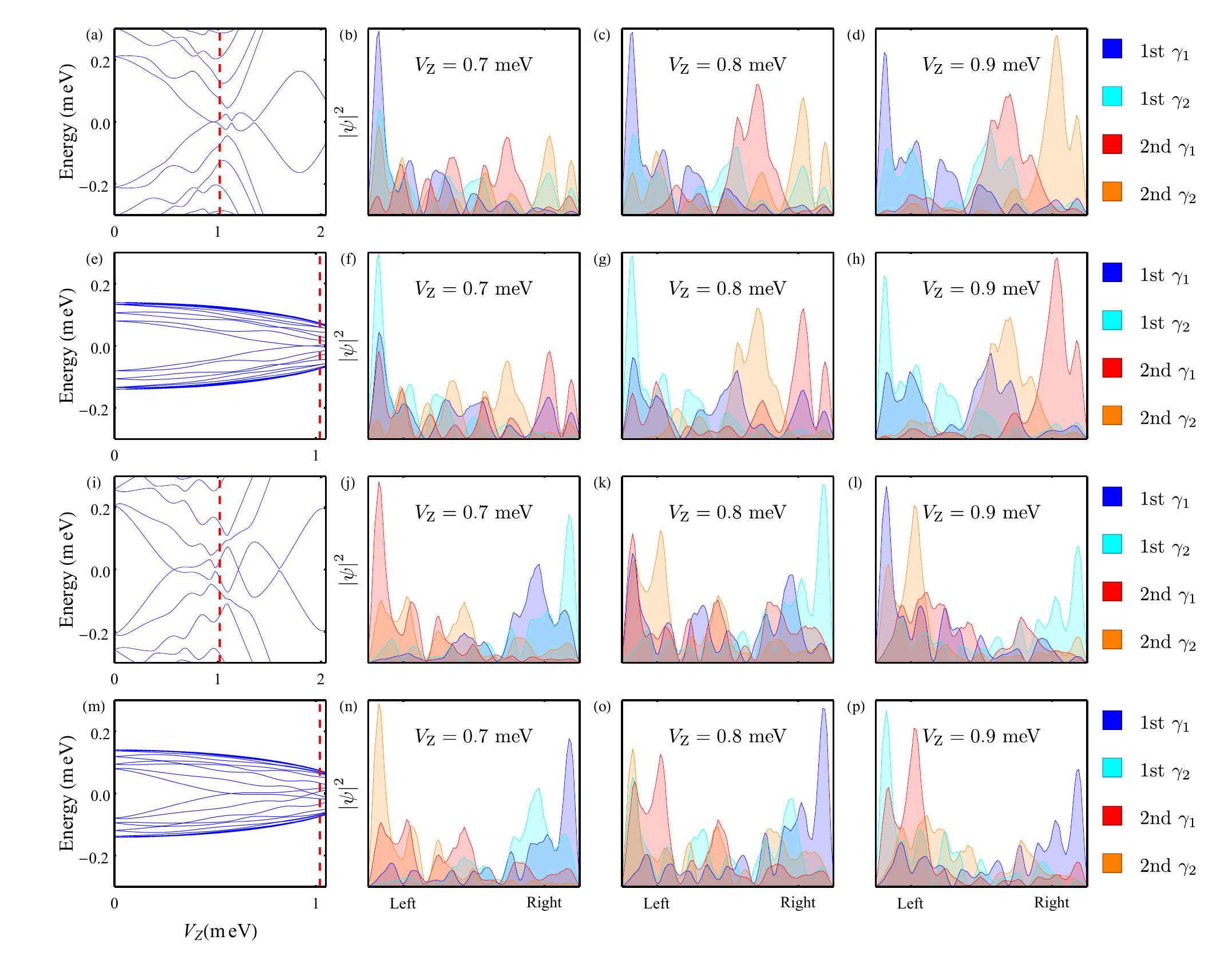}
	\caption{(a), (b), (c), (d) correspond to Figs.~\ref{fig:muVar}(a) and~\ref{fig:muVar}(b). (e), (f), (g), and (h) correspond to Figs.~\ref{fig:muVar}(c) and~\ref{fig:muVar}(d). (i), (j), (k), (l) correspond to Figs.~\ref{fig:muVar}(e) and~\ref{fig:muVar}(f). (m), (n), (o), and (p) correspond to Figs.~\ref{fig:muVar}(g) and~\ref{fig:muVar}(h).}
	\label{fig:wfmuVar}	
\end{figure*}
\begin{figure*}[h]
	\centering
	\includegraphics[width=6.8in]{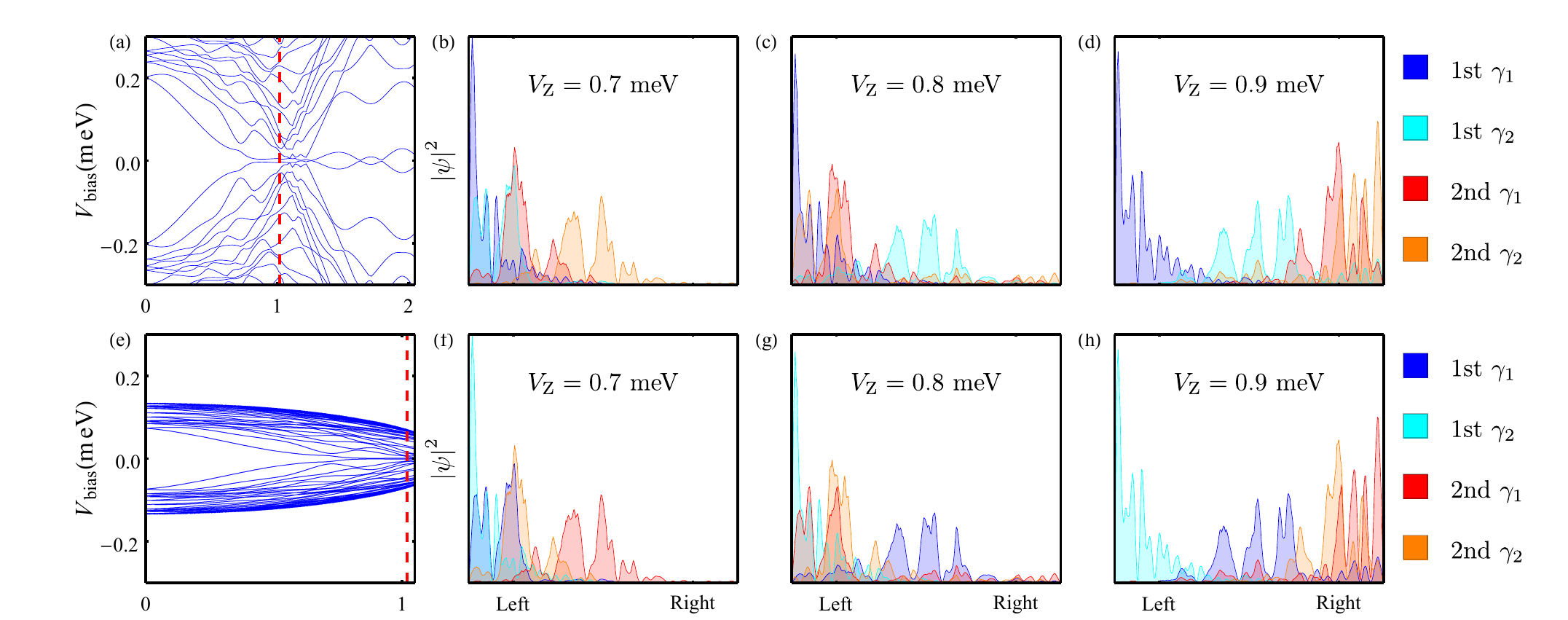}
	\caption{(a), (b), (c), and (d) correspond to Figs.~\ref{fig:muVar}(i) and~\ref{fig:muVar}(j). (e), (f), (g), and (h) correspond to Figs.~\ref{fig:muVar}(k) and~\ref{fig:muVar}(l).}
	\label{fig:wfmuVar3}	
\end{figure*}
\begin{figure*}[h]
	\centering
	\includegraphics[width=6.8in]{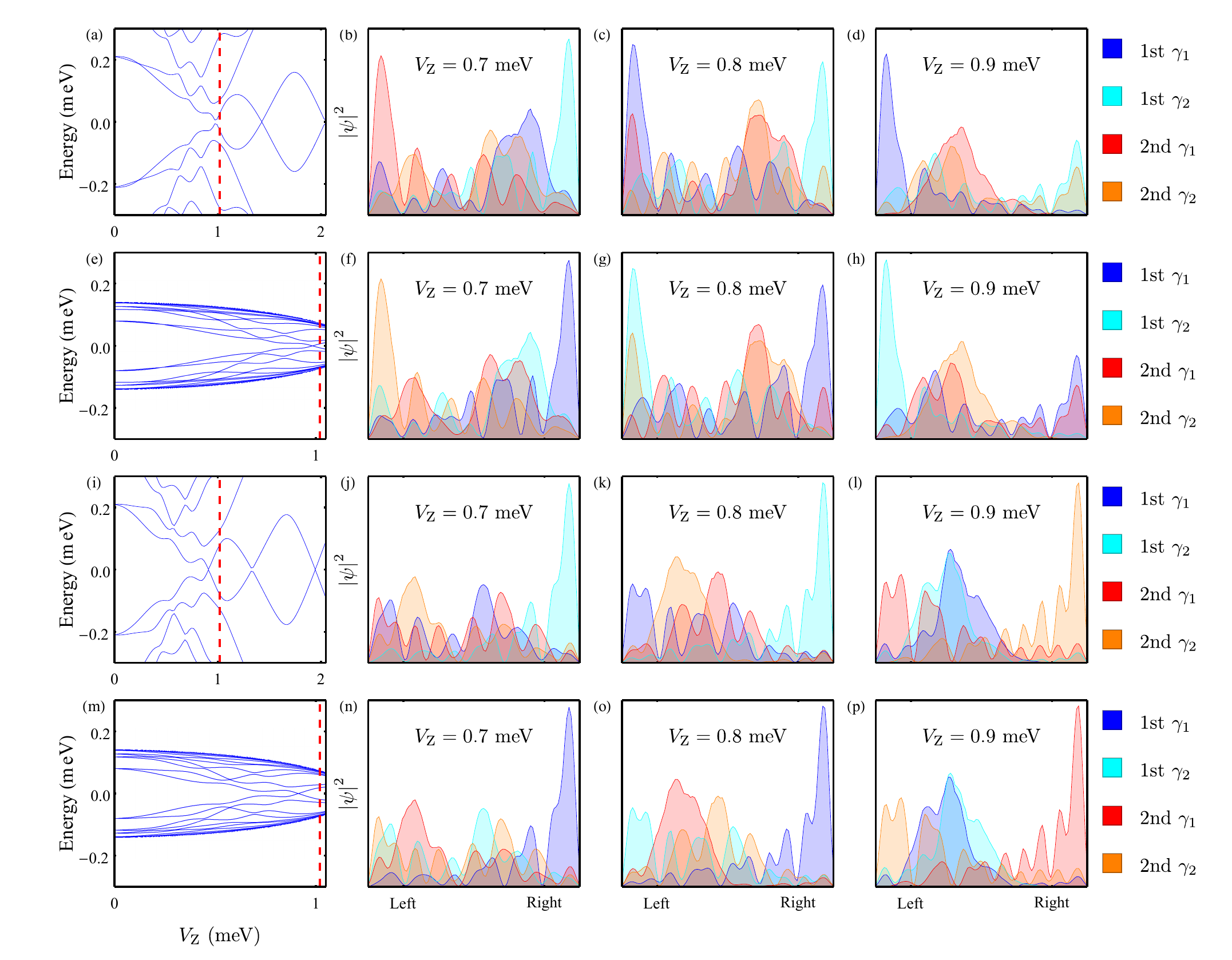}
	\caption{(a), (b), (c), and (d) correspond to Figs.~\ref{fig:gVar}(a) and~\ref{fig:gVar}(b). (e), (f), (g), and (h) correspond to Figs.~\ref{fig:gVar}(c) and~\ref{fig:gVar}(d). (i), (j), (k), and (l) correspond to Figs.~\ref{fig:gVar}(e) and~\ref{fig:gVar}(f). (m), (n), (o), and (p) correspond to Figs.~\ref{fig:gVar}(g) and~\ref{fig:gVar}(h).}
	\label{fig:wfgVar}	
\end{figure*}
\begin{figure*}[h]
	\centering
	\includegraphics[width=6.8in]{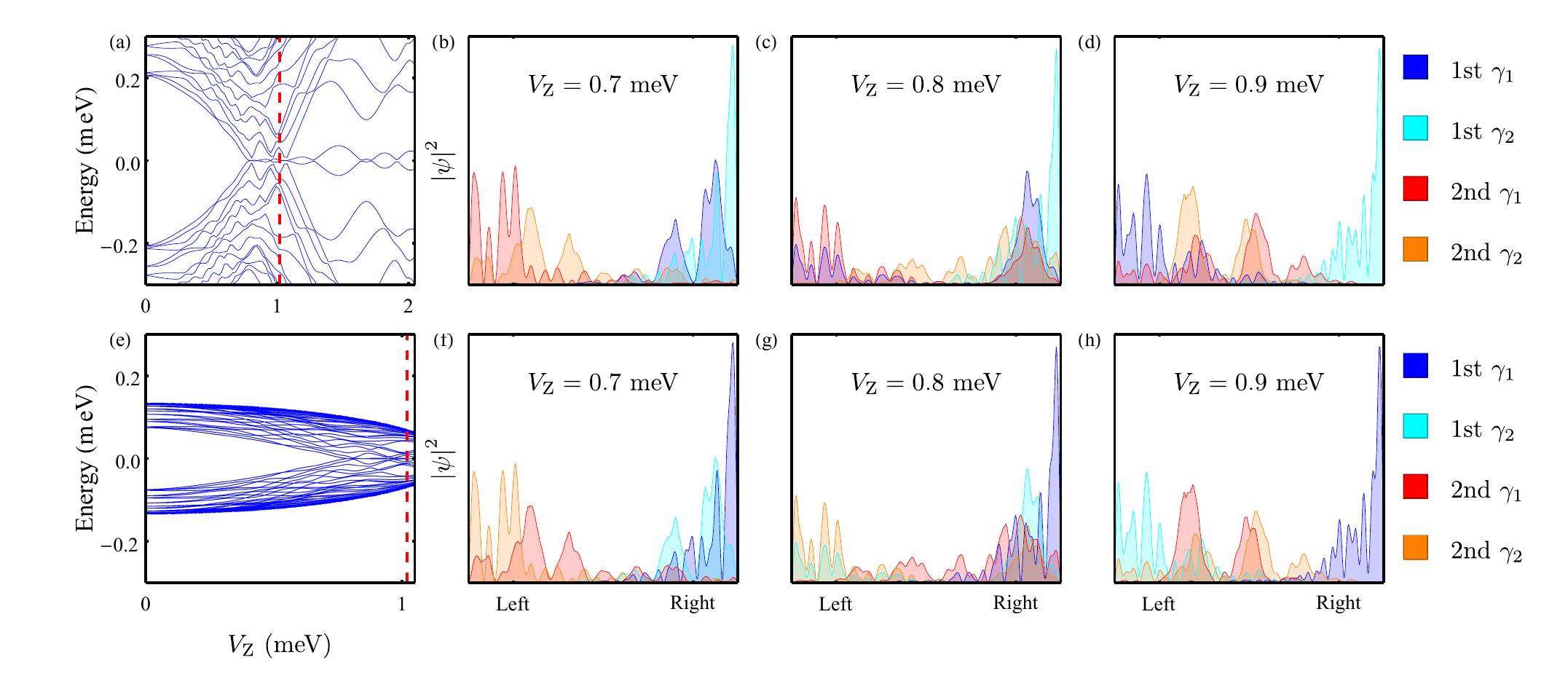}
	\caption{(a), (b), (c), and (d) correspond to Figs.~\ref{fig:gVar}(i) and~\ref{fig:gVar}(j). (e), (f), (g), and (h) correspond to Figs.~\ref{fig:gVar}(k) and~\ref{fig:gVar}(l).}
	\label{fig:wfgVar3}	
\end{figure*}
\begin{figure*}[h]
	\centering
	\includegraphics[width=6.8in]{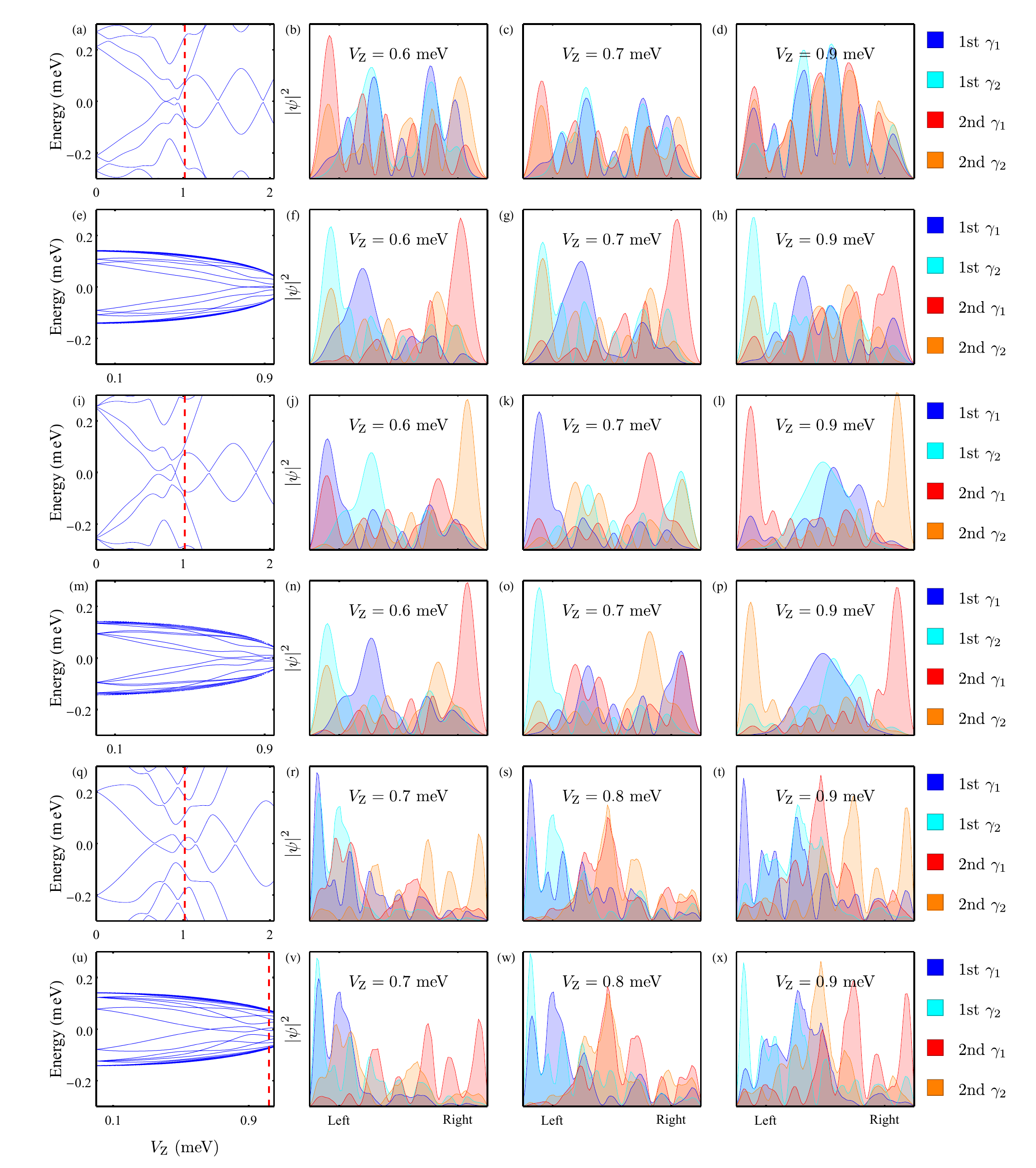}
	\caption{(a), (b), (c), and (d) correspond to Figs.~\ref{fig:uncorr}(a) and~\ref{fig:uncorr}(b). 
		(e), (f), (g), and (h) correspond to Figs.~\ref{fig:uncorr}(c) and~\ref{fig:uncorr}(d).
		 (i), (j), (k), and (l) correspond to Figs.~\ref{fig:uncorr}(e) and~\ref{fig:uncorr}(f). 
		 (m), (n), (o), and (p) correspond to Figs.~\ref{fig:uncorr}(g) and~\ref{fig:uncorr}(h). 
	(q), (r), (s), and (t) correspond to Figs.~\ref{fig:uncorr}(i) and~\ref{fig:uncorr}(j). 
(u), (v), (w), and (x) correspond to Figs.~\ref{fig:uncorr}(k) and~\ref{fig:uncorr}(l).}
\label{fig:wfuncorr}
\end{figure*}
\clearpage
\twocolumngrid
\bibliography{goodbadugly}

\end{document}